\documentclass[aps,
prc,
twocolumn,
nofootinbib,
noshowkeys,
noshowpacs,
superscriptaddress,
floatfix
]{revtex4-2}
\pdfoutput=1

\usepackage{dcolumn}
\usepackage{bm}
\usepackage{slashed}
\usepackage{graphicx}
\usepackage[colorlinks=true,linktocpage=true,linkcolor=blue,citecolor=red,urlcolor=blue]{hyperref}
\usepackage{float}
\usepackage{nicefrac}
\usepackage[normalem]{ulem}
\usepackage{amsmath}
\usepackage{subfigure} 
\usepackage{mathtools,amssymb}
\usepackage{slashed}

\newcommand{\beq}{\begin{eqnarray}}
	\newcommand{\eeq}{\end{eqnarray}}

\def\a{\alpha}
\def\b{\beta}
\def\g{\gamma}
\def\d{\delta}

\def\Aone{{ \cal A}_1 }
\def\Atwo{{ \cal A}_2 }
\def\Athree{{ \cal A}_3 }

\def\nn{\nonumber}

\graphicspath{{./figs/}}
\begin{document}
	
	\title{ 
		Spin dynamics with realistic hydrodynamic background \\ for relativistic heavy-ion collisions}
	\author{Sushant K. Singh} 
	\email{sushant7557@gmail.com}
	\affiliation{Department of Physics \& Astronomy, University of Florence, Via G. Sansone 1, 50019 Sesto Fiorentino, Florence, Italy}
	\affiliation{Variable Energy Cyclotron Centre, 1/AF, Bidhan Nagar, Kolkata 700064, India}
	\author{Radoslaw Ryblewski}
	\email{radoslaw.ryblewski@ifj.edu.pl}
	\affiliation{Institute of Nuclear Physics Polish Academy of Sciences, PL-31342 Krakow, Poland}
	\author{Wojciech Florkowski} 
	\email{wojciech.florkowski@uj.edu.pl}
	\affiliation{Institute of Theoretical Physics, Jagiellonian University, ul. St. \L ojasiewicza 11, 30-348 Krakow, Poland}
	\date{\today} 
	\bigskip
\begin{abstract}
The equations of perfect spin hydrodynamics are solved for the first time using a realistic (3+1)-dimensional hydrodynamic background, calibrated to reproduce a comprehensive set of hadronic observables, including rapidity distributions, transverse momentum spectra, and elliptic flow coefficients for Au+Au collisions at the beam energy of $\sqrt{s_{\rm NN}} = 200$~GeV. The spin dynamics is governed by the conservation of the spin tensor, describing spin-\nicefrac{1}{2} particles, with particle mass in the spin tensor treated as an effective parameter. We investigate several scenarios, varying both the effective mass and the initial evolution time for the spin polarization tensor. The model predictions are then compared with experimental measurements of global and longitudinal spin polarization of $\Lambda$ hyperons. Our results indicate that a successful description of the data requires a delayed initial evolution time for the perfect spin hydrodynamics of about 4~fm/\textit{c} (in contrast to the standard initial time of 1 fm/\textit{c} used for the hydrodynamic background). This delay marks a transition from the phase where spin-orbit interaction is significant to the regime where spin-conserving processes dominate. Our findings suggest that the spin-orbit dissipative interaction plays a significant role only in the very early stages of the system's evolution.
\end{abstract}
	
\date{\today}  
\pacs{ }
	
\maketitle
	
\section{Introduction}
	
The measurements of the spin polarization of the $\Lambda$ hyperons and vector mesons has opened completely new possibilities for studying the properties of hot and dense matter produced in heavy-ion collisions~\cite{HADES:2014ttv, STAR:2017ckg, STAR:2018gyt, STAR:2019erd, ALICE:2019aid, STAR:2021beb, ALICE:2019onw, Niida:2024ntm} (see Ref.~\cite{Becattini2024_IJMPE} for a theory review). However,  despite the increasing availability of data and plans for further measurements~\cite{Bondar:2024uqk,Sorensen:2023zkk}, there remains a lack of consensus on the theoretical side regarding potential explanations for the observed polarization effects. The most widely adopted approach originates from quantum kinetic theory, resulting in a framework where polarization effects are determined solely by gradients of standard hydrodynamic variables at freeze-out. Notably, at least two significantly different implementations of this scheme yield satisfactory descriptions of the $\Lambda$ polarization data~\cite{Liu:2021uhn, Fu:2021pok, Becattini:2021suc, Becattini:2021iol, Alzhrani:2022dpi}, leaving uncertainity about the true physical mechanism responsible for the observed polarization phenomena.
	
An alternative approach to those mentioned above is the framework of spin hydrodynamics, which has been developed over several years by independent research groups worldwide, employing a variety of physical concepts~\cite{Becattini:2009wh,Becattini:2011zz, Becattini:2013fla, Becattini:2013vja, Karpenko:2016jyx, Becattini:2016gvu, Becattini:2017gcx, Liu:2021uhn, Fu:2021pok, Becattini:2021suc, Becattini:2021iol, Alzhrani:2022dpi, Palermo:2024tza, Florkowski:2017ruc, Florkowski:2017dyn, Florkowski:2018ahw, Florkowski:2018fap,Montenegro:2018bcf, Kumar:2018iud, Kumar:2018lok,Becattini:2018duy,Florkowski:2019qdp,Montenegro:2020paq,Garbiso:2020puw,Bhadury:2020puc, Bhadury:2020cop, Bhadury:2022ulr,Bhadury:2023vjx, Singh:2022ltu, Weickgenannt:2019dks, Weickgenannt:2021cuo, Weickgenannt:2020aaf, Wagner:2022amr, Weickgenannt:2022zxs, Weickgenannt:2023nge,Weickgenannt:2023bss,Weickgenannt:2024ibf, Wagner:2024fhf, Peng:2021ago,Cartwright:2021qpp, Hu:2021pwh, Li:2020eon, Shi:2020htn,  Singh:2022ltu, Hattori:2019lfp, Fukushima:2020ucl, Daher:2022xon, Daher:2022wzf, Sarwar:2022yzs, Peng:2021ago,Yi:2021unq,Wang:2021ngp,Wang:2021wqq, Biswas:2022bht,Cao:2022aku,Weickgenannt:2022jes,Weickgenannt:2022qvh,Weickgenannt:2023btk, Biswas:2023qsw, Xie:2023gbo, Daher:2024ixz, Xie:2023gbo, Ren:2024pur, Daher:2024bah, Fang:2024sym,Dey:2024cwo,She:2024rnx,Gallegos:2021bzp,Gallegos:2022jow, Hongo:2021ona, Kumar:2023ojl, She:2021lhe, Montenegro:2017rbu,Montenegro:2020paq,Torrieri:2022ogj, Florkowski:2024bfw,Becattini:2023ouz,Kiamari:2023fbe,Tiwari:2024trl, Drogosz:2024gzv,Drogosz:2024rbd}.~Thus far, efforts have primarily focused on establishing the foundational framework of spin hydrodynamics, assessing its stability and causality properties\mbox{~\cite{Daher:2022wzf, Sarwar:2022yzs, Xie:2023gbo, Daher:2024ixz, Ren:2024pur, Daher:2024bah}}, and deriving quasi-analytic solutions for simple expansion geometries, particularly Bjorken flow~\cite{Wang:2021ngp, Biswas:2022bht, Xie:2023gbo,Drogosz:2024lkx}. 
	
The application of spin hydrodynamics to describe the spin polarization data appears natural, given the remarkable success of conventional relativistic hydrodynamics in explaining hadronic observables, particularly flow coefficients (see, for example, Ref.~\cite{Schenke:2021mxx}). The ongoing development of spin hydrodynamics aims to incorporate spin polarization effects within the well-established framework of relativistic hydrodynamics. 
	
In this work, we employ the formulation of perfect spin hydrodynamics as outlined in Refs.~\cite{Florkowski:2017ruc, Florkowski:2017dyn, Florkowski:2018fap}. Key features of this approach include the use of a symmetric energy-momentum tensor together with a conserved spin tensor to describe spin-\nicefrac{1}{2} particles. For small values of the spin polarization tensor (a dimensionless quantity in natural units), the spin dynamics decouples from the background~\cite{Florkowski:2018ahw} i.e. the spin dynamics can be treated as externally determined by the spinless hydrodynamic background --- specifically, a solution of the standard equations of relativistic dissipative hydrodynamics. Here, for the first time, we solve the equations of perfect spin hydrodynamics using a realistic (3+1)-dimensional [(3+1)D] hydrodynamic background. This background is calibrated to reproduce bulk hadronic observables, including rapidity distributions, transverse momentum spectra, and elliptic flow coefficients. 

We examine several scenarios by varying the effective mass and initial evolution time of the spin polarization tensor and compare the model predictions with experimental data. Our results indicate that a successful description of the data requires a moderately delayed evolution time for spin hydrodynamics of about 4 fm/\textit{c}, compared to the standard initial timescale of 1 fm/\textit{c}. This timescale marks the transition from the phase where the spin-orbit interaction is significant to the regime dominated by spin-conserving processes.

The structure of our paper is as follows: In Sec.~\ref{sec:theory}, we introduce the hydrodynamic equations for both the background and the spin tensor. The numerical methods used to solve these equations are described in Sec.~\ref{sec:numfram}. Our results are presented in Sec.~\ref{sec:results}, where we compare them with experimental data and other theoretical models. Finally, we summarize and conclude in Sec.~\ref{sec:summary}. Technical details are provided in five appendices. 

\medskip
{\it Conventions:} We work in natural units where $\hbar=k_B=c=1$. The metric is defined as $g_{\mu\nu}=\text{diag}(1,-1,-1,-1)$. We use Milne coordinates $(\tau,x,y,\eta_s)$, where the longitudinal proper time $\tau$ and spacetime rapidity $\eta_s$ are defined as $\tau =\sqrt{t^2-z^2}$ and $\eta_s =\frac{1}{2}\log \left(\frac{t+z}{t-z}\right)$, respectively. For the Levi-Civita tensor $\epsilon^{\mu\nu\alpha\beta}$, we follow the sign convention $\epsilon^{0123} =-\epsilon_{0123} = +1$. We also adopt the notation $A^{[\mu}B^{\nu]}=(A^\mu B^\nu - A^\nu B^\mu)/2$ and $A^{(\mu}B^{\nu)}=(A^\mu B^\nu + A^\nu B^\mu)/2$ for the antisymmetric and symmetric parts of tensors. 

\section{Theoretical framework}
\label{sec:theory}

\subsection{Background hydrodynamics}

The equations of relativistic viscous hydrodynamics arise from the conservation laws for the energy-momentum tensor, $T^{\a\b}$, and the net baryon current, $N^{\a}$,
\beq
D_\alpha T^{\a\b} (x) &=& 0\,,
\label{eq:T_cons}\\
D_\a N^\a (x) &=& 0\,,
\label{eq:N_cons}
\eeq
where $D_\a$ denotes the covariant derivative~\footnote{In this work, we use curvilinear coordinates, which introduce non-zero Christoffel symbols. However, the spacetime remains flat.}.
	
We define the flow velocity $u^\a$ using Landau's convention, which specifies that in the local rest frame of the fluid cell, the energy flux should vanish
\beq
T^{\a\b} u_\b= \varepsilon\,u^\a.
\label{eq:Lframe}
\eeq
In this case, the constitutive relation for the energy-momentum tensor takes the form
\beq
T^{\a\b} = \varepsilon\,u^\a \, u^\b - (P_{\rm eq}+\Pi) \, \Delta^{\a\b}\,+ \pi^{\a\b},
\label{eq:T_curr}
\eeq
where $\varepsilon$ is the energy density, $P_{\rm eq}$ is the corresponding equilibrium pressure, $\pi^{\a\b}$ is the shear stress tensor, $\Pi$ is the bulk pressure, and $\Delta^{\a\b}=g^{\a\b}-u^\a u^\b$ is the projector onto the space orthogonal to the flow velocity.
	
The baryon current is expressed as
\beq
N^{\a} =  n\,u^\a + n^\a,
\label{eq:N_curr}
\eeq
where $n$ represents the baryon number density. For this analysis, we assume that dissipative corrections from charge diffusion vanish, setting $n^\a=0$.
	
At second order in spacetime gradients, the time evolution of the shear stress tensor $\pi^{\a\b}$ and bulk pressure $\Pi$ follows from the relaxation-type equations derived in the Denicol-Niemi-Molnar-Rischke (DNMR) framework~\cite{Denicol:2012cn,Denicol:2015transcoeff}:
\beq
\dot{\Pi}&=&\frac{\Pi_{\rm N S}-\Pi}{\tau_{\Pi}}-\frac{\delta_{\Pi \Pi}}{\tau_{\Pi}} \Pi\, \theta+\frac{\lambda_{\Pi \pi}}{\tau_{\Pi}} \pi^{\a\b} \sigma_{\a\b} \label{eq:Pi}\,,\\
\dot{\pi}^{\langle\a\b\rangle}&=&\frac{\pi_{\rm N S}^{\a \b}-\pi^{\a\b}}{\tau_\pi}
-\frac{\delta_{\pi \pi}}{\tau_\pi} \pi^{\a \b} \theta+\frac{\lambda_{\pi \Pi}}{\tau_\pi} \Pi \,\sigma^{\a\b}\nonumber\\
&&~~~~~~-\frac{\tau_{\pi \pi}}{\tau_\pi} \pi_\g^{\langle\a} \sigma^{\b\rangle \g}+\frac{\phi_7}{\tau_\pi} \pi_\g^{\langle\a} \pi^{\b\rangle \g}\,\label{eq:pi}.
\eeq
Here, $\dot{(~)}\equiv u \cdot D$ denotes the comoving derivative, $\theta \equiv D \cdot u$ is the expansion scalar, and $\sigma^{\a\b} \equiv D^{\langle\a} u^{\b\rangle}\equiv \Delta_{\g \d}^{\a \b} D^\g u^\d$ is the shear flow tensor. The projector $\Delta_{\g \d}^{\a \b} \equiv \frac{1}{2}\left[\Delta_\g^\a \Delta_\d^\b+\Delta_\d^\a\Delta_\g^\b -(2 / 3) \Delta^{\a \b} \Delta_{\g \d}\right]$ isolates the symmetric, traceless, and orthogonal part of a rank-2 tensor relative to $u^\a$. 

The Navier-Stokes forms of the dissipative currents in Eqs.~\eqref{eq:Pi}--\eqref{eq:pi} are $\Pi_{\rm NS} = - \zeta \theta$ and $\pi_{\rm NS}^{\a \b} = 2 \eta \,\sigma^{\a \b}$, where the bulk and shear viscosities are chosen as in~\cite{Denicol:2015transcoeff,Denicol:2018wdp},
\beq
\label{eq:etazeta}
\eta  = C_\eta \frac{\varepsilon_{\rm eq}+P_{\rm eq}}{T} , \quad \zeta = 75\eta \left(\frac{1}{3}-c_s^2\right)^2
\eeq
with
\beq
c_s^2=\left.\frac{\partial P_{\rm eq}}{\partial \varepsilon_{\rm eq}}\right|_{n_{\rm eq}}
+\left. \frac{n_{\rm eq}}{\varepsilon_{\rm eq}+P_{\rm eq}} \frac{\partial P_{\rm eq}}{\partial n_{\rm eq}}\right|_{\varepsilon_{\rm eq}}
\eeq
representing the speed of sound squared. The coefficient $C_\eta$ is set equal to 0.12.
	
The relaxation times for bulk and shear corrections as well as the second-order transport coefficients, derived using the 14-moment approximation in massless limit, are given by~\cite{Denicol:2015transcoeff,Transcoeff2014_p7}:
\begin{align}
&\tau_\pi = \frac{5\eta}{\varepsilon_{\rm eq}+P_{\rm eq}} \,,\quad \tau_\Pi = \frac{\zeta}{15\left(\frac{1}{3}-c_s^2\right)^2(\varepsilon_{\rm eq}+P_{\rm eq})},\\
&\frac{\delta_{\Pi\Pi}}{\tau_\Pi} = \frac{2}{3}	\,, \quad  \frac{\lambda_{\Pi\pi}}{\tau_\Pi} = \frac{8}{5}\left(\frac{1}{3}-c_s^2\right)\,,\quad \frac{\delta_{\pi\pi}}{\tau_\pi} = \frac{4}{3},\\
& \frac{\lambda_{\pi\Pi}}{\tau_\pi} = \frac{6}{5}\,,\quad \frac{\tau_{\pi\pi}}{\tau_\pi} = \frac{10}{7}\,,\quad \frac{\phi_7}{\tau_\pi} = \frac{9}{70 P_{\rm eq} \tau_\pi}\,.\label{eq:coef}
\end{align}
The choice of $\eta$ and $\zeta$ in Eq.~(\ref{eq:etazeta}) implies $\tau_\pi = \tau_\Pi = \frac{5C_\eta}{T}$.

%
	 	
\subsection{Spin hydrodynamics}
	
For a symmetric energy-momentum tensor, such as the one considered in Eq.~\eqref{eq:T_curr}, the conservation law for total angular momentum, $D_\a J^{\a,\b \g}_{\rm } = 0$, implies the conservation of the spin current
\beq
D_\a S^{\a,\b \g}_{\rm } (x) = 0\,.
\label{eq:S_cons}
\eeq
The spin current is a rank-3 tensor that is antisymmetric in its last two indices. In the de~Groot-van~Leeuven-van~Weert (GLW) pseudogauge, the spin current is expressed as~\cite{Florkowski:2018ahw,Florkowski:2021wvk}
\beq
S^{\a,\b\g} 
&=&  \Aone u^\a \omega^{\b\g}+\Atwo u^\a u^{[\b}\omega^{\g]\d}u_\d\nonumber\\
&&~~~~~~+\Athree \left(u^{[\b}\omega^{\g] \a}+g^{\a [\b}\omega^{\g]\d}u_\d\right),
\label{eq:S_curr}
\eeq
where $\omega^{\b\g}$ is the spin polarization tensor, which is antisymmetric: $\omega^{\b\g}=-\omega^{\g\b}$.

The coefficient functions ${\cal A}$ are given by the following expressions:
\beq
\Aone &=&  \cosh\left({\mu_B \over T}\right) \frac{T^3}{\pi^2}\left[\left(4+\frac{z^2}{2}\right)K_2\left(z\right)+z K_1\left(z\right)\right]\nn \label{A1},\\
\Atwo &=&2 \cosh\left({\mu_B \over T}\right)  \frac{ T^3}{\pi^2}\left[\left(12+\frac{z^2}{2}\right)K_2\left(z\right)+3z K_1\left(z\right)\right]  \nn\label{A2},\\
\Athree &=& \frac{1}{2}\left( \Aone-\frac{\Atwo}{2}\right),    
\eeq
where $\mu_B$ is the baryon chemical potential,  $z\equiv m/T$ is the ratio of the particle mass $m$ to the temperature $T$, and $K_n(z)$ denotes the modified Bessel functions of the second kind of order $n$. Even though the spin degrees of freedom evolve independently, the spin-polarized particles interact with the medium. As a result, their in-medium effective mass can differ from the vacuum one. Therefore, we treat $m$ as an effective parameter.

By substituting Eq.~\eqref{eq:S_curr} into Eq.~\eqref{eq:S_cons}, one obtains a system of six evolution equations for the components of the spin polarization tensor $\omega_{\b\g}$, provided that the temperature $T$, baryon chemical potential $\mu_B$, and four-velocity $u^\a$ are known.
	
\section{Numerical framework}
\label{sec:numfram}
	
\subsection{Background hydrodynamics}
\label{sec:bkgnum}

Equations~\eqref{eq:T_cons}, \eqref{eq:N_cons}, \eqref{eq:Pi}, and \eqref{eq:pi}, supplemented by an equation of state relating pressure to energy density and baryon density, $P_{\rm eq}\!=\!P_{\rm eq}(\varepsilon,n)$, form a closed set of eleven differential equations for eleven unknown hydrodynamic fields: $\varepsilon$, $n$, $u^\mu$, $\Pi$, and $\pi^{\mu\nu}$~\footnote{The flow vector is normalized as $u^\mu u_\mu=1$ and the shear stress tensor is symmetric, traceless, and orthogonal to the flow vector, hence, there are three independent components in $u^\mu$ and five in $\pi^{\mu\nu}$.}. It is important to note that, within the model, we consider the limit of small polarization~\cite{Florkowski:2018ahw}, where Eqs.~\eqref{eq:T_cons}, \eqref{eq:N_cons}, \eqref{eq:Pi}, and \eqref{eq:pi} do not receive feedback from spin equations of motion via $\omega$. Therefore, we solve these equations numerically in (3+1) dimensions in Milne coordinates $(\tau,x,y,\eta_s)$ using the code developed in Ref.~\cite{Singh2023hdo}, which employs the Godunov-type relativistic Harten-Lax-van Leer-Einfeldt (HLLE) approximate Riemann solver, as described in Ref.~\cite{Karpenko:2013wva}. Additionally, we extend the code from Ref.~\cite{Singh2023hdo} to include the cross terms in Eqs.~(\ref{eq:Pi})--(\ref{eq:pi}), associated with the non-vanishing transport coefficients $\lambda_{\Pi\pi}$, $\lambda_{\pi\Pi}$, $\tau_{\pi\pi}$, and $\phi_7$. 

In contrast to Ref.~\cite{Singh2023hdo}, in the present study, we use the lattice-QCD-based equation of state (EOS) at finite net baryon density, NEOS-BQS~\cite{Monnai2019eos,eosweblink}, which exhibits a crossover phase transition across the entire parametric space of the phase diagram. 

Herein, we focus on the Au+Au collisions at the top RHIC energy of $\sqrt{s_{\rm NN}}=200$ GeV. We initialize the background evolution at the proper time $\tau_0=1$~fm. The initial energy density and baryon density profiles are determined according to the model based on the Glauber collision geometry with local energy-momentum conservation~\cite{Shen:2020jwv,Ryu:2021lnx}. However, unlike in Refs.~\cite{Shen:2020jwv,Ryu:2021lnx}, to compute the thickness functions and wounded nucleon densities we use the optical limit of the Glauber model. The initial transverse flow components are chosen to be zero whereas the longitudinal flow is numerically obtained from the initial energy-momentum tensor components. For simplicity, we also assume that the dissipative corrections initially vanish, i.e. , $\pi^{\mu\nu}(\tau_0)=0$ and $\Pi(\tau_0)=0$.  Further details of the initial state model used for the energy-momentum and baryon currents are provided in Appendix \ref{sec:ic}. 

The initial conditions obtained within the optical limit of the Glauber model involve a fixed impact parameter. Its value, corresponding to a given centrality range, is determined using the two-component Glauber model. The procedure for determining the impact parameter is detailed in Appendix~\ref{sec:impactpar}, and the values obtained for the centrality classes considered in this work are listed in Table~\ref{tab:cent}.
	
\begin{figure}[t]
\subfigure{\includegraphics[width=0.4\textwidth]{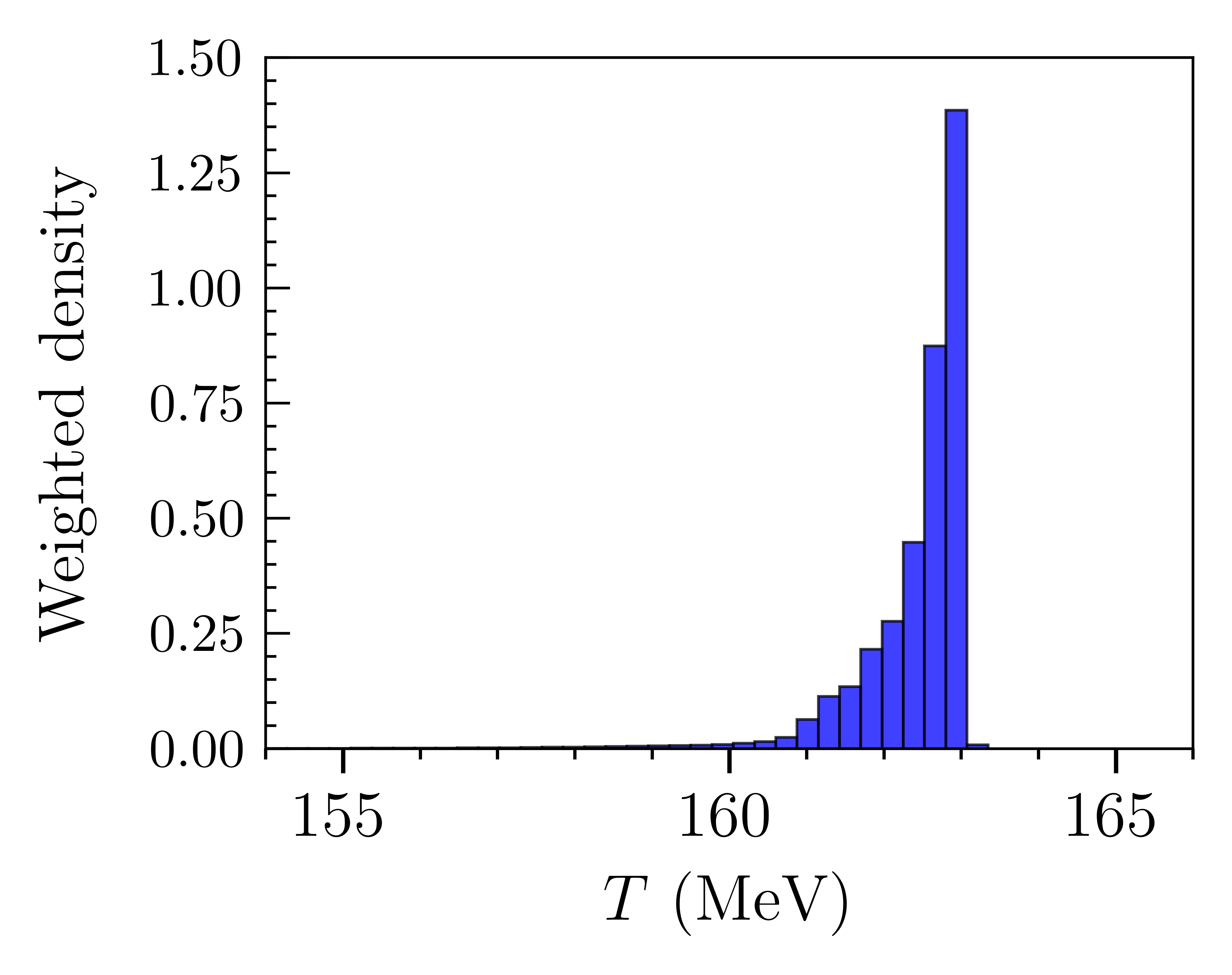}}\\
\hspace*{0.6cm}
\subfigure{\includegraphics[width=0.4\textwidth]{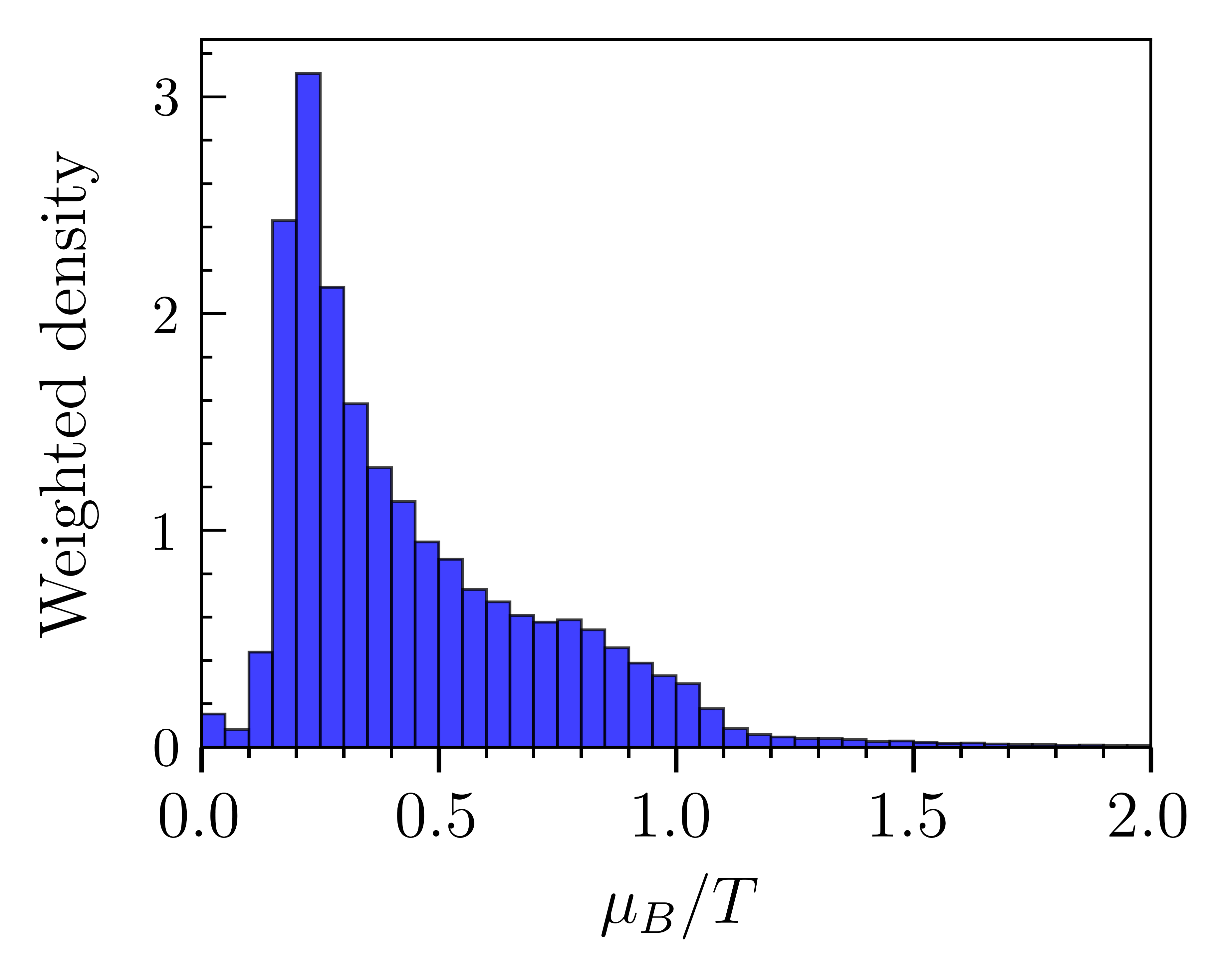}}
\caption{Area-weighted distributions of temperature $T$ and scaled baryon chemical potential $\mu_B/T$ on the switching hypersurface $\Sigma$.}
\label{fig:fodist}
\end{figure}

We evolve the hydrodynamic equations until the energy density in the system decreases below the threshold value $\varepsilon_{\rm sw}= $~0.5 GeV/fm$^{3}$. The condition $\varepsilon(T,\mu_B)=\varepsilon_{\rm sw}$ defines the switching hypersurface $\Sigma$, which is extracted using the CORNELIUS code~\cite{Huovinen2012}. The hydrodynamic fields on the hypersurface are passed to a hadron sampler~\cite{Karpenko:2015xea,Schafer2022,samplerweblink}, which generates particles from fluid elements. The resulting particle set serves as input to the SMASH transport model~\cite{SMASH2016prc,wergieluk_2024_10707746}, which describes subsequent hadron interactions and decays.

\subsection{Spin hydrodynamics}
\label{sec:spinnum}
	
Using the hydrodynamic background obtained from Eqs.~\eqref{eq:T_cons}, \eqref{eq:N_cons}, \eqref{eq:Pi}, and \eqref{eq:pi} (as described in the previous section), we determine the effective temperature $T(x)$ and baryon chemical potential $\mu_B(x)$. These, together with the four-velocity field, allow us, through Eq.~\eqref{eq:S_curr}, to compute the dynamics of the spin polarization tensor $\omega$ from Eq.~\eqref{eq:S_cons}. For this purpose, we further extend the code to incorporate the spin conservation equation~\eqref{eq:S_cons}. The respective numerical implementation, using also HLLE algorithm, is explained in detail in Appendix~\ref{sec:numspin}.

\begin{figure}[t]
\subfigure{\includegraphics[width=0.4 \textwidth]{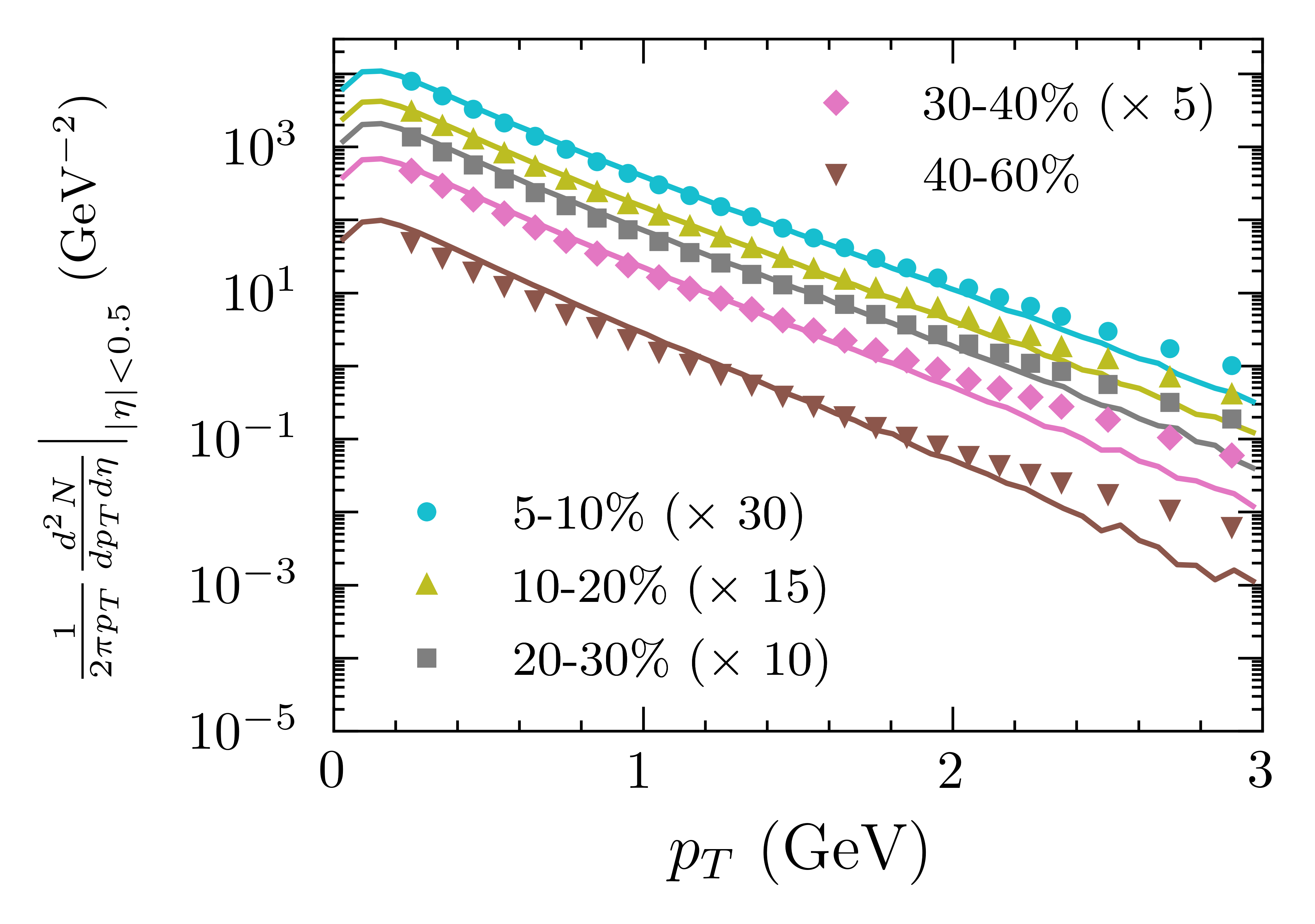}}\\
\subfigure{\includegraphics[width=0.4 \textwidth]{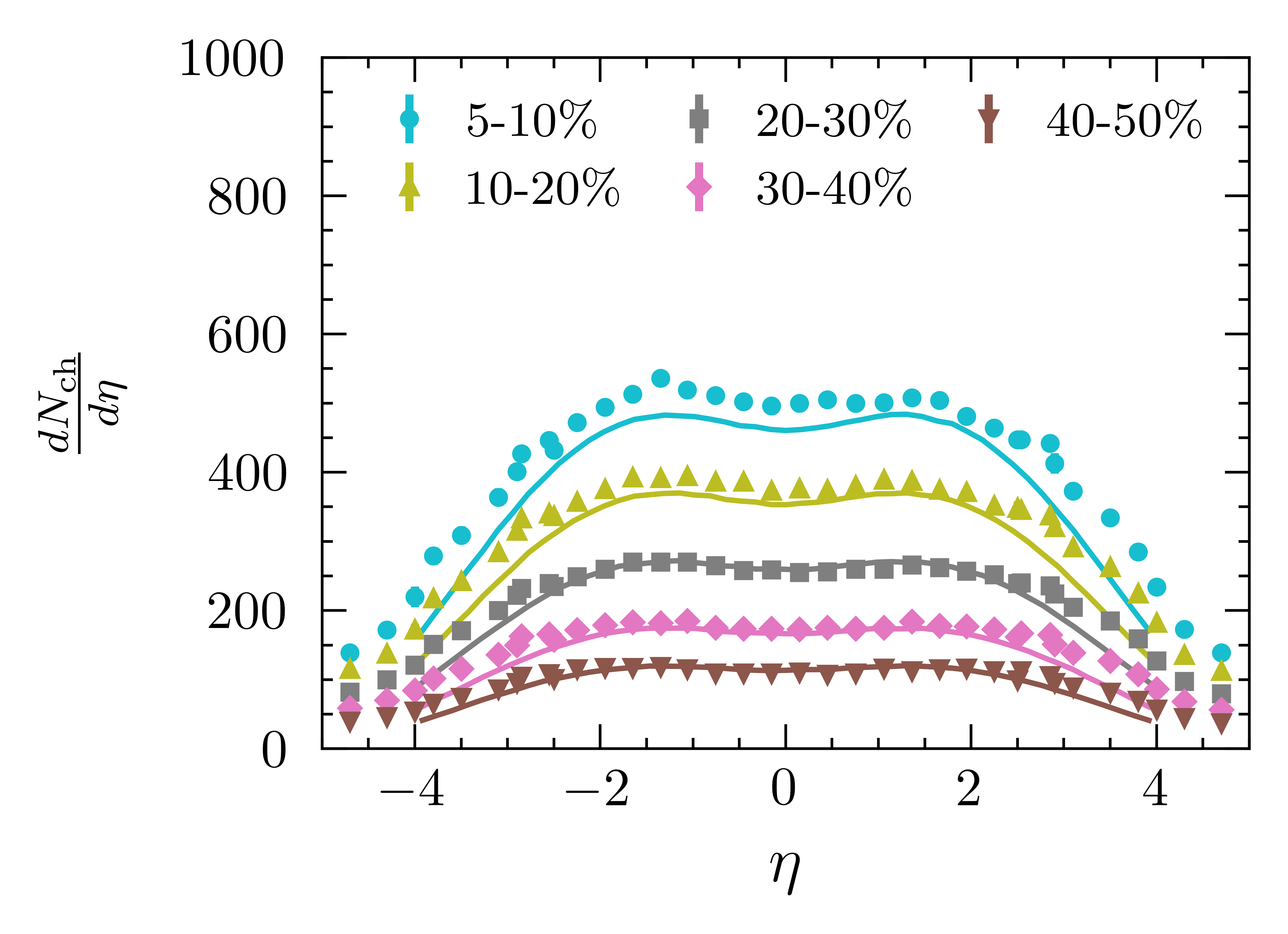}}\\
\hspace*{0.6cm}
\includegraphics[width=0.4 \textwidth]{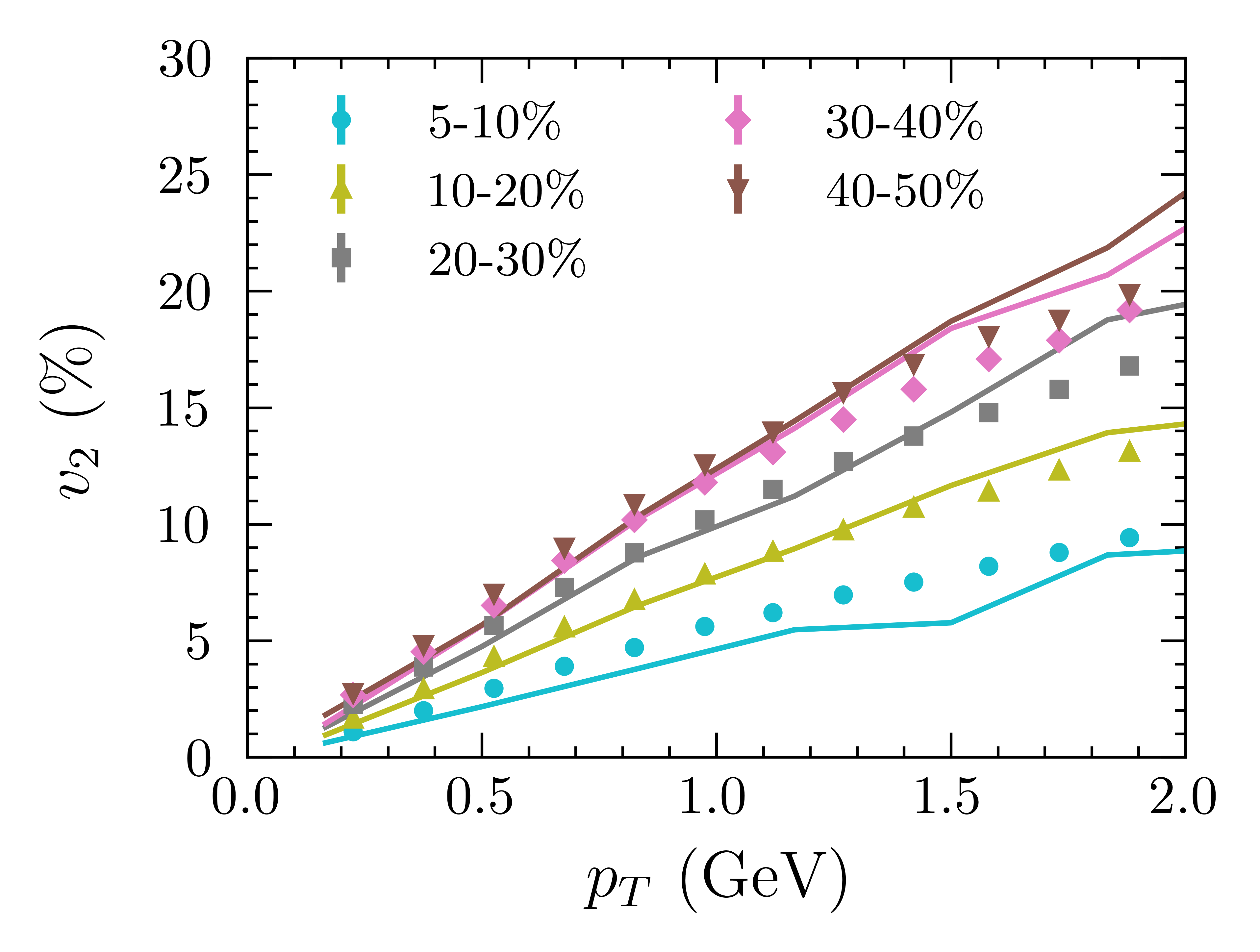}
\caption{Bulk observables for Au+Au collisions at $\sqrt{s_{\rm NN}}=200$ GeV. The solid lines represent our numerical results, while the different symbols correspond to experimental data taken from Ref.~\cite{STAR2003_pT} in the top panel, from Ref.~\cite{BRAHMS2002_dNdeta} in the middle panel, and from Ref.~\cite{STAR200_v2} in the bottom panel.}
\label{fig:s200}
\end{figure}

Motivated by Refs.~\cite{Becattini:2021suc,Liu:2021uhn,Matteo2022glw,Wagner:2024fry}, we use the following initial condition for the spin polarization tensor
\begin{equation}
    \label{eqn:ic1}
    \omega_{\mu \nu}(\tau_0^s)  =\varpi^{\text{iso}} _{\mu\nu} + 4\hat{\tau}_{[\mu}  \xi^{\text{iso}}_{\nu]\rho}u^\rho, 
\end{equation}
where the right-hand side is understood to be evaluated at the initial time of the spin evolution $\tau_0^s$, which we treat as a free parameter of the model to be extracted from the experimental data. In Eq.~\eqref{eqn:ic1}, $\varpi _{\mu\nu}^{\text{iso}} = \frac{1}{T}\partial_{[\nu}u_{\mu]}$ is the isothermal part of thermal vorticity $\varpi _{\mu\nu}= \partial_{[\nu}\beta_{\mu]}$, namely,  $\varpi _{\mu\nu} =\varpi _{\mu\nu}^{\text{iso}} + \varpi _{\mu\nu}^{\text{T}}$~\footnote{The isothermal part of thermal vorticity is related to the so-called kinematic vorticity, $\varpi _{\mu\nu}^{\text{k}}$, by $\varpi _{\mu\nu}^{\text{k}}=T \varpi _{\mu\nu}^{\text{iso}}$ \cite{Wu:2019eyi}.}, where $\varpi_{\mu\nu}^{\text{T}} =  \frac{1}{T}u_{[\mu}\partial_{\nu]} \ln T$ and $\beta^\mu$ is the ratio of the flow velocity $u^\mu$ to local temperature $T$, $\beta^\mu=u^\mu/T$. Similarly, we introduce $\xi_{\mu\nu}^{\text{iso}}=\frac{1}{T}\partial_{( \nu}u_{\mu )}$, which represents the isothermal part of the thermal shear tensor $\xi _{\mu\nu} = \partial_{(\nu}\beta_{\mu)}$. The part $\varpi_{\mu\nu}^{\text{T}}$ (as well as the respective part of the thermal shear tensor) vanishes when temperature gradients are absent.  Note that $\hat{\tau}^\mu=(1,0,0,0)$ is a fixed timelike vector which in Milne coordinates is normal to the constant-$\tau$ hypersurface.

As an alternative, we also consider the analog of the initial condition~\eqref{eqn:ic1} which includes all temperature gradients present in the system
\begin{equation}
    \label{eqn:ic2}
    \omega_{\mu \nu}(\tau_0^s)  =\varpi _{\mu\nu} + 4\hat{\tau}_{[\mu}  \xi_{\nu]\rho}u^\rho.
\end{equation}    
We note that by using Eqs.~\eqref{eqn:ic1} or \eqref{eqn:ic2}, we intend to account for equilibration of spin degrees of freedom resulting from strong spin-orbit interactions occurring in the early stages of heavy-ion collisions, before the system reaches the regime where perfect spin hydrodynamics is applicable. 
	 
\begin{figure}[t]
\includegraphics[width=0.4 \textwidth]{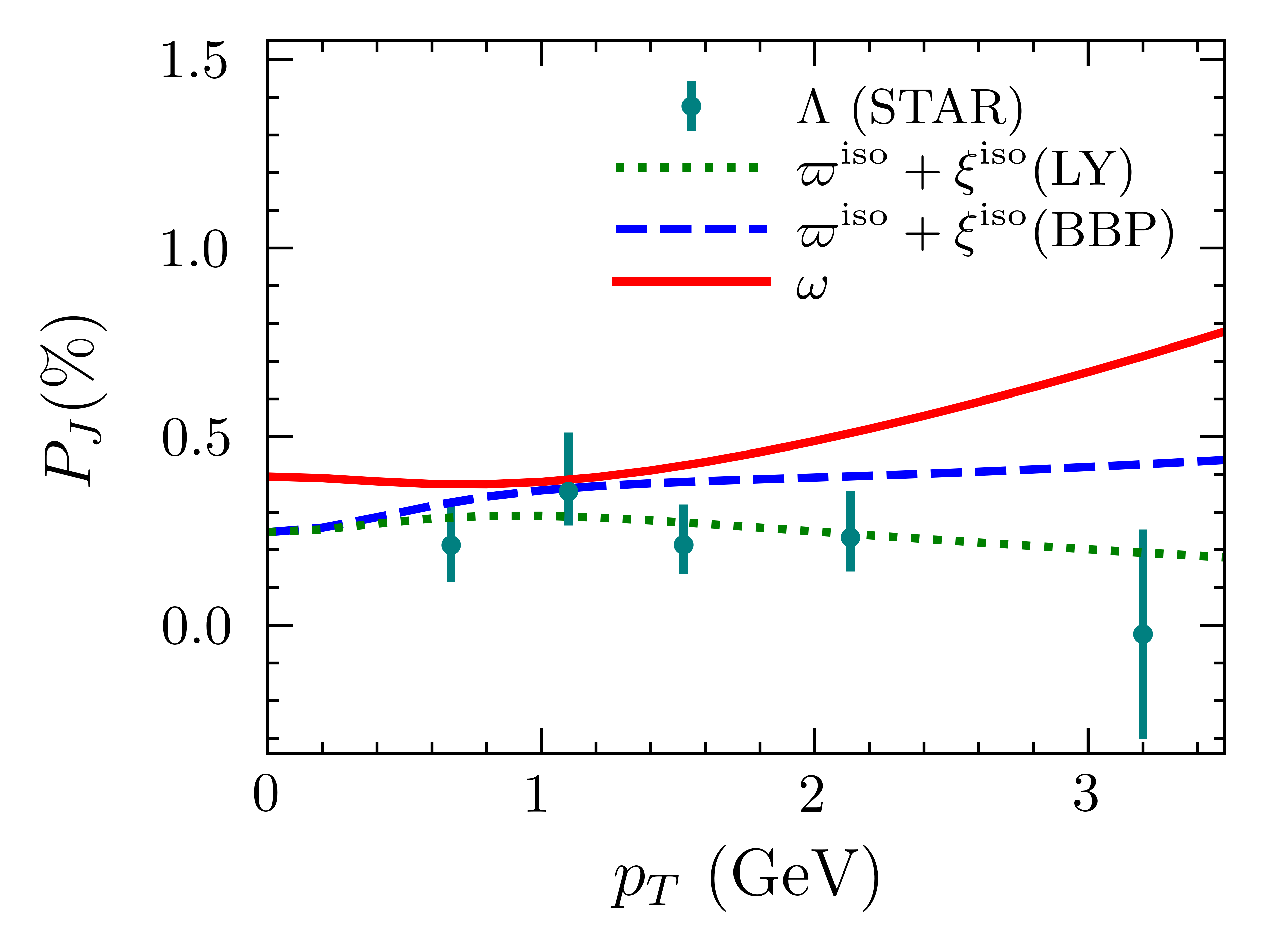}
\includegraphics[width=0.4 \textwidth]{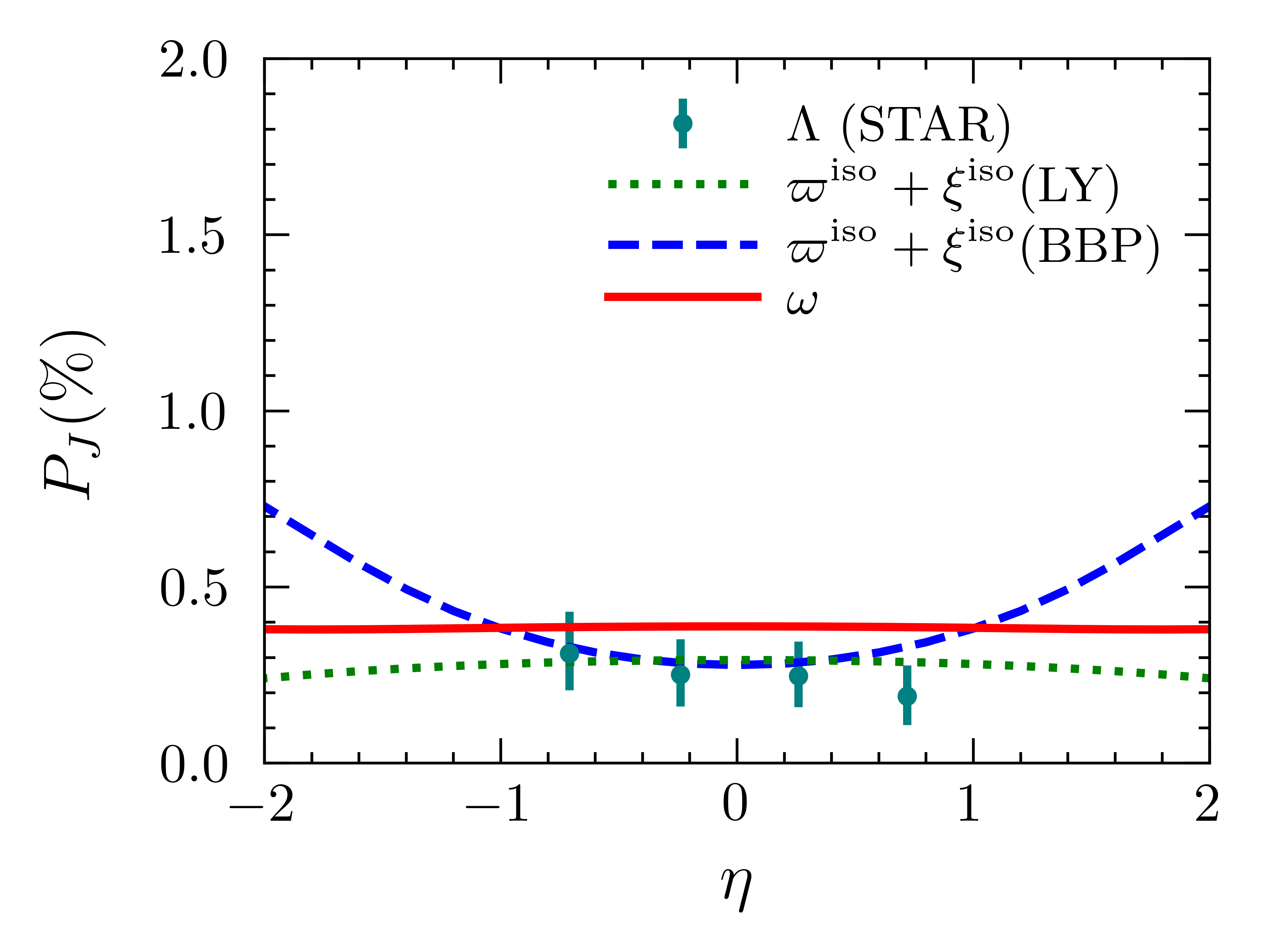}
\caption{Component of $\Lambda$ polarization along the orbital angular momentum direction, plotted as a function of transverse momemtum, $p_T$ (top), and pseudorapidity, $\eta$ (bottom). Experimental data is taken from Ref.~\cite{STAR:2018gyt} with updated decay parameter ($\alpha_\Lambda = 0.732$). Solid lines represent our results from spin hydrodynamics with the initial condition model given by Eq.~(\ref{eqn:ic1}), using the parameters, $\tau_0^s = 4$ fm and $m=m_\Lambda$.}
\label{fig:spintimePJ}
\end{figure}

\begin{figure}[t]
\includegraphics[width=0.4 \textwidth]{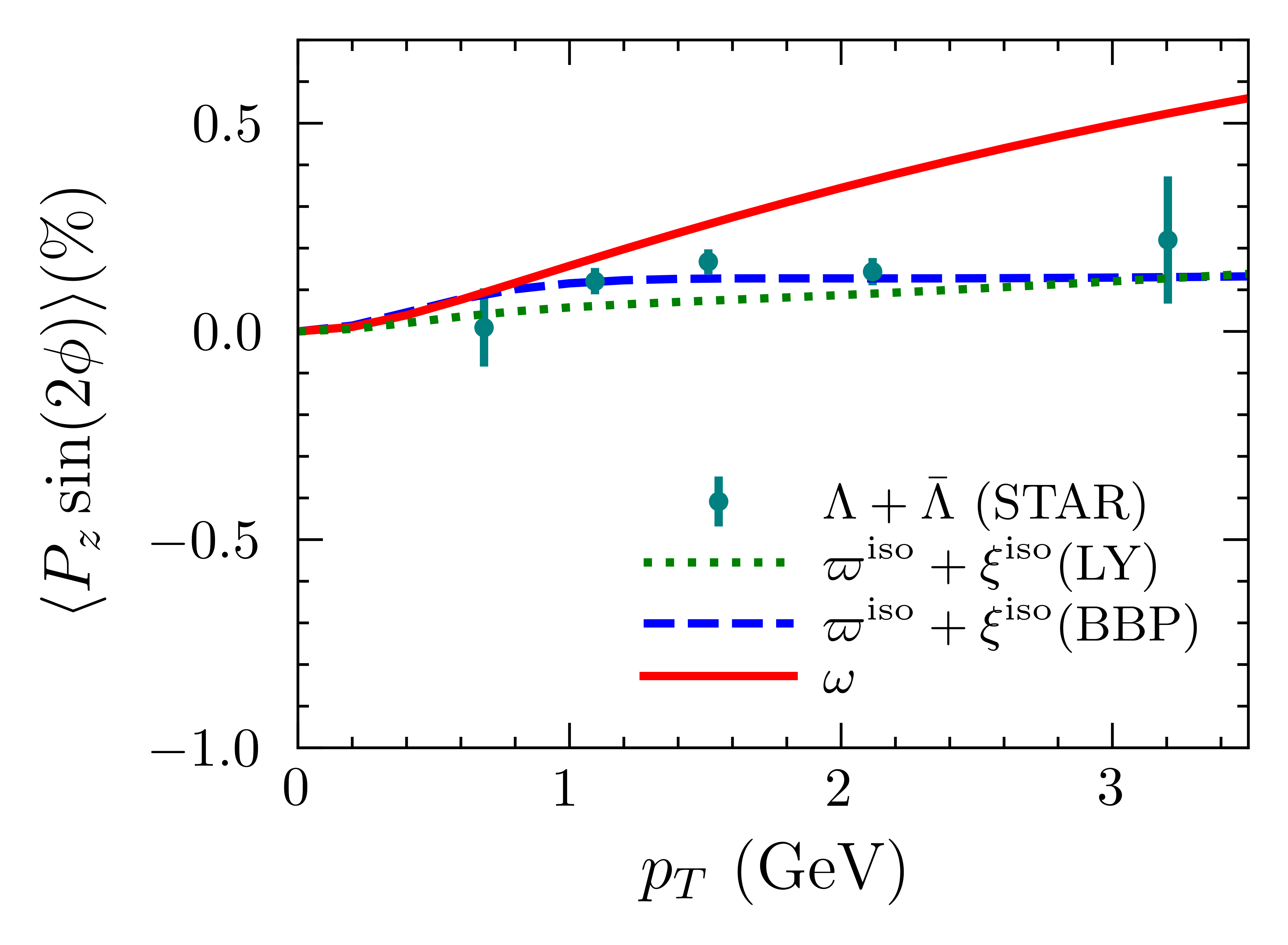}
\includegraphics[width=0.4 \textwidth]{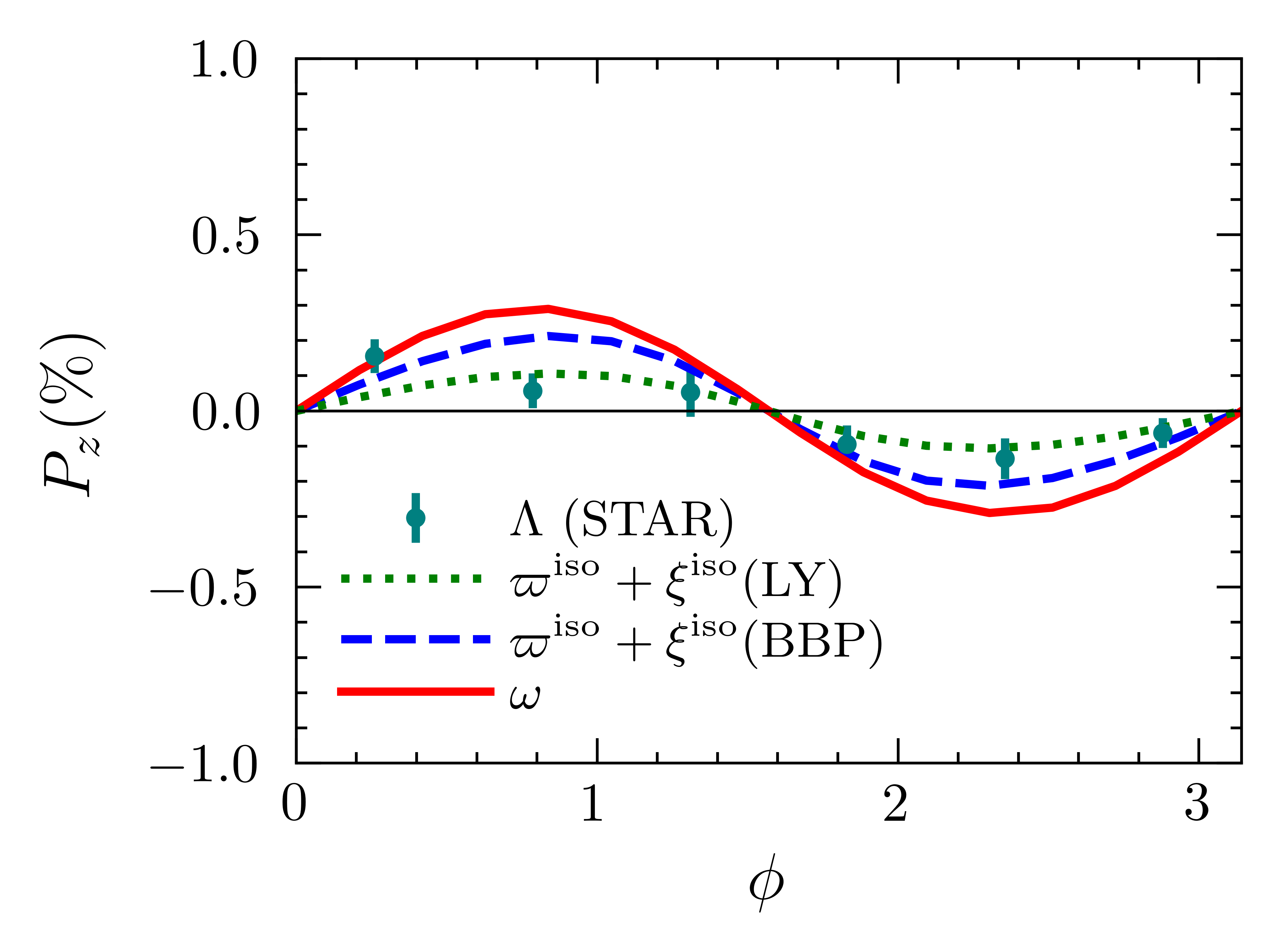}
\caption{Longitudinal $\Lambda$ polarization, plotted as a function of transverse momentum, $p_T$ (top), and azimuthal angle, $\phi$ (bottom). Experimental data are taken from Ref.~\cite{STAR:2019erd}with updated decay parameter ($\alpha_\Lambda = 0.732$). Solid lines represent our results from spin hydrodynamics with the initial condition model given by Eq.~(\ref{eqn:ic1}), using the parameters $\tau_0^s = 4$ fm and $m=m_\Lambda$.}
\label{fig:spintimePz}
\end{figure}

\subsection{Spin polarization of Lambdas}
\label{sec:polvec}

In this work, we focus on polarization observables for $\Lambda$ hyperons. After determining the spacetime dynamics of the spin polarization tensor $\omega(x)$, we calculate the components of the polarization vector $S^\mu(p)$ for $\Lambda$ particles with momenta $p$. For the GLW pseudogauge used in the present study, the polarization vector is given by the formula~\cite{Matteo2022glw} 
\beq
 \label{eq:Sglw}
\!\!\! S^\mu(p) = -\frac{1}{8m_\Lambda}\epsilon^{\mu\nu\rho \sigma}p_\sigma \frac{\int d\Sigma\cdot p \ n_F(1-n_F) \omega_{\nu\rho}}{\int d\Sigma\cdot p \ n_F},
\eeq
where $n_F=n_F(T, \mu_B, p\cdot u; m_\Lambda)$ is the Fermi-Dirac distribution function, which depends on the background hydrodynamic variables and the $\Lambda$ mass, $m_\Lambda$. The integral is performed over the switching hypersurface $\Sigma$ determined as described in Sec.~\ref{sec:bkgnum}.  

Our results are compared with those obtained using the spin polarization formula obtained at first order of thermodynamic gradients~\cite{Becattini:2021suc,Matteo2022glw},
\beq
S^\mu(p) = S^\mu_{\varpi}(p) + S^\mu_{\xi}(p),
\eeq
where the term containing contribution from thermal vorticity, $S^\mu_{\varpi}(p)$, is given by~\cite{Becattini:2015ska}
\beq
\!\!\!\!\! S_{\varpi}^\mu(p) &=& -\frac{1}{8m_\Lambda}\epsilon^{\mu\nu\rho \sigma}p_\sigma \frac{\int d\Sigma\cdot p \ n_F(1-n_F) \varpi_{\nu\rho}}{\int d\Sigma\cdot p \ n_F}.
\label{eq:SvFB}
\eeq
For the contribution from thermal shear, there are currently two prescriptions available in the literature: one proposed in Ref.~\cite{Becattini:2021suc} (henceforth denoted as BBP), given by
\beq
S^\mu_{\xi,\text{BBP}}(p) &=& -\frac{\epsilon^{\mu\nu\rho \sigma}}{4m_\Lambda} \frac{p_\sigma p^\lambda}{p\cdot \hat{t}}\frac{\int d\Sigma\cdot p \ n_F(1-n_F) \hat{t}_\nu \xi_{\lambda\rho}}{\int d\Sigma\cdot p \ n_F},
\nn \\
\label{eq:SsFB}
\eeq
and another proposed in Ref.~\cite{Liu:2021uhn} (henceforth denoted as LY), given by
\beq
S^\mu_{\xi,\text{LY}}(p) &=& -\frac{\epsilon^{\mu\nu\rho \sigma}}{4m_\Lambda} p_\sigma p^\lambda\frac{\int d\Sigma\cdot p \ n_F(1-n_F) \frac{\beta u_\nu}{p\cdot u} \partial ^\perp _{(\rho}u_{\lambda)}}{\int d\Sigma\cdot p \ n_F}\nn \\
&=& -\frac{\epsilon^{\mu\nu\rho \sigma}}{4m_\Lambda}p_\sigma \frac{\int d\Sigma\cdot p \ n_F(1-n_F) \frac{p^\lambda_\perp u_\nu}{p\cdot u} \xi_{\rho\lambda}}{\int d\Sigma\cdot p \ n_F},
\label{eq:SsLY}
\eeq
where $\partial ^\perp_\mu\equiv \Delta_\mu^{~\nu}\partial_\nu$ and $p^\perp_\mu\equiv \Delta_\mu^{~\nu}p_\nu$.

Following recent works, we use Eqs.~\eqref{eq:SvFB}--\eqref{eq:SsLY} in two ways: with and without temperature gradients included. In the latter case, we replace $\varpi$ and $\xi$ by $\varpi^{\text{iso}}$ and $\xi^{\text{iso}}$, respectively. The approach with neglected temperature gradients has been shown to be crucial for correctly reproducing the longitudinal component of the spin vector in Refs.~\cite{Becattini:2021iol,Ryu:2021lnx,Alzhrani:2022dpi,Palermo:2024tza}. 

It is important to note that since we perform calculations on the switching hypersurface $\Sigma$ defined by a constant energy density, our hypersurface does not exactly correspond to a constant value of $T$. However, as shown in Fig.~\ref{fig:fodist}, the temperature is nearly constant on $\Sigma$. On the other hand, the ratio $\mu_B/T$ exhibits a rather broad distribution.
	
\section{Results}
\label{sec:results}

\subsection{Background hydrodynamics}
\label{sec:bkgres}

Our numerical results describing charged hadron observables are presented in Fig.~\ref{fig:s200}. The upper panel shows the transverse-momentum spectra, the middle panel displays the pseudorapidity distributions, and the lower panel illustrates the elliptic flow. These results are compared with experimental data from Au+Au collisions at $\sqrt{s_{\rm NN}} = 200$ GeV across various centrality bins. Overall, we observe good agreement between the model calculations and the experimental data. 
  
\begin{figure}[t]
  \includegraphics[width=0.4 \textwidth]{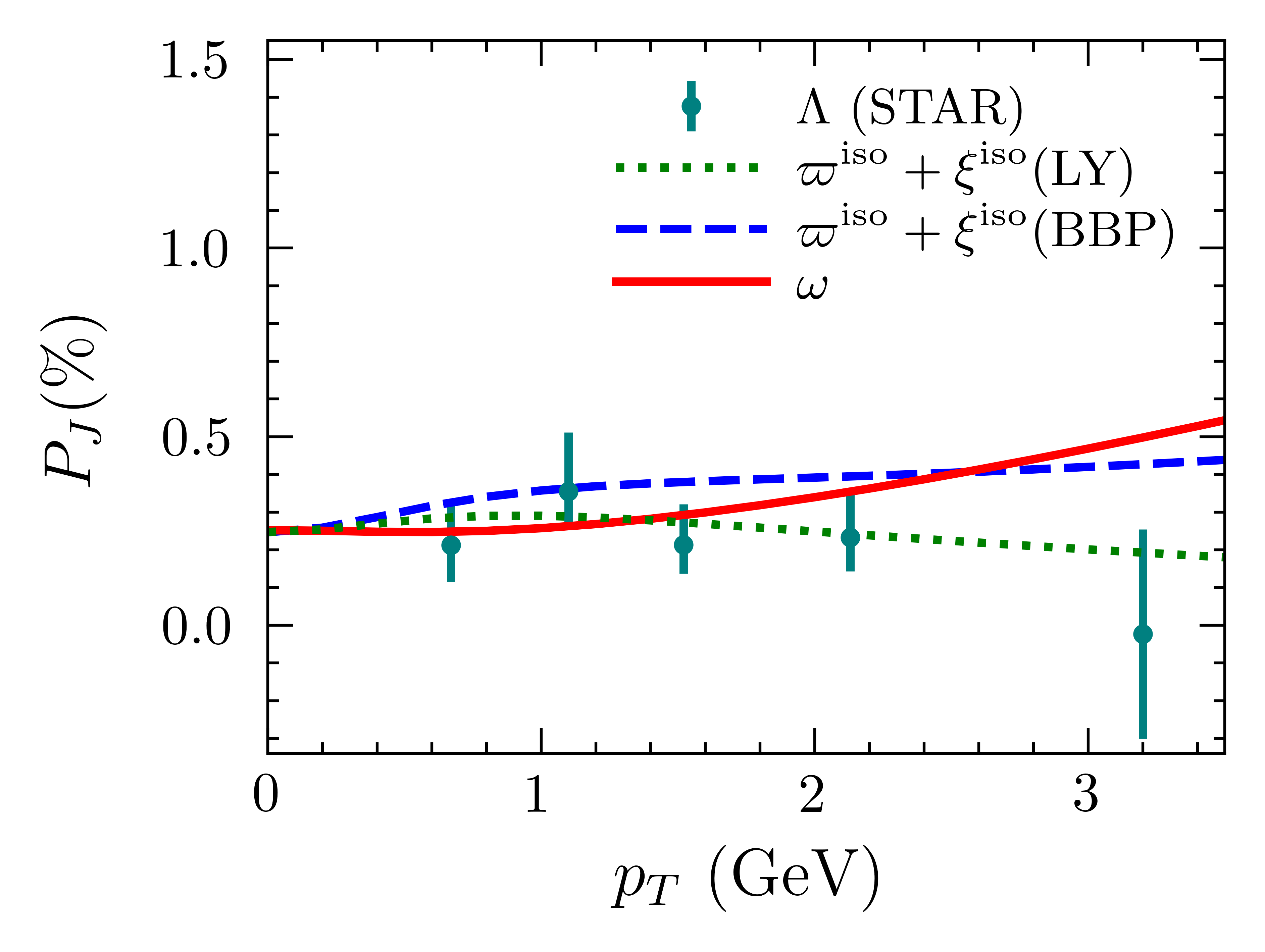}
  \includegraphics[width=0.4 \textwidth]{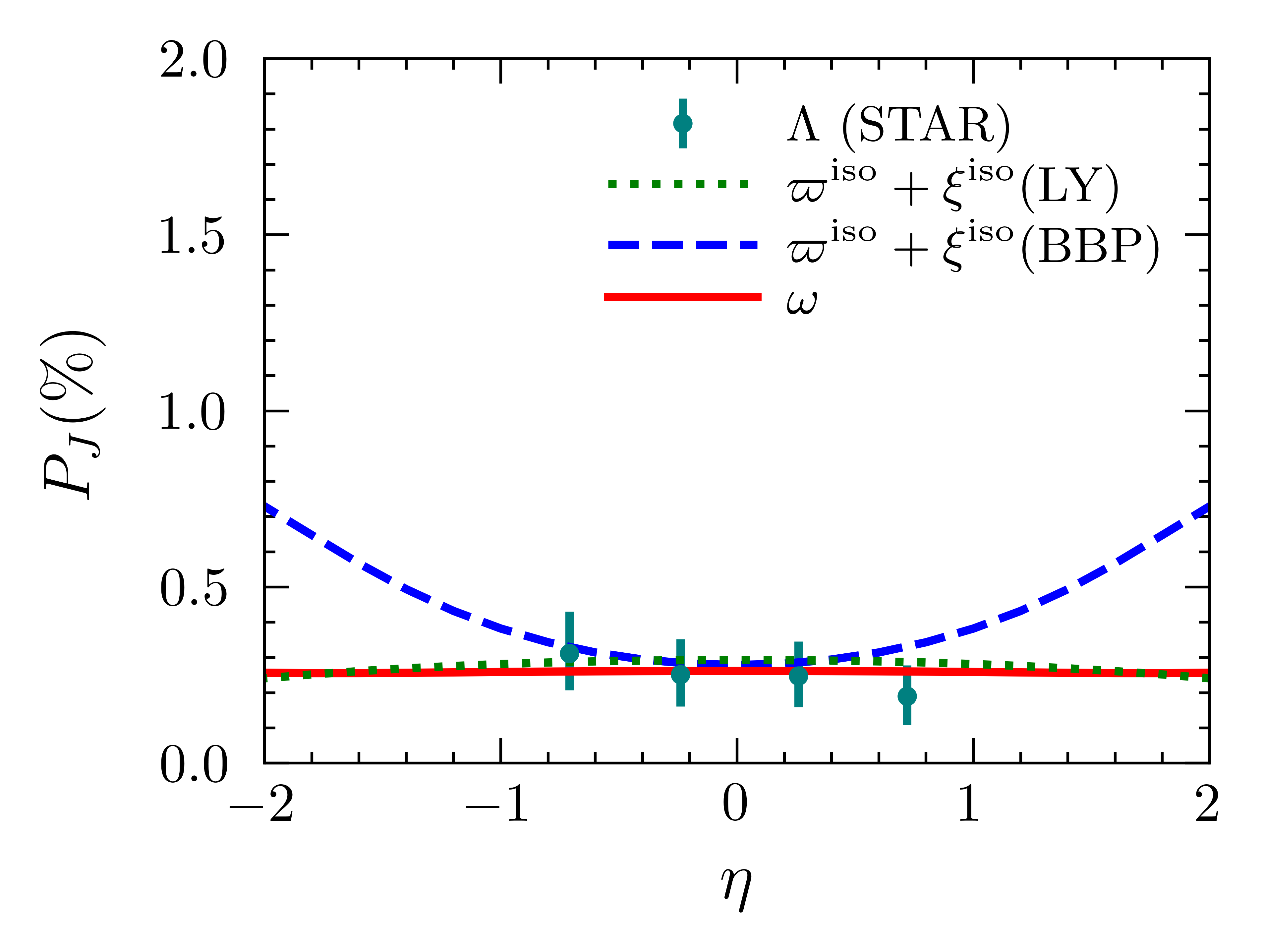}
  \caption{Same as Fig.~\ref{fig:spintimePJ} but with   $m=300$ MeV.}
  \label{fig:spintimePJ300}
\end{figure}

\begin{figure}[t]
  \includegraphics[width=0.4 \textwidth]{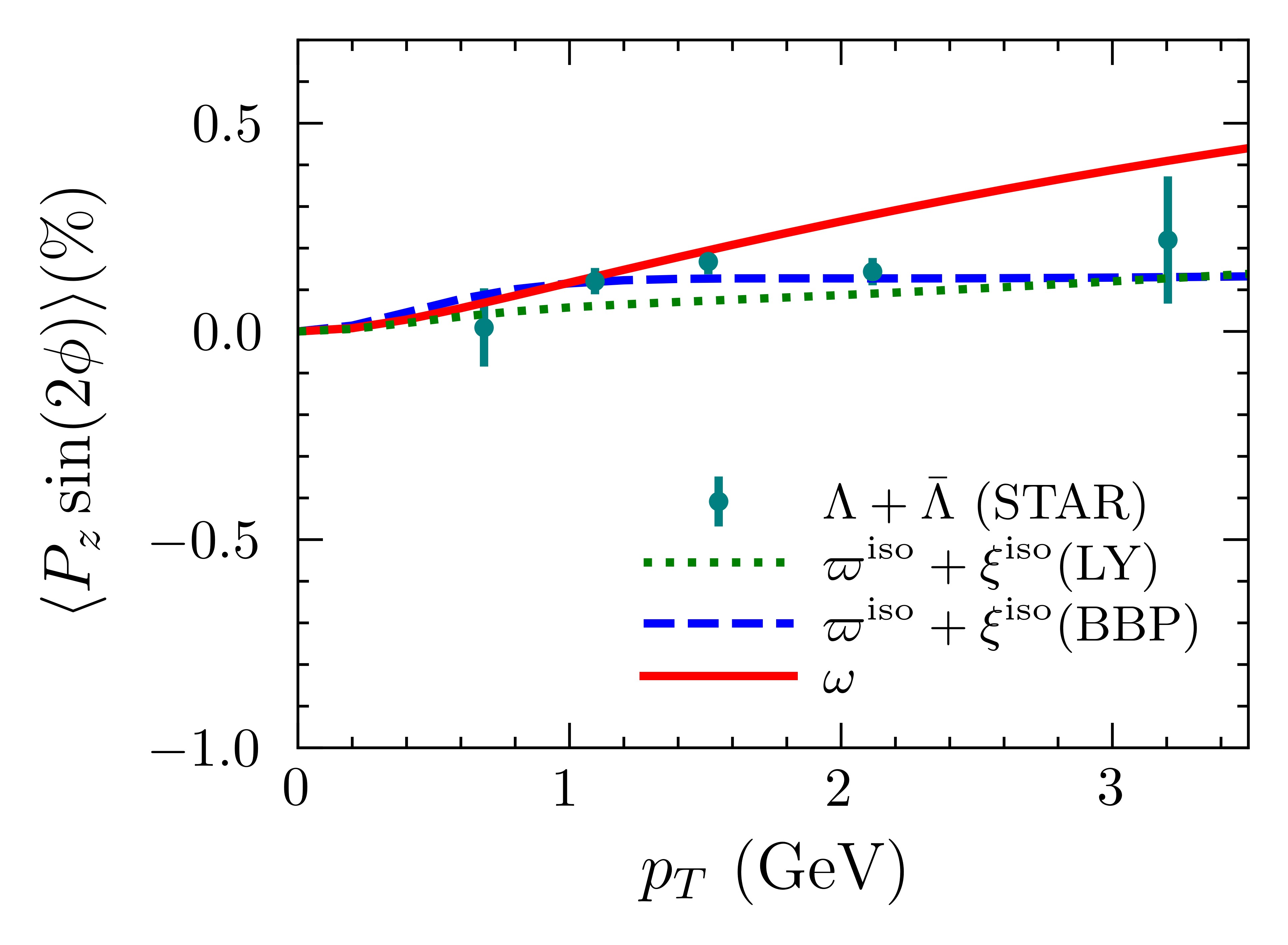}
  \includegraphics[width=0.4 \textwidth]{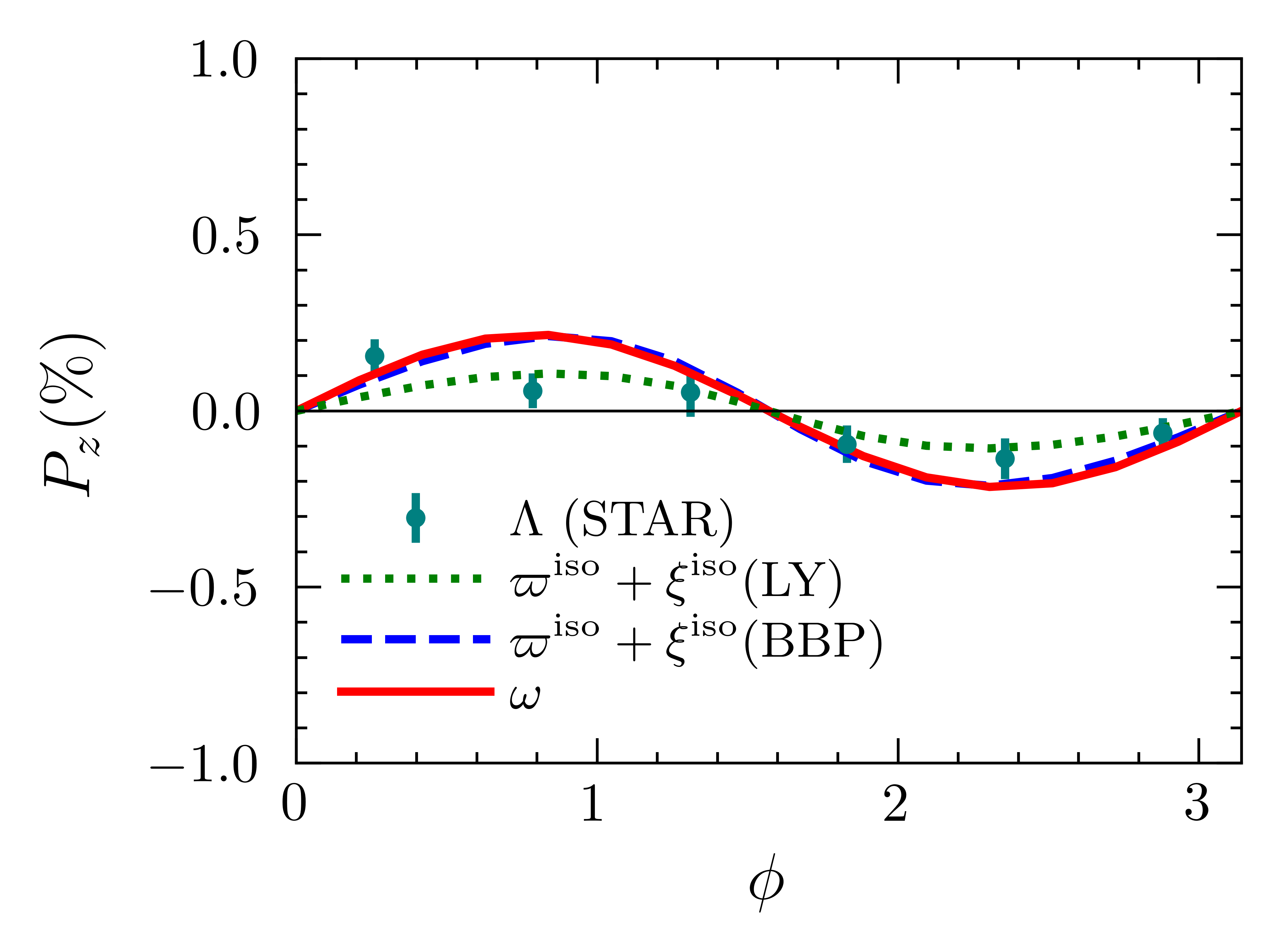}
  \caption{Same as Fig.~\ref{fig:spintimePz} but with   $m=300$ MeV.}
  \label{fig:spintimePz300}
\end{figure}
    
\subsection{Spin hydrodynamics}

\begin{figure}[t]
  \includegraphics[width=0.4 \textwidth]{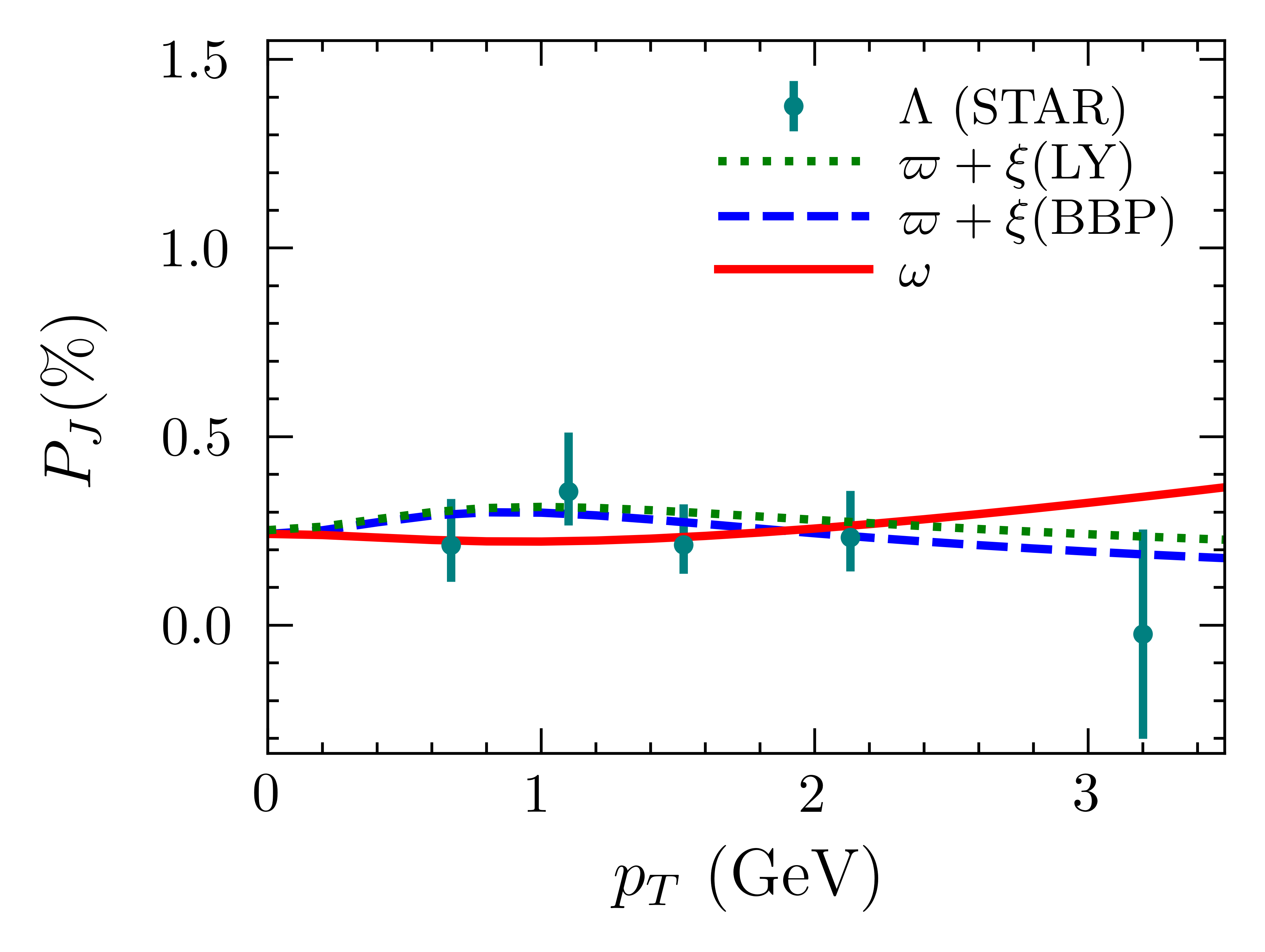}
  \includegraphics[width=0.4 \textwidth]{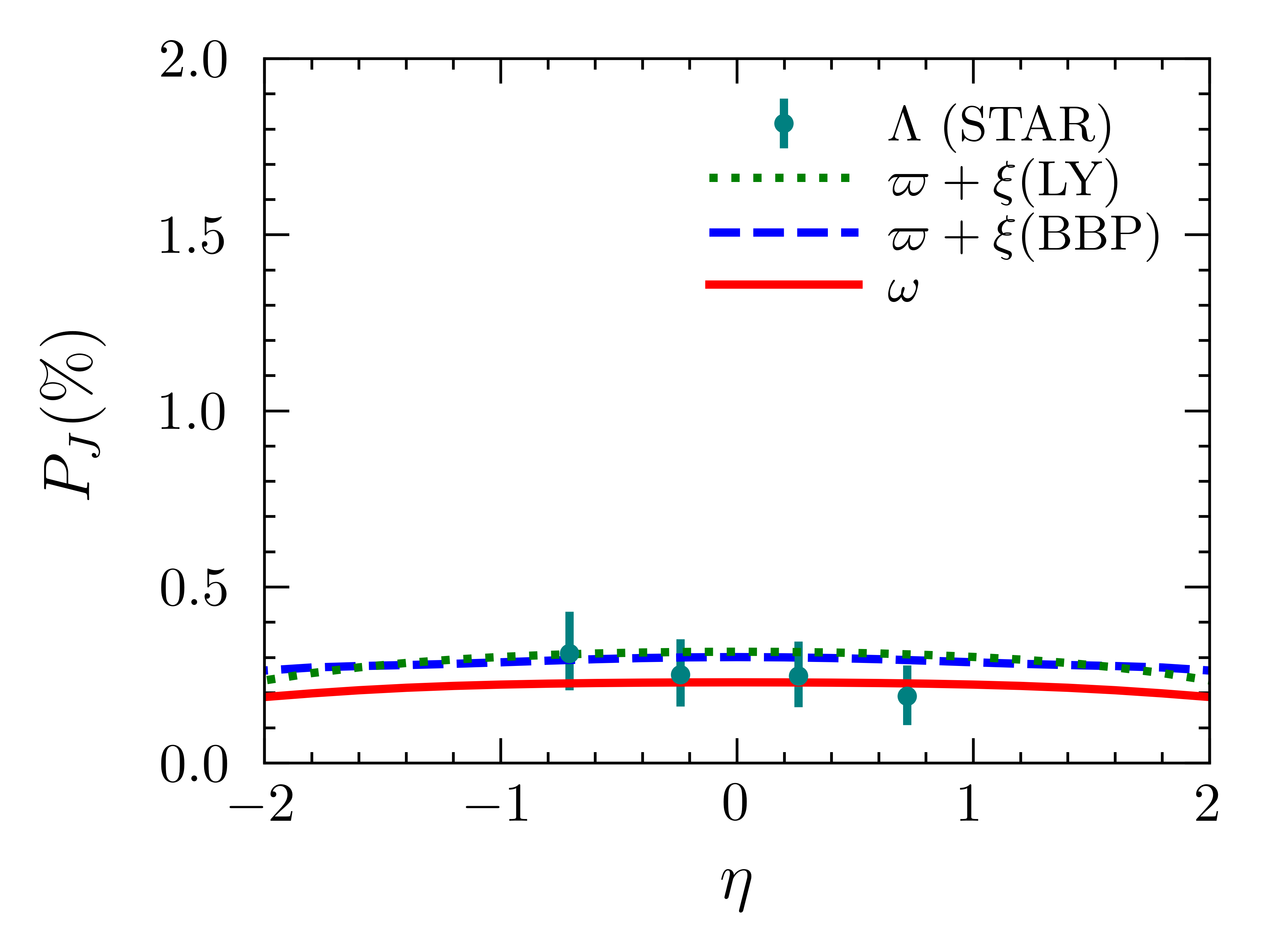}
  \caption{Same as Fig.~\ref{fig:spintimePJ} but with initial condition (\ref{eqn:ic2}) and parameters $\tau_0^s = 4$ fm and $m=300$ MeV.}
  \label{fig:spintimePJ300T}
\end{figure}

\begin{figure}[t]
  \includegraphics[width=0.4 \textwidth]{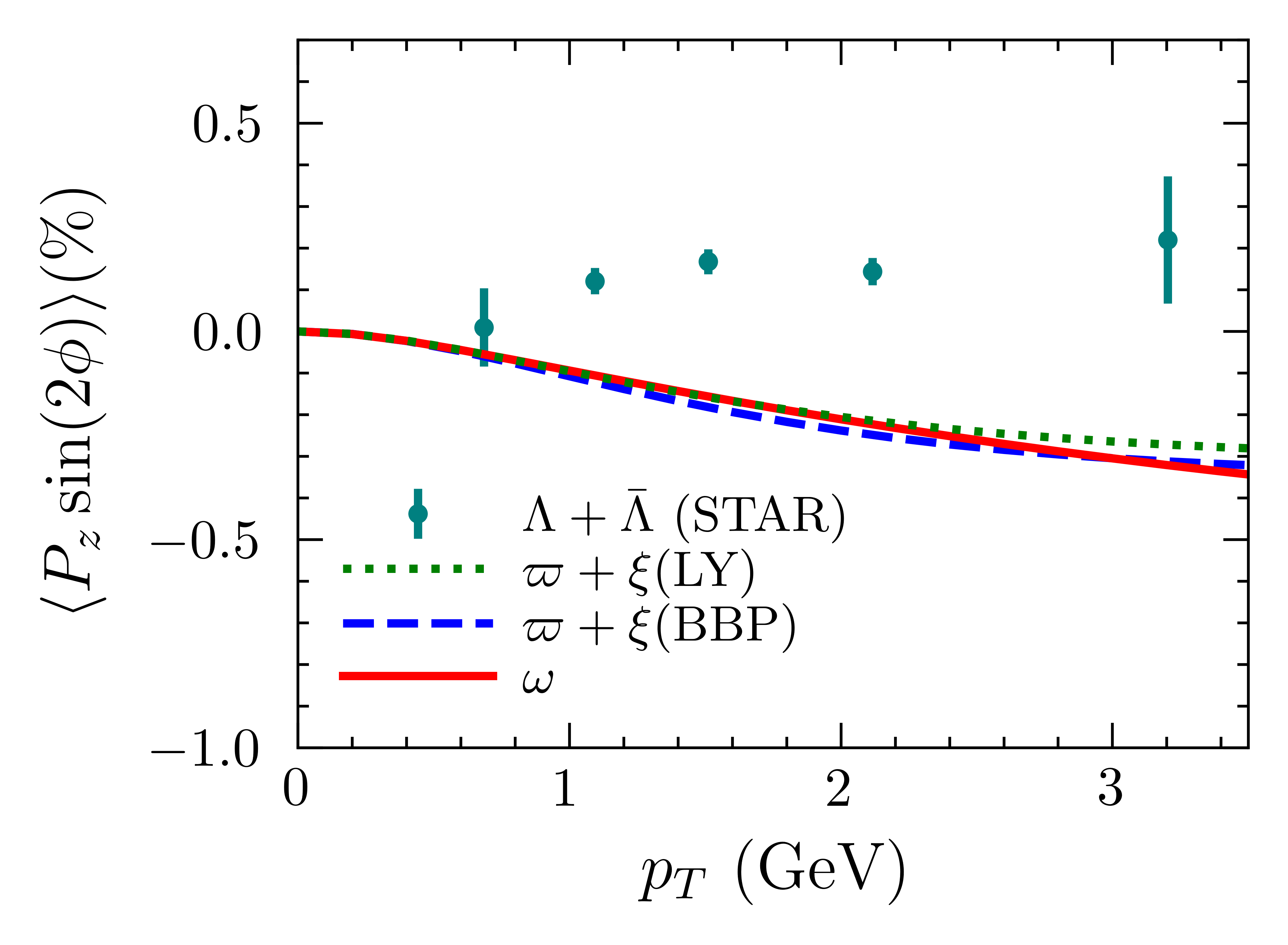}
  \includegraphics[width=0.4 \textwidth]{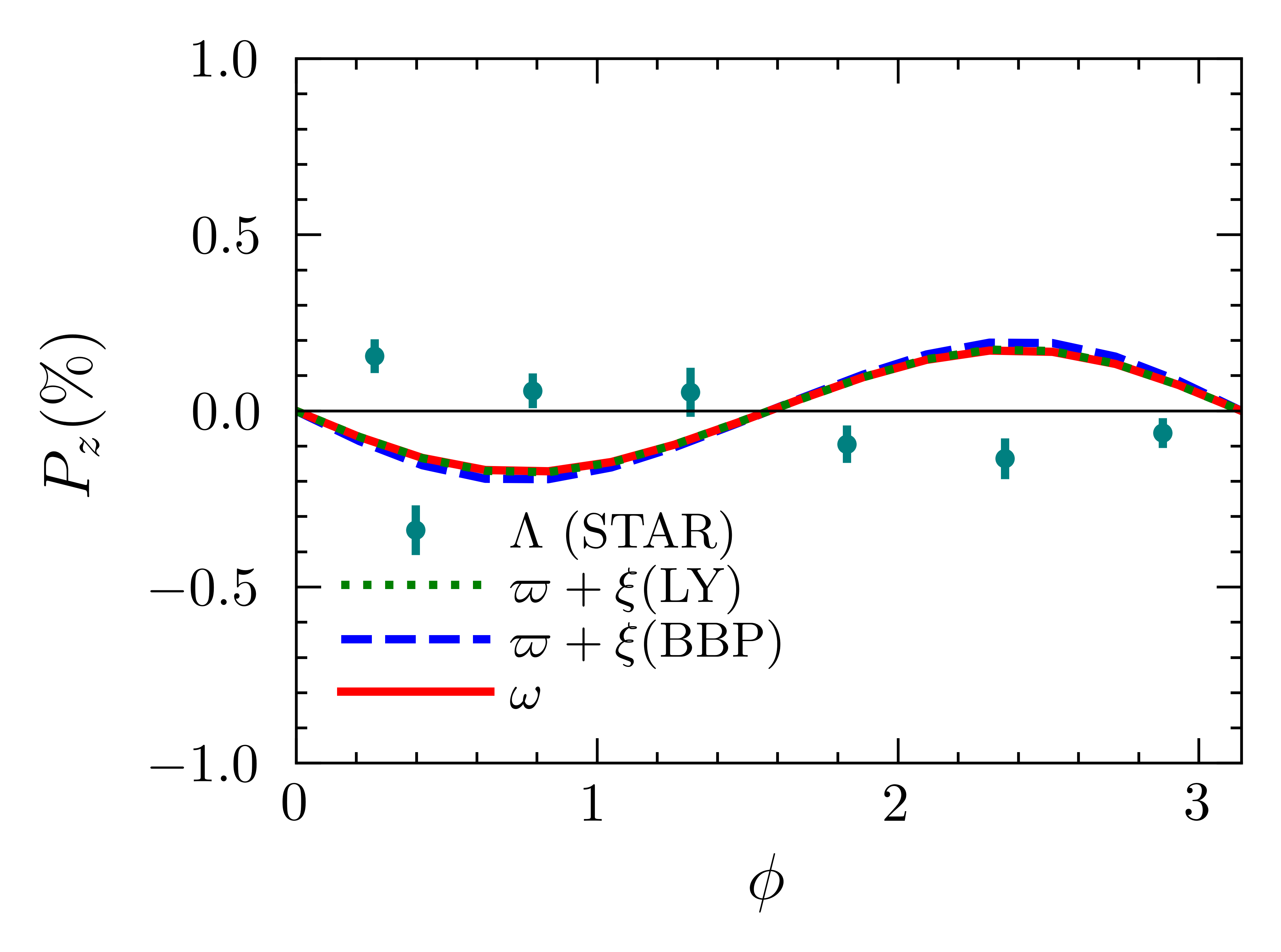}
  \caption{Same as Fig.~\ref{fig:spintimePz} but with initial condition (\ref{eqn:ic2}) and parameters $\tau_0^s = 4$ fm and $m=300$ MeV.}
  \label{fig:spintimePz300T}
\end{figure}

\begin{figure}[t]
  \includegraphics[width=0.4 \textwidth]{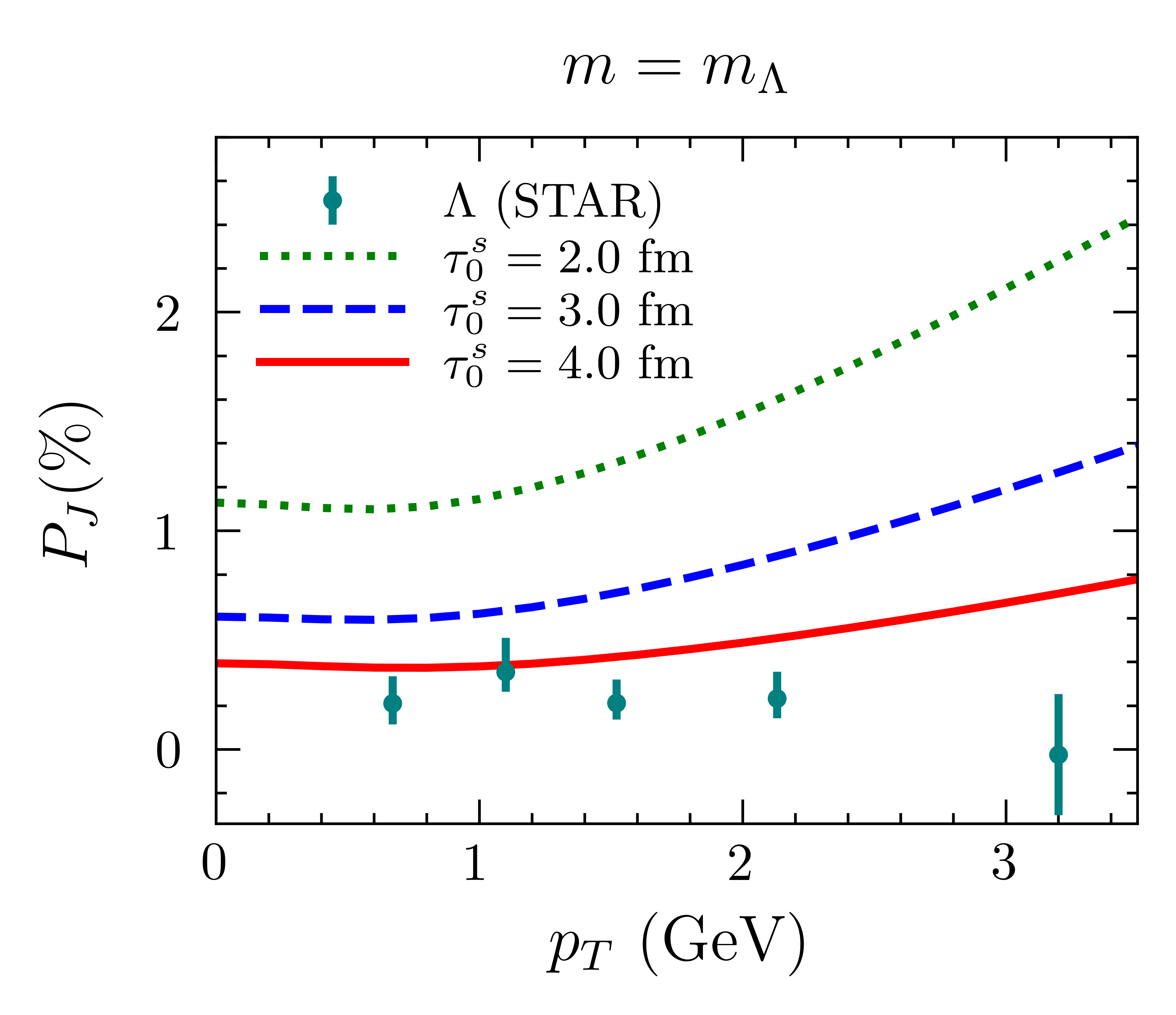}
  \includegraphics[width=0.42 \textwidth]{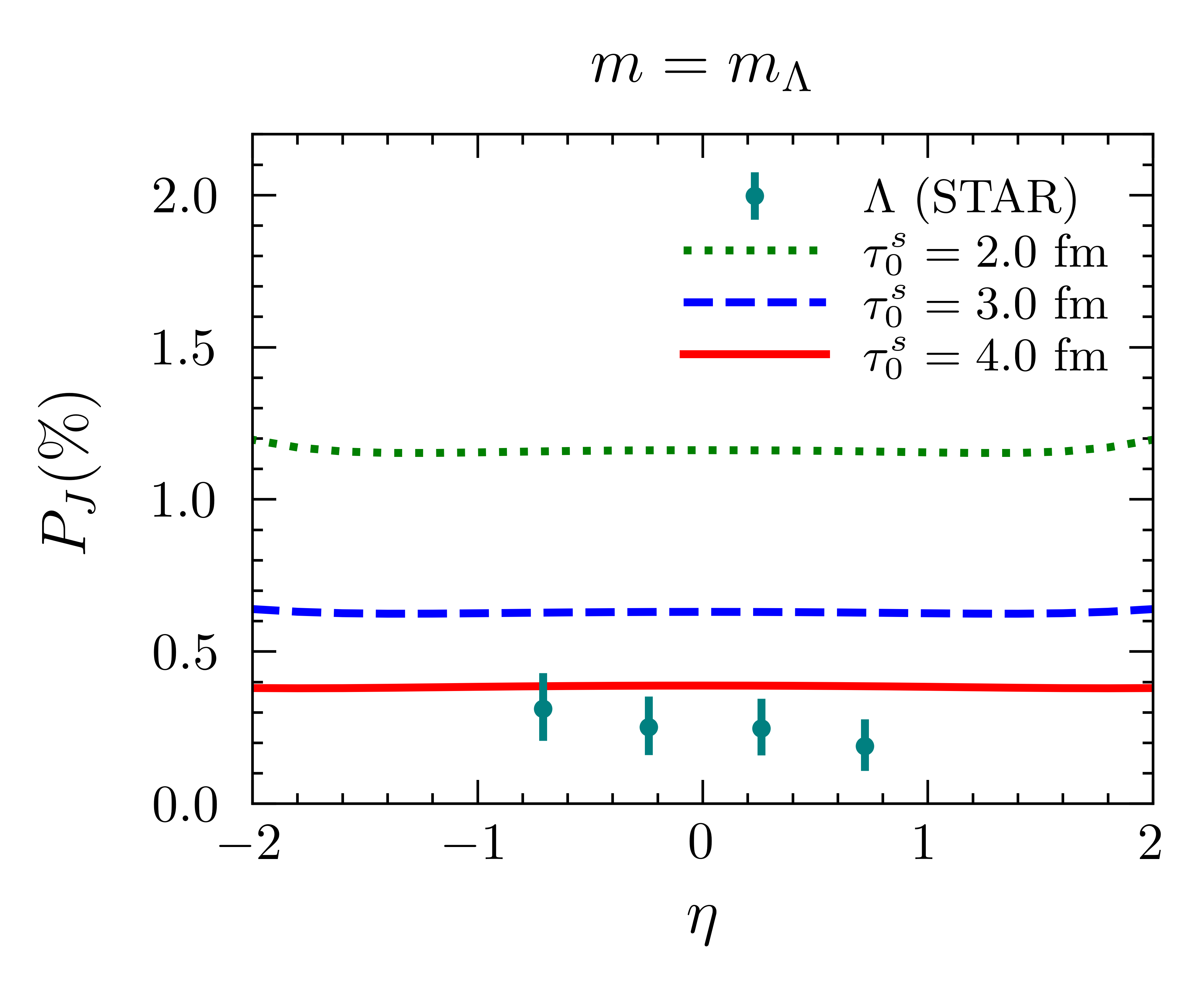}
  \caption{Our numerical results for the component of $\Lambda$ polarization along the orbital angular momentum direction for different initial time of spin evolution $\tau_0^s$ keeping mass fixed as $m_\Lambda$. Experimental data is taken from Ref.~\cite{STAR:2018gyt}  with updated decay parameter ($\alpha_\Lambda = 0.732$).}
  \label{fig:difftau0sPJ}
\end{figure}

\begin{figure}[t]
  \includegraphics[width=0.4 \textwidth]{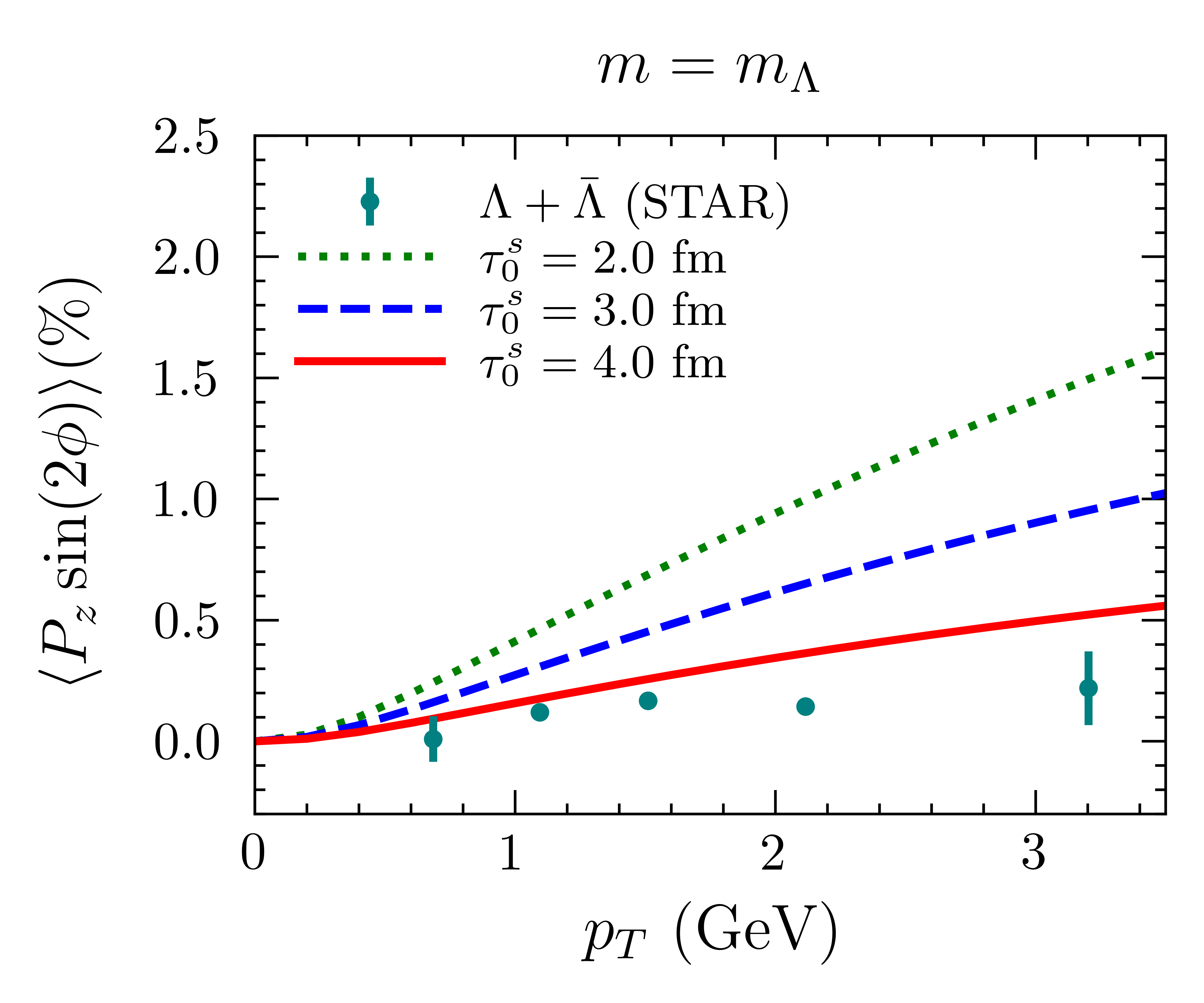}
  \includegraphics[width=0.42 \textwidth]{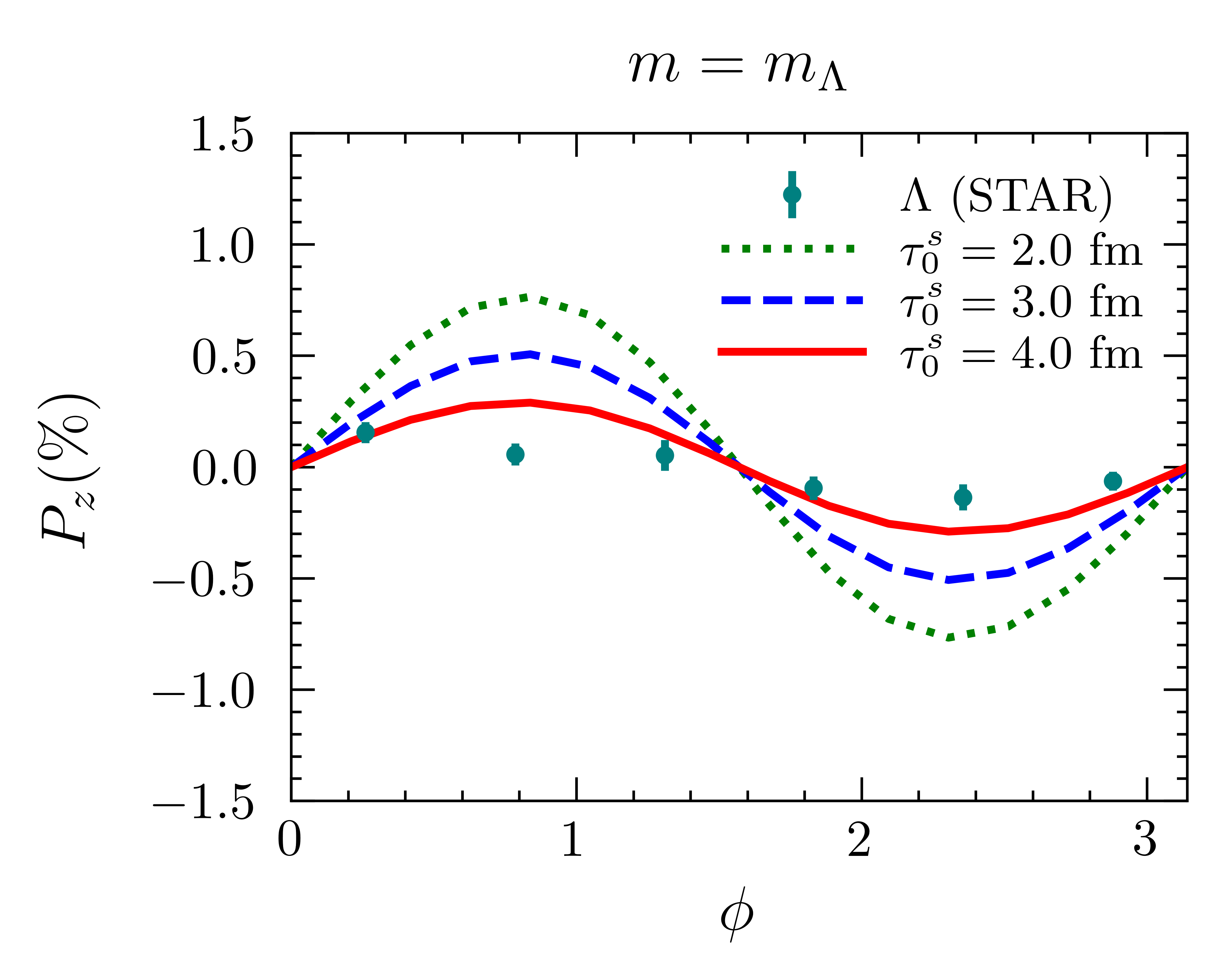}
  \caption{Our numerical results for longitudinal $\Lambda$ polarization for different initial time of spin evolution $\tau_0^s$ keeping mass fixed as $m_\Lambda$. Experimental data are taken from Ref.~\cite{STAR:2019erd} with updated decay parameter ($\alpha_\Lambda = 0.732$).}
    \label{fig:difftau0sPz}
\end{figure}

Figures \ref{fig:spintimePJ} -- \ref{fig:difftau0sPz} present a collection of our numerical results for the spin polarization of $\Lambda$ hyperons, obtained using various model parameters and initial conditions. We varied the initial time for spin hydrodynamics, $\tau_0^s$, the effective mass appearing in the spin tensor, $m$, and the form of the initial conditions by switching between Eqs.~(\ref{eqn:ic1}) and (\ref{eqn:ic2}).

Figures~\ref{fig:spintimePJ} and~\ref{fig:spintimePz} show our results (solid red lines) for the global and longitudinal polarizations, respectively, using $\tau_0^s = 4$~fm and $m=m_\Lambda$. Experimental data are represented by teal dots with vertical lines marking the experimental errors. For comparison, we also show the results obtained within the frameworks proposed in Ref.~\cite{Becattini:2021suc} (blue dashed lines , denoted as BBP) and Ref.~\cite{Liu:2021uhn} (green dotted lines, denoted as LY), using Eqs.~\eqref{eq:SsFB} and \eqref{eq:SsLY}, respectively. In the case of the global polarization, Fig.~\ref{fig:spintimePJ}, our prediction for the $p_T$ dependence of $P_J$ agrees with the experimental data for $p_T$ smaller than 1~GeV. On the other hand, our results for the rapidity dependence of $P_J$ are very close to the data in the whole measured range. Similar trends can be observed for the longitudinal polarization shown in Fig.~\ref{fig:spintimePz}. The results presented in Figs.~\ref{fig:spintimePJ} and~\ref{fig:spintimePz} indicate that the LY and BBP frameworks describe the data better, especially the LY approach, which accurately reproduces all the experimental points, including those at large $p_T$ for $P_J$. 

To examine the effect of reducing the effective mass, we performed calculations within our framework using $m = 300$~MeV. We stress that we use the reduced mass in the calculations that determine the spacetime evolution of the spin tensor only. To calculate the spin polarization we use the physical $\Lambda$ mass. Our results with the reduced mass (with initial time and form of the initial condition for the spin polarization tensor unchanged) are shown in Figs.~\ref{fig:spintimePJ300} and~\ref{fig:spintimePz300}. Here, we observe a clear improvement of the agreement with the data compared to the case where the value $m=m_\Lambda$ is used throughout the calculations. In this case, our results closely align with the BBP predictions.
	
To further explore the sensitivity of our results to the form of the initial conditions, we have replaced the form \eqref{eqn:ic1} by \eqref{eqn:ic2}, thereby including temperature gradients in the initial conditions. Again, with the reduced mass but unchanged initial time for spin hydrodynamics we obtain the results shown in Figs.~\ref{fig:spintimePJ300T} and~\ref{fig:spintimePz300T}. While the magnitude of $P_J$ is only slightly changed, the longitudinal polarization changes sign --- an effect known from early attempts to describe $P_z$ with thermal vorticity, Eq.~\eqref{eq:SvFB}, only and dubbed the ``sign problem''.

The results presented so far are consistent with earlier findings suggesting that omitting temperature gradients or reducing the mass improves agreement with data.  Our framework offers, however, new insights into the spin evolution of matter produced in heavy-ion collisions. Since our framework is based on the equation that describes conservation of the spin part of the total angular momentum, its ability to reproduce the experimental data suggests that the last stages of the evolution of matter are dominated by the processes that do not exchange the spin and orbital parts of the angular momentum. 

To determine when the initial stage with a strong spin-orbit coupling transitions into a stage where this coupling is weak, we initiated our spin evolution at different times, specifically $\tau_0^s = 2, 3$, and 4~fm. The results of these calculations (using $m=m_\Lambda$) are shown in Figs.~\ref{fig:difftau0sPJ} and~\ref{fig:difftau0sPz}. Inspection of these figures indicates that the initialization time for spin hydrodynamics cannot be set too early. The acceptable agreement with the data is only achieved with $\tau_0^s = 4$~fm. In a way, the exclusion of early times from our analysis of spin dynamics seems reasonable, since the formalism of spin hydrodynamics used in this work refers to well-defined particles (quasiparticles) with spin~\nicefrac{1}{2}. In the very early stages of heavy-ion collisions, the system is very strongly interacting, the concept of quasiparticles can be questioned, and the currently developed methods of spin hydrodynamics might be not applicable. 
	
\section{Summary and outlook}
\label{sec:summary}
	
In this work, we solved the equations of perfect spin hydrodynamics for the first time using a realistic 3+1D hydrodynamic background, calibrated to reproduce a wide range of hadronic observables measured in Au+Au collisions at the beam energy of $\sqrt{s_{\rm NN}} = 200$~GeV. The spin dynamics of spin 1/2 particles has been determined from the conservation law for the spin part of the total angular momentum. The mass appearing in the spin tensor has been treated as an effective parameter.  We have analyzed several scenarios by varying the effective mass, the initial evolution time, and the form of the initial spin polarization tensor.  Our model predictions have been compared with the data on global and longitudinal spin polarization of $\Lambda$ hyperons. The results indicate that a successful description of the data requires a delayed initial evolution time for the perfect spin hydrodynamics of about 4~fm (compared to the standard initial time of 1 fm used for the hydrodynamic background). This delay signals a transition from the phase where the spin-orbit interaction is significant to the regime where spin-conserving processes dominate. Our results suggest that the spin-orbit dissipative interaction plays a significant role only in the very early stages of the system evolution.
	
\begin{acknowledgments}
The authors (S.K.S and W.F.) thank ECT* for support allowing participation in the workshop ``Spin and quantum features of the QCD plasma" where some part of this work was done. This research was supported in part by the Polish National Science Center (NCN) Grants No.~2018/30/E/ST2/00432 (R.R., S.K.S.) and 2022/47/B/ST2/01372 (W.F.). S.K.S is supported by ICSC - \emph{Centro Nazionale di Ricerca in High Performance Computing, Big Data and Quantum Computing}, funded by the European Union - NextGenerationEU
\end{acknowledgments}
 \newpage

\bibliographystyle{apsrev4-2}
\bibliography{spinhydro}

\begin{thebibliography}{121}%
\makeatletter
\providecommand \@ifxundefined [1]{%
 \@ifx{#1\undefined}
}%
\providecommand \@ifnum [1]{%
 \ifnum #1\expandafter \@firstoftwo
 \else \expandafter \@secondoftwo
 \fi
}%
\providecommand \@ifx [1]{%
 \ifx #1\expandafter \@firstoftwo
 \else \expandafter \@secondoftwo
 \fi
}%
\providecommand \natexlab [1]{#1}%
\providecommand \enquote  [1]{``#1''}%
\providecommand \bibnamefont  [1]{#1}%
\providecommand \bibfnamefont [1]{#1}%
\providecommand \citenamefont [1]{#1}%
\providecommand \href@noop [0]{\@secondoftwo}%
\providecommand \href [0]{\begingroup \@sanitize@url \@href}%
\providecommand \@href[1]{\@@startlink{#1}\@@href}%
\providecommand \@@href[1]{\endgroup#1\@@endlink}%
\providecommand \@sanitize@url [0]{\catcode `\\12\catcode `\$12\catcode
  `\&12\catcode `\#12\catcode `\^12\catcode `\_12\catcode `\%12\relax}%
\providecommand \@@startlink[1]{}%
\providecommand \@@endlink[0]{}%
\providecommand \url  [0]{\begingroup\@sanitize@url \@url }%
\providecommand \@url [1]{\endgroup\@href {#1}{\urlprefix }}%
\providecommand \urlprefix  [0]{URL }%
\providecommand \Eprint [0]{\href }%
\providecommand \doibase [0]{https://doi.org/}%
\providecommand \selectlanguage [0]{\@gobble}%
\providecommand \bibinfo  [0]{\@secondoftwo}%
\providecommand \bibfield  [0]{\@secondoftwo}%
\providecommand \translation [1]{[#1]}%
\providecommand \BibitemOpen [0]{}%
\providecommand \bibitemStop [0]{}%
\providecommand \bibitemNoStop [0]{.\EOS\space}%
\providecommand \EOS [0]{\spacefactor3000\relax}%
\providecommand \BibitemShut  [1]{\csname bibitem#1\endcsname}%
\let\auto@bib@innerbib\@empty
\bibitem [{\citenamefont {Agakishiev}\ \emph {et~al.}(2014)\citenamefont
  {Agakishiev} \emph {et~al.}}]{HADES:2014ttv}%
  \BibitemOpen
  \bibfield  {author} {\bibinfo {author} {\bibfnamefont {G.}~\bibnamefont
  {Agakishiev}} \emph {et~al.} (\bibinfo {collaboration} {HADES}),\ }\href
  {https://doi.org/10.1140/epja/i2014-14081-2} {\bibfield  {journal} {\bibinfo
  {journal} {Eur. Phys. J. A}\ }\textbf {\bibinfo {volume} {50}},\ \bibinfo
  {pages} {81} (\bibinfo {year} {2014})},\ \Eprint
  {https://arxiv.org/abs/1404.3014} {arXiv:1404.3014 [nucl-ex]} \BibitemShut
  {NoStop}%
\bibitem [{\citenamefont {Adamczyk}\ \emph {et~al.}(2017)\citenamefont
  {Adamczyk} \emph {et~al.}}]{STAR:2017ckg}%
  \BibitemOpen
  \bibfield  {author} {\bibinfo {author} {\bibfnamefont {L.}~\bibnamefont
  {Adamczyk}} \emph {et~al.} (\bibinfo {collaboration} {STAR}),\ }\href
  {https://doi.org/10.1038/nature23004} {\bibfield  {journal} {\bibinfo
  {journal} {Nature}\ }\textbf {\bibinfo {volume} {548}},\ \bibinfo {pages}
  {62} (\bibinfo {year} {2017})},\ \Eprint {https://arxiv.org/abs/1701.06657}
  {arXiv:1701.06657 [nucl-ex]} \BibitemShut {NoStop}%
\bibitem [{\citenamefont {Adam}\ \emph {et~al.}(2018)\citenamefont {Adam} \emph
  {et~al.}}]{STAR:2018gyt}%
  \BibitemOpen
  \bibfield  {author} {\bibinfo {author} {\bibfnamefont {J.}~\bibnamefont
  {Adam}} \emph {et~al.} (\bibinfo {collaboration} {STAR}),\ }\href
  {https://doi.org/10.1103/PhysRevC.98.014910} {\bibfield  {journal} {\bibinfo
  {journal} {Phys. Rev. C}\ }\textbf {\bibinfo {volume} {98}},\ \bibinfo
  {pages} {014910} (\bibinfo {year} {2018})},\ \Eprint
  {https://arxiv.org/abs/1805.04400} {arXiv:1805.04400 [nucl-ex]} \BibitemShut
  {NoStop}%
\bibitem [{\citenamefont {Adam}\ \emph {et~al.}(2019)\citenamefont {Adam} \emph
  {et~al.}}]{STAR:2019erd}%
  \BibitemOpen
  \bibfield  {author} {\bibinfo {author} {\bibfnamefont {J.}~\bibnamefont
  {Adam}} \emph {et~al.} (\bibinfo {collaboration} {STAR}),\ }\href
  {https://doi.org/10.1103/PhysRevLett.123.132301} {\bibfield  {journal}
  {\bibinfo  {journal} {Phys. Rev. Lett.}\ }\textbf {\bibinfo {volume} {123}},\
  \bibinfo {pages} {132301} (\bibinfo {year} {2019})},\ \Eprint
  {https://arxiv.org/abs/1905.11917} {arXiv:1905.11917 [nucl-ex]} \BibitemShut
  {NoStop}%
\bibitem [{\citenamefont {Acharya}\ \emph
  {et~al.}(2020{\natexlab{a}})\citenamefont {Acharya} \emph
  {et~al.}}]{ALICE:2019aid}%
  \BibitemOpen
  \bibfield  {author} {\bibinfo {author} {\bibfnamefont {S.}~\bibnamefont
  {Acharya}} \emph {et~al.} (\bibinfo {collaboration} {ALICE}),\ }\href
  {https://doi.org/10.1103/PhysRevLett.125.012301} {\bibfield  {journal}
  {\bibinfo  {journal} {Phys. Rev. Lett.}\ }\textbf {\bibinfo {volume} {125}},\
  \bibinfo {pages} {012301} (\bibinfo {year} {2020}{\natexlab{a}})},\ \Eprint
  {https://arxiv.org/abs/1910.14408} {arXiv:1910.14408 [nucl-ex]} \BibitemShut
  {NoStop}%
\bibitem [{\citenamefont {Abdallah}\ \emph {et~al.}(2021)\citenamefont
  {Abdallah} \emph {et~al.}}]{STAR:2021beb}%
  \BibitemOpen
  \bibfield  {author} {\bibinfo {author} {\bibfnamefont {M.~S.}\ \bibnamefont
  {Abdallah}} \emph {et~al.} (\bibinfo {collaboration} {STAR}),\ }\href
  {https://doi.org/10.1103/PhysRevC.104.L061901} {\bibfield  {journal}
  {\bibinfo  {journal} {Phys. Rev. C}\ }\textbf {\bibinfo {volume} {104}},\
  \bibinfo {pages} {L061901} (\bibinfo {year} {2021})},\ \Eprint
  {https://arxiv.org/abs/2108.00044} {arXiv:2108.00044 [nucl-ex]} \BibitemShut
  {NoStop}%
\bibitem [{\citenamefont {Acharya}\ \emph
  {et~al.}(2020{\natexlab{b}})\citenamefont {Acharya} \emph
  {et~al.}}]{ALICE:2019onw}%
  \BibitemOpen
  \bibfield  {author} {\bibinfo {author} {\bibfnamefont {S.}~\bibnamefont
  {Acharya}} \emph {et~al.} (\bibinfo {collaboration} {ALICE}),\ }\href
  {https://doi.org/10.1103/PhysRevC.101.044611} {\bibfield  {journal} {\bibinfo
   {journal} {Phys. Rev. C}\ }\textbf {\bibinfo {volume} {101}},\ \bibinfo
  {pages} {044611} (\bibinfo {year} {2020}{\natexlab{b}})},\ \bibinfo {note}
  {[Erratum: Phys.Rev.C 105, 029902 (2022)]},\ \Eprint
  {https://arxiv.org/abs/1909.01281} {arXiv:1909.01281 [nucl-ex]} \BibitemShut
  {NoStop}%
\bibitem [{\citenamefont {Niida}\ and\ \citenamefont
  {Voloshin}(2024)}]{Niida:2024ntm}%
  \BibitemOpen
  \bibfield  {author} {\bibinfo {author} {\bibfnamefont {T.}~\bibnamefont
  {Niida}}\ and\ \bibinfo {author} {\bibfnamefont {S.~A.}\ \bibnamefont
  {Voloshin}},\ }\href {https://doi.org/10.1142/S0218301324300108} {\bibfield
  {journal} {\bibinfo  {journal} {Int. J. Mod. Phys. E}\ }\textbf {\bibinfo
  {volume} {33}},\ \bibinfo {pages} {2430010} (\bibinfo {year} {2024})},\
  \Eprint {https://arxiv.org/abs/2404.11042} {arXiv:2404.11042 [nucl-ex]}
  \BibitemShut {NoStop}%
\bibitem [{\citenamefont {Becattini}\ \emph
  {et~al.}(2024{\natexlab{a}})\citenamefont {Becattini}, \citenamefont
  {Buzzegoli}, \citenamefont {Niida}, \citenamefont {Pu}, \citenamefont
  {Tang},\ and\ \citenamefont {Wang}}]{Becattini2024_IJMPE}%
  \BibitemOpen
  \bibfield  {author} {\bibinfo {author} {\bibfnamefont {F.}~\bibnamefont
  {Becattini}}, \bibinfo {author} {\bibfnamefont {M.}~\bibnamefont
  {Buzzegoli}}, \bibinfo {author} {\bibfnamefont {T.}~\bibnamefont {Niida}},
  \bibinfo {author} {\bibfnamefont {S.}~\bibnamefont {Pu}}, \bibinfo {author}
  {\bibfnamefont {A.-H.}\ \bibnamefont {Tang}},\ and\ \bibinfo {author}
  {\bibfnamefont {Q.}~\bibnamefont {Wang}},\ }\href
  {https://doi.org/10.1142/S0218301324300066} {\bibfield  {journal} {\bibinfo
  {journal} {International Journal of Modern Physics E}\ }\textbf {\bibinfo
  {volume} {33}},\ \bibinfo {pages} {2430006} (\bibinfo {year}
  {2024}{\natexlab{a}})},\ \Eprint {https://arxiv.org/abs/2402.04540}
  {arXiv:2402.04540 [nucl-th]} \BibitemShut {NoStop}%
\bibitem [{\citenamefont {Bondar}\ and\ \citenamefont
  {Florkowski}(2024)}]{Bondar:2024uqk}%
  \BibitemOpen
  \bibfield  {author} {\bibinfo {author} {\bibfnamefont {Y.}~\bibnamefont
  {Bondar}}\ and\ \bibinfo {author} {\bibfnamefont {W.}~\bibnamefont
  {Florkowski}},\ }\href {https://doi.org/10.5506/APhysPolB.55.9-A1} {\bibfield
   {journal} {\bibinfo  {journal} {Acta Phys. Polon. B}\ }\textbf {\bibinfo
  {volume} {55}},\ \bibinfo {pages} {9} (\bibinfo {year} {2024})},\ \Eprint
  {https://arxiv.org/abs/2408.11050} {arXiv:2408.11050 [hep-ph]} \BibitemShut
  {NoStop}%
\bibitem [{\citenamefont {Sorensen}\ \emph {et~al.}(2024)\citenamefont
  {Sorensen} \emph {et~al.}}]{Sorensen:2023zkk}%
  \BibitemOpen
  \bibfield  {author} {\bibinfo {author} {\bibfnamefont {A.}~\bibnamefont
  {Sorensen}} \emph {et~al.},\ }\href
  {https://doi.org/10.1016/j.ppnp.2023.104080} {\bibfield  {journal} {\bibinfo
  {journal} {Prog. Part. Nucl. Phys.}\ }\textbf {\bibinfo {volume} {134}},\
  \bibinfo {pages} {104080} (\bibinfo {year} {2024})},\ \Eprint
  {https://arxiv.org/abs/2301.13253} {arXiv:2301.13253 [nucl-th]} \BibitemShut
  {NoStop}%
\bibitem [{\citenamefont {Liu}\ and\ \citenamefont {Yin}(2021)}]{Liu:2021uhn}%
  \BibitemOpen
  \bibfield  {author} {\bibinfo {author} {\bibfnamefont {S.~Y.~F.}\
  \bibnamefont {Liu}}\ and\ \bibinfo {author} {\bibfnamefont {Y.}~\bibnamefont
  {Yin}},\ }\href {https://doi.org/10.1007/JHEP07(2021)188} {\bibfield
  {journal} {\bibinfo  {journal} {JHEP}\ }\textbf {\bibinfo {volume} {07}},\
  \bibinfo {pages} {188}},\ \Eprint {https://arxiv.org/abs/2103.09200}
  {arXiv:2103.09200 [hep-ph]} \BibitemShut {NoStop}%
\bibitem [{\citenamefont {Fu}\ \emph {et~al.}(2021)\citenamefont {Fu},
  \citenamefont {Liu}, \citenamefont {Pang}, \citenamefont {Song},\ and\
  \citenamefont {Yin}}]{Fu:2021pok}%
  \BibitemOpen
  \bibfield  {author} {\bibinfo {author} {\bibfnamefont {B.}~\bibnamefont
  {Fu}}, \bibinfo {author} {\bibfnamefont {S.~Y.~F.}\ \bibnamefont {Liu}},
  \bibinfo {author} {\bibfnamefont {L.}~\bibnamefont {Pang}}, \bibinfo {author}
  {\bibfnamefont {H.}~\bibnamefont {Song}},\ and\ \bibinfo {author}
  {\bibfnamefont {Y.}~\bibnamefont {Yin}},\ }\href
  {https://doi.org/10.1103/PhysRevLett.127.142301} {\bibfield  {journal}
  {\bibinfo  {journal} {Phys. Rev. Lett.}\ }\textbf {\bibinfo {volume} {127}},\
  \bibinfo {pages} {142301} (\bibinfo {year} {2021})},\ \Eprint
  {https://arxiv.org/abs/2103.10403} {arXiv:2103.10403 [hep-ph]} \BibitemShut
  {NoStop}%
\bibitem [{\citenamefont {Becattini}\ \emph
  {et~al.}(2021{\natexlab{a}})\citenamefont {Becattini}, \citenamefont
  {Buzzegoli},\ and\ \citenamefont {Palermo}}]{Becattini:2021suc}%
  \BibitemOpen
  \bibfield  {author} {\bibinfo {author} {\bibfnamefont {F.}~\bibnamefont
  {Becattini}}, \bibinfo {author} {\bibfnamefont {M.}~\bibnamefont
  {Buzzegoli}},\ and\ \bibinfo {author} {\bibfnamefont {A.}~\bibnamefont
  {Palermo}},\ }\href {https://doi.org/10.1016/j.physletb.2021.136519}
  {\bibfield  {journal} {\bibinfo  {journal} {Phys. Lett. B}\ }\textbf
  {\bibinfo {volume} {820}},\ \bibinfo {pages} {136519} (\bibinfo {year}
  {2021}{\natexlab{a}})},\ \Eprint {https://arxiv.org/abs/2103.10917}
  {arXiv:2103.10917 [nucl-th]} \BibitemShut {NoStop}%
\bibitem [{\citenamefont {Becattini}\ \emph
  {et~al.}(2021{\natexlab{b}})\citenamefont {Becattini}, \citenamefont
  {Buzzegoli}, \citenamefont {Inghirami}, \citenamefont {Karpenko},\ and\
  \citenamefont {Palermo}}]{Becattini:2021iol}%
  \BibitemOpen
  \bibfield  {author} {\bibinfo {author} {\bibfnamefont {F.}~\bibnamefont
  {Becattini}}, \bibinfo {author} {\bibfnamefont {M.}~\bibnamefont
  {Buzzegoli}}, \bibinfo {author} {\bibfnamefont {G.}~\bibnamefont
  {Inghirami}}, \bibinfo {author} {\bibfnamefont {I.}~\bibnamefont
  {Karpenko}},\ and\ \bibinfo {author} {\bibfnamefont {A.}~\bibnamefont
  {Palermo}},\ }\href {https://doi.org/10.1103/PhysRevLett.127.272302}
  {\bibfield  {journal} {\bibinfo  {journal} {Phys. Rev. Lett.}\ }\textbf
  {\bibinfo {volume} {127}},\ \bibinfo {pages} {272302} (\bibinfo {year}
  {2021}{\natexlab{b}})},\ \Eprint {https://arxiv.org/abs/2103.14621}
  {arXiv:2103.14621 [nucl-th]} \BibitemShut {NoStop}%
\bibitem [{\citenamefont {Alzhrani}\ \emph {et~al.}(2022)\citenamefont
  {Alzhrani}, \citenamefont {Ryu},\ and\ \citenamefont
  {Shen}}]{Alzhrani:2022dpi}%
  \BibitemOpen
  \bibfield  {author} {\bibinfo {author} {\bibfnamefont {S.}~\bibnamefont
  {Alzhrani}}, \bibinfo {author} {\bibfnamefont {S.}~\bibnamefont {Ryu}},\ and\
  \bibinfo {author} {\bibfnamefont {C.}~\bibnamefont {Shen}},\ }\href
  {https://doi.org/10.1103/PhysRevC.106.014905} {\bibfield  {journal} {\bibinfo
   {journal} {Phys. Rev. C}\ }\textbf {\bibinfo {volume} {106}},\ \bibinfo
  {pages} {014905} (\bibinfo {year} {2022})},\ \Eprint
  {https://arxiv.org/abs/2203.15718} {arXiv:2203.15718 [nucl-th]} \BibitemShut
  {NoStop}%
\bibitem [{\citenamefont {Becattini}\ and\ \citenamefont
  {Tinti}(2010)}]{Becattini:2009wh}%
  \BibitemOpen
  \bibfield  {author} {\bibinfo {author} {\bibfnamefont {F.}~\bibnamefont
  {Becattini}}\ and\ \bibinfo {author} {\bibfnamefont {L.}~\bibnamefont
  {Tinti}},\ }\href {https://doi.org/10.1016/j.aop.2010.03.007} {\bibfield
  {journal} {\bibinfo  {journal} {Annals Phys.}\ }\textbf {\bibinfo {volume}
  {325}},\ \bibinfo {pages} {1566} (\bibinfo {year} {2010})},\ \Eprint
  {https://arxiv.org/abs/0911.0864} {arXiv:0911.0864 [gr-qc]} \BibitemShut
  {NoStop}%
\bibitem [{\citenamefont {Becattini}(2011)}]{Becattini:2011zz}%
  \BibitemOpen
  \bibfield  {author} {\bibinfo {author} {\bibfnamefont {F.}~\bibnamefont
  {Becattini}},\ }\href {https://doi.org/10.1134/S1547477111080036} {\bibfield
  {journal} {\bibinfo  {journal} {Phys. Part. Nucl. Lett.}\ }\textbf {\bibinfo
  {volume} {8}},\ \bibinfo {pages} {801} (\bibinfo {year} {2011})}\BibitemShut
  {NoStop}%
\bibitem [{\citenamefont {Becattini}\ \emph
  {et~al.}(2013{\natexlab{a}})\citenamefont {Becattini}, \citenamefont
  {Chandra}, \citenamefont {Del~Zanna},\ and\ \citenamefont
  {Grossi}}]{Becattini:2013fla}%
  \BibitemOpen
  \bibfield  {author} {\bibinfo {author} {\bibfnamefont {F.}~\bibnamefont
  {Becattini}}, \bibinfo {author} {\bibfnamefont {V.}~\bibnamefont {Chandra}},
  \bibinfo {author} {\bibfnamefont {L.}~\bibnamefont {Del~Zanna}},\ and\
  \bibinfo {author} {\bibfnamefont {E.}~\bibnamefont {Grossi}},\ }\href
  {https://doi.org/10.1016/j.aop.2013.07.004} {\bibfield  {journal} {\bibinfo
  {journal} {Annals Phys.}\ }\textbf {\bibinfo {volume} {338}},\ \bibinfo
  {pages} {32} (\bibinfo {year} {2013}{\natexlab{a}})},\ \Eprint
  {https://arxiv.org/abs/1303.3431} {arXiv:1303.3431 [nucl-th]} \BibitemShut
  {NoStop}%
\bibitem [{\citenamefont {Becattini}\ \emph
  {et~al.}(2013{\natexlab{b}})\citenamefont {Becattini}, \citenamefont
  {Csernai},\ and\ \citenamefont {Wang}}]{Becattini:2013vja}%
  \BibitemOpen
  \bibfield  {author} {\bibinfo {author} {\bibfnamefont {F.}~\bibnamefont
  {Becattini}}, \bibinfo {author} {\bibfnamefont {L.}~\bibnamefont {Csernai}},\
  and\ \bibinfo {author} {\bibfnamefont {D.~J.}\ \bibnamefont {Wang}},\ }\href
  {https://doi.org/10.1103/PhysRevC.88.034905} {\bibfield  {journal} {\bibinfo
  {journal} {Phys. Rev. C}\ }\textbf {\bibinfo {volume} {88}},\ \bibinfo
  {pages} {034905} (\bibinfo {year} {2013}{\natexlab{b}})},\ \bibinfo {note}
  {[Erratum: Phys.Rev.C 93, 069901 (2016)]},\ \Eprint
  {https://arxiv.org/abs/1304.4427} {arXiv:1304.4427 [nucl-th]} \BibitemShut
  {NoStop}%
\bibitem [{\citenamefont {Karpenko}\ and\ \citenamefont
  {Becattini}(2017)}]{Karpenko:2016jyx}%
  \BibitemOpen
  \bibfield  {author} {\bibinfo {author} {\bibfnamefont {I.}~\bibnamefont
  {Karpenko}}\ and\ \bibinfo {author} {\bibfnamefont {F.}~\bibnamefont
  {Becattini}},\ }\href {https://doi.org/10.1140/epjc/s10052-017-4765-1}
  {\bibfield  {journal} {\bibinfo  {journal} {Eur. Phys. J. C}\ }\textbf
  {\bibinfo {volume} {77}},\ \bibinfo {pages} {213} (\bibinfo {year} {2017})},\
  \Eprint {https://arxiv.org/abs/1610.04717} {arXiv:1610.04717 [nucl-th]}
  \BibitemShut {NoStop}%
\bibitem [{\citenamefont {Becattini}\ \emph {et~al.}(2017)\citenamefont
  {Becattini}, \citenamefont {Karpenko}, \citenamefont {Lisa}, \citenamefont
  {Upsal},\ and\ \citenamefont {Voloshin}}]{Becattini:2016gvu}%
  \BibitemOpen
  \bibfield  {author} {\bibinfo {author} {\bibfnamefont {F.}~\bibnamefont
  {Becattini}}, \bibinfo {author} {\bibfnamefont {I.}~\bibnamefont {Karpenko}},
  \bibinfo {author} {\bibfnamefont {M.}~\bibnamefont {Lisa}}, \bibinfo {author}
  {\bibfnamefont {I.}~\bibnamefont {Upsal}},\ and\ \bibinfo {author}
  {\bibfnamefont {S.}~\bibnamefont {Voloshin}},\ }\href
  {https://doi.org/10.1103/PhysRevC.95.054902} {\bibfield  {journal} {\bibinfo
  {journal} {Phys. Rev. C}\ }\textbf {\bibinfo {volume} {95}},\ \bibinfo
  {pages} {054902} (\bibinfo {year} {2017})},\ \Eprint
  {https://arxiv.org/abs/1610.02506} {arXiv:1610.02506 [nucl-th]} \BibitemShut
  {NoStop}%
\bibitem [{\citenamefont {Becattini}\ and\ \citenamefont
  {Karpenko}(2018)}]{Becattini:2017gcx}%
  \BibitemOpen
  \bibfield  {author} {\bibinfo {author} {\bibfnamefont {F.}~\bibnamefont
  {Becattini}}\ and\ \bibinfo {author} {\bibfnamefont {I.}~\bibnamefont
  {Karpenko}},\ }\href {https://doi.org/10.1103/PhysRevLett.120.012302}
  {\bibfield  {journal} {\bibinfo  {journal} {Phys. Rev. Lett.}\ }\textbf
  {\bibinfo {volume} {120}},\ \bibinfo {pages} {012302} (\bibinfo {year}
  {2018})},\ \Eprint {https://arxiv.org/abs/1707.07984} {arXiv:1707.07984
  [nucl-th]} \BibitemShut {NoStop}%
\bibitem [{\citenamefont {Palermo}\ \emph {et~al.}(2024)\citenamefont
  {Palermo}, \citenamefont {Grossi}, \citenamefont {Karpenko},\ and\
  \citenamefont {Becattini}}]{Palermo:2024tza}%
  \BibitemOpen
  \bibfield  {author} {\bibinfo {author} {\bibfnamefont {A.}~\bibnamefont
  {Palermo}}, \bibinfo {author} {\bibfnamefont {E.}~\bibnamefont {Grossi}},
  \bibinfo {author} {\bibfnamefont {I.}~\bibnamefont {Karpenko}},\ and\
  \bibinfo {author} {\bibfnamefont {F.}~\bibnamefont {Becattini}},\ }\href
  {https://doi.org/10.1140/epjc/s10052-024-13229-z} {\bibfield  {journal}
  {\bibinfo  {journal} {Eur. Phys. J. C}\ }\textbf {\bibinfo {volume} {84}},\
  \bibinfo {pages} {920} (\bibinfo {year} {2024})},\ \Eprint
  {https://arxiv.org/abs/2404.14295} {arXiv:2404.14295 [nucl-th]} \BibitemShut
  {NoStop}%
\bibitem [{\citenamefont {Florkowski}\ \emph
  {et~al.}(2018{\natexlab{a}})\citenamefont {Florkowski}, \citenamefont
  {Friman}, \citenamefont {Jaiswal},\ and\ \citenamefont
  {Speranza}}]{Florkowski:2017ruc}%
  \BibitemOpen
  \bibfield  {author} {\bibinfo {author} {\bibfnamefont {W.}~\bibnamefont
  {Florkowski}}, \bibinfo {author} {\bibfnamefont {B.}~\bibnamefont {Friman}},
  \bibinfo {author} {\bibfnamefont {A.}~\bibnamefont {Jaiswal}},\ and\ \bibinfo
  {author} {\bibfnamefont {E.}~\bibnamefont {Speranza}},\ }\href
  {https://doi.org/10.1103/PhysRevC.97.041901} {\bibfield  {journal} {\bibinfo
  {journal} {Phys. Rev. C}\ }\textbf {\bibinfo {volume} {97}},\ \bibinfo
  {pages} {041901} (\bibinfo {year} {2018}{\natexlab{a}})},\ \Eprint
  {https://arxiv.org/abs/1705.00587} {arXiv:1705.00587 [nucl-th]} \BibitemShut
  {NoStop}%
\bibitem [{\citenamefont {Florkowski}\ \emph
  {et~al.}(2018{\natexlab{b}})\citenamefont {Florkowski}, \citenamefont
  {Friman}, \citenamefont {Jaiswal}, \citenamefont {Ryblewski},\ and\
  \citenamefont {Speranza}}]{Florkowski:2017dyn}%
  \BibitemOpen
  \bibfield  {author} {\bibinfo {author} {\bibfnamefont {W.}~\bibnamefont
  {Florkowski}}, \bibinfo {author} {\bibfnamefont {B.}~\bibnamefont {Friman}},
  \bibinfo {author} {\bibfnamefont {A.}~\bibnamefont {Jaiswal}}, \bibinfo
  {author} {\bibfnamefont {R.}~\bibnamefont {Ryblewski}},\ and\ \bibinfo
  {author} {\bibfnamefont {E.}~\bibnamefont {Speranza}},\ }\href
  {https://doi.org/10.1103/PhysRevD.97.116017} {\bibfield  {journal} {\bibinfo
  {journal} {Phys. Rev. D}\ }\textbf {\bibinfo {volume} {97}},\ \bibinfo
  {pages} {116017} (\bibinfo {year} {2018}{\natexlab{b}})},\ \Eprint
  {https://arxiv.org/abs/1712.07676} {arXiv:1712.07676 [nucl-th]} \BibitemShut
  {NoStop}%
\bibitem [{\citenamefont {Florkowski}\ \emph
  {et~al.}(2018{\natexlab{c}})\citenamefont {Florkowski}, \citenamefont
  {Kumar},\ and\ \citenamefont {Ryblewski}}]{Florkowski:2018ahw}%
  \BibitemOpen
  \bibfield  {author} {\bibinfo {author} {\bibfnamefont {W.}~\bibnamefont
  {Florkowski}}, \bibinfo {author} {\bibfnamefont {A.}~\bibnamefont {Kumar}},\
  and\ \bibinfo {author} {\bibfnamefont {R.}~\bibnamefont {Ryblewski}},\ }\href
  {https://doi.org/10.1103/PhysRevC.98.044906} {\bibfield  {journal} {\bibinfo
  {journal} {Phys. Rev. C}\ }\textbf {\bibinfo {volume} {98}},\ \bibinfo
  {pages} {044906} (\bibinfo {year} {2018}{\natexlab{c}})},\ \Eprint
  {https://arxiv.org/abs/1806.02616} {arXiv:1806.02616 [hep-ph]} \BibitemShut
  {NoStop}%
\bibitem [{\citenamefont {Florkowski}\ \emph
  {et~al.}(2019{\natexlab{a}})\citenamefont {Florkowski}, \citenamefont
  {Kumar},\ and\ \citenamefont {Ryblewski}}]{Florkowski:2018fap}%
  \BibitemOpen
  \bibfield  {author} {\bibinfo {author} {\bibfnamefont {W.}~\bibnamefont
  {Florkowski}}, \bibinfo {author} {\bibfnamefont {A.}~\bibnamefont {Kumar}},\
  and\ \bibinfo {author} {\bibfnamefont {R.}~\bibnamefont {Ryblewski}},\ }\href
  {https://doi.org/10.1016/j.ppnp.2019.07.001} {\bibfield  {journal} {\bibinfo
  {journal} {Prog. Part. Nucl. Phys.}\ }\textbf {\bibinfo {volume} {108}},\
  \bibinfo {pages} {103709} (\bibinfo {year} {2019}{\natexlab{a}})},\ \Eprint
  {https://arxiv.org/abs/1811.04409} {arXiv:1811.04409 [nucl-th]} \BibitemShut
  {NoStop}%
\bibitem [{\citenamefont {Montenegro}\ and\ \citenamefont
  {Torrieri}(2019)}]{Montenegro:2018bcf}%
  \BibitemOpen
  \bibfield  {author} {\bibinfo {author} {\bibfnamefont {D.}~\bibnamefont
  {Montenegro}}\ and\ \bibinfo {author} {\bibfnamefont {G.}~\bibnamefont
  {Torrieri}},\ }\href {https://doi.org/10.1103/PhysRevD.100.056011} {\bibfield
   {journal} {\bibinfo  {journal} {Phys. Rev. D}\ }\textbf {\bibinfo {volume}
  {100}},\ \bibinfo {pages} {056011} (\bibinfo {year} {2019})},\ \Eprint
  {https://arxiv.org/abs/1807.02796} {arXiv:1807.02796 [hep-th]} \BibitemShut
  {NoStop}%
\bibitem [{\citenamefont {Kumar}(2018{\natexlab{a}})}]{Kumar:2018iud}%
  \BibitemOpen
  \bibfield  {author} {\bibinfo {author} {\bibfnamefont {A.}~\bibnamefont
  {Kumar}},\ }in\ \href {https://doi.org/10.5506/APhysPolBSupp.12.393} {\emph
  {\bibinfo {booktitle} {{13th Workshop on Particle Correlations and
  Femtoscopy}}}}\ (\bibinfo {year} {2018})\ \Eprint
  {https://arxiv.org/abs/1810.05364} {arXiv:1810.05364 [nucl-th]} \BibitemShut
  {NoStop}%
\bibitem [{\citenamefont {Kumar}(2018{\natexlab{b}})}]{Kumar:2018lok}%
  \BibitemOpen
  \bibfield  {author} {\bibinfo {author} {\bibfnamefont {A.}~\bibnamefont
  {Kumar}},\ }\href {https://doi.org/10.22323/1.336.0281} {\bibfield  {journal}
  {\bibinfo  {journal} {PoS}\ }\textbf {\bibinfo {volume} {Confinement2018}},\
  \bibinfo {pages} {281} (\bibinfo {year} {2018}{\natexlab{b}})},\ \Eprint
  {https://arxiv.org/abs/1812.00217} {arXiv:1812.00217 [nucl-th]} \BibitemShut
  {NoStop}%
\bibitem [{\citenamefont {Becattini}\ \emph {et~al.}(2019)\citenamefont
  {Becattini}, \citenamefont {Florkowski},\ and\ \citenamefont
  {Speranza}}]{Becattini:2018duy}%
  \BibitemOpen
  \bibfield  {author} {\bibinfo {author} {\bibfnamefont {F.}~\bibnamefont
  {Becattini}}, \bibinfo {author} {\bibfnamefont {W.}~\bibnamefont
  {Florkowski}},\ and\ \bibinfo {author} {\bibfnamefont {E.}~\bibnamefont
  {Speranza}},\ }\href {https://doi.org/10.1016/j.physletb.2018.12.016}
  {\bibfield  {journal} {\bibinfo  {journal} {Phys. Lett. B}\ }\textbf
  {\bibinfo {volume} {789}},\ \bibinfo {pages} {419} (\bibinfo {year}
  {2019})},\ \Eprint {https://arxiv.org/abs/1807.10994} {arXiv:1807.10994
  [hep-th]} \BibitemShut {NoStop}%
\bibitem [{\citenamefont {Florkowski}\ \emph
  {et~al.}(2019{\natexlab{b}})\citenamefont {Florkowski}, \citenamefont
  {Kumar}, \citenamefont {Ryblewski},\ and\ \citenamefont
  {Singh}}]{Florkowski:2019qdp}%
  \BibitemOpen
  \bibfield  {author} {\bibinfo {author} {\bibfnamefont {W.}~\bibnamefont
  {Florkowski}}, \bibinfo {author} {\bibfnamefont {A.}~\bibnamefont {Kumar}},
  \bibinfo {author} {\bibfnamefont {R.}~\bibnamefont {Ryblewski}},\ and\
  \bibinfo {author} {\bibfnamefont {R.}~\bibnamefont {Singh}},\ }\href
  {https://doi.org/10.1103/PhysRevC.99.044910} {\bibfield  {journal} {\bibinfo
  {journal} {Phys. Rev. C}\ }\textbf {\bibinfo {volume} {99}},\ \bibinfo
  {pages} {044910} (\bibinfo {year} {2019}{\natexlab{b}})},\ \Eprint
  {https://arxiv.org/abs/1901.09655} {arXiv:1901.09655 [hep-ph]} \BibitemShut
  {NoStop}%
\bibitem [{\citenamefont {Montenegro}\ and\ \citenamefont
  {Torrieri}(2020)}]{Montenegro:2020paq}%
  \BibitemOpen
  \bibfield  {author} {\bibinfo {author} {\bibfnamefont {D.}~\bibnamefont
  {Montenegro}}\ and\ \bibinfo {author} {\bibfnamefont {G.}~\bibnamefont
  {Torrieri}},\ }\href {https://doi.org/10.1103/PhysRevD.102.036007} {\bibfield
   {journal} {\bibinfo  {journal} {Phys. Rev. D}\ }\textbf {\bibinfo {volume}
  {102}},\ \bibinfo {pages} {036007} (\bibinfo {year} {2020})},\ \Eprint
  {https://arxiv.org/abs/2004.10195} {arXiv:2004.10195 [hep-th]} \BibitemShut
  {NoStop}%
\bibitem [{\citenamefont {Garbiso}\ and\ \citenamefont
  {Kaminski}(2020)}]{Garbiso:2020puw}%
  \BibitemOpen
  \bibfield  {author} {\bibinfo {author} {\bibfnamefont {M.}~\bibnamefont
  {Garbiso}}\ and\ \bibinfo {author} {\bibfnamefont {M.}~\bibnamefont
  {Kaminski}},\ }\href {https://doi.org/10.1007/JHEP12(2020)112} {\bibfield
  {journal} {\bibinfo  {journal} {JHEP}\ }\textbf {\bibinfo {volume} {12}},\
  \bibinfo {pages} {112}},\ \Eprint {https://arxiv.org/abs/2007.04345}
  {arXiv:2007.04345 [hep-th]} \BibitemShut {NoStop}%
\bibitem [{\citenamefont {Bhadury}\ \emph
  {et~al.}(2021{\natexlab{a}})\citenamefont {Bhadury}, \citenamefont
  {Florkowski}, \citenamefont {Jaiswal}, \citenamefont {Kumar},\ and\
  \citenamefont {Ryblewski}}]{Bhadury:2020puc}%
  \BibitemOpen
  \bibfield  {author} {\bibinfo {author} {\bibfnamefont {S.}~\bibnamefont
  {Bhadury}}, \bibinfo {author} {\bibfnamefont {W.}~\bibnamefont {Florkowski}},
  \bibinfo {author} {\bibfnamefont {A.}~\bibnamefont {Jaiswal}}, \bibinfo
  {author} {\bibfnamefont {A.}~\bibnamefont {Kumar}},\ and\ \bibinfo {author}
  {\bibfnamefont {R.}~\bibnamefont {Ryblewski}},\ }\href
  {https://doi.org/10.1016/j.physletb.2021.136096} {\bibfield  {journal}
  {\bibinfo  {journal} {Phys. Lett. B}\ }\textbf {\bibinfo {volume} {814}},\
  \bibinfo {pages} {136096} (\bibinfo {year} {2021}{\natexlab{a}})},\ \Eprint
  {https://arxiv.org/abs/2002.03937} {arXiv:2002.03937 [hep-ph]} \BibitemShut
  {NoStop}%
\bibitem [{\citenamefont {Bhadury}\ \emph
  {et~al.}(2021{\natexlab{b}})\citenamefont {Bhadury}, \citenamefont
  {Florkowski}, \citenamefont {Jaiswal}, \citenamefont {Kumar},\ and\
  \citenamefont {Ryblewski}}]{Bhadury:2020cop}%
  \BibitemOpen
  \bibfield  {author} {\bibinfo {author} {\bibfnamefont {S.}~\bibnamefont
  {Bhadury}}, \bibinfo {author} {\bibfnamefont {W.}~\bibnamefont {Florkowski}},
  \bibinfo {author} {\bibfnamefont {A.}~\bibnamefont {Jaiswal}}, \bibinfo
  {author} {\bibfnamefont {A.}~\bibnamefont {Kumar}},\ and\ \bibinfo {author}
  {\bibfnamefont {R.}~\bibnamefont {Ryblewski}},\ }\href
  {https://doi.org/10.1103/PhysRevD.103.014030} {\bibfield  {journal} {\bibinfo
   {journal} {Phys. Rev. D}\ }\textbf {\bibinfo {volume} {103}},\ \bibinfo
  {pages} {014030} (\bibinfo {year} {2021}{\natexlab{b}})},\ \Eprint
  {https://arxiv.org/abs/2008.10976} {arXiv:2008.10976 [nucl-th]} \BibitemShut
  {NoStop}%
\bibitem [{\citenamefont {Bhadury}\ \emph {et~al.}(2022)\citenamefont
  {Bhadury}, \citenamefont {Florkowski}, \citenamefont {Jaiswal}, \citenamefont
  {Kumar},\ and\ \citenamefont {Ryblewski}}]{Bhadury:2022ulr}%
  \BibitemOpen
  \bibfield  {author} {\bibinfo {author} {\bibfnamefont {S.}~\bibnamefont
  {Bhadury}}, \bibinfo {author} {\bibfnamefont {W.}~\bibnamefont {Florkowski}},
  \bibinfo {author} {\bibfnamefont {A.}~\bibnamefont {Jaiswal}}, \bibinfo
  {author} {\bibfnamefont {A.}~\bibnamefont {Kumar}},\ and\ \bibinfo {author}
  {\bibfnamefont {R.}~\bibnamefont {Ryblewski}},\ }\href
  {https://doi.org/10.1103/PhysRevLett.129.192301} {\bibfield  {journal}
  {\bibinfo  {journal} {Phys. Rev. Lett.}\ }\textbf {\bibinfo {volume} {129}},\
  \bibinfo {pages} {192301} (\bibinfo {year} {2022})},\ \Eprint
  {https://arxiv.org/abs/2204.01357} {arXiv:2204.01357 [nucl-th]} \BibitemShut
  {NoStop}%
\bibitem [{\citenamefont {Bhadury}\ \emph {et~al.}(2024)\citenamefont
  {Bhadury}, \citenamefont {Das}, \citenamefont {Florkowski}, \citenamefont
  {K.},\ and\ \citenamefont {Ryblewski}}]{Bhadury:2023vjx}%
  \BibitemOpen
  \bibfield  {author} {\bibinfo {author} {\bibfnamefont {S.}~\bibnamefont
  {Bhadury}}, \bibinfo {author} {\bibfnamefont {A.}~\bibnamefont {Das}},
  \bibinfo {author} {\bibfnamefont {W.}~\bibnamefont {Florkowski}}, \bibinfo
  {author} {\bibfnamefont {G.~K.}\ \bibnamefont {K.}},\ and\ \bibinfo {author}
  {\bibfnamefont {R.}~\bibnamefont {Ryblewski}},\ }\href
  {https://doi.org/10.1016/j.physletb.2024.138464} {\bibfield  {journal}
  {\bibinfo  {journal} {Phys. Lett. B}\ }\textbf {\bibinfo {volume} {849}},\
  \bibinfo {pages} {138464} (\bibinfo {year} {2024})},\ \Eprint
  {https://arxiv.org/abs/2307.12436} {arXiv:2307.12436 [hep-ph]} \BibitemShut
  {NoStop}%
\bibitem [{\citenamefont {Singh}\ \emph {et~al.}(2023)\citenamefont {Singh},
  \citenamefont {Shokri},\ and\ \citenamefont {Mehr}}]{Singh:2022ltu}%
  \BibitemOpen
  \bibfield  {author} {\bibinfo {author} {\bibfnamefont {R.}~\bibnamefont
  {Singh}}, \bibinfo {author} {\bibfnamefont {M.}~\bibnamefont {Shokri}},\ and\
  \bibinfo {author} {\bibfnamefont {S.~M. A.~T.}\ \bibnamefont {Mehr}},\ }\href
  {https://doi.org/10.1016/j.nuclphysa.2023.122656} {\bibfield  {journal}
  {\bibinfo  {journal} {Nucl. Phys. A}\ }\textbf {\bibinfo {volume} {1035}},\
  \bibinfo {pages} {122656} (\bibinfo {year} {2023})},\ \Eprint
  {https://arxiv.org/abs/2202.11504} {arXiv:2202.11504 [hep-ph]} \BibitemShut
  {NoStop}%
\bibitem [{\citenamefont {Weickgenannt}\ \emph {et~al.}(2019)\citenamefont
  {Weickgenannt}, \citenamefont {Sheng}, \citenamefont {Speranza},
  \citenamefont {Wang},\ and\ \citenamefont {Rischke}}]{Weickgenannt:2019dks}%
  \BibitemOpen
  \bibfield  {author} {\bibinfo {author} {\bibfnamefont {N.}~\bibnamefont
  {Weickgenannt}}, \bibinfo {author} {\bibfnamefont {X.-L.}\ \bibnamefont
  {Sheng}}, \bibinfo {author} {\bibfnamefont {E.}~\bibnamefont {Speranza}},
  \bibinfo {author} {\bibfnamefont {Q.}~\bibnamefont {Wang}},\ and\ \bibinfo
  {author} {\bibfnamefont {D.~H.}\ \bibnamefont {Rischke}},\ }\href
  {https://doi.org/10.1103/PhysRevD.100.056018} {\bibfield  {journal} {\bibinfo
   {journal} {Phys. Rev. D}\ }\textbf {\bibinfo {volume} {100}},\ \bibinfo
  {pages} {056018} (\bibinfo {year} {2019})},\ \Eprint
  {https://arxiv.org/abs/1902.06513} {arXiv:1902.06513 [hep-ph]} \BibitemShut
  {NoStop}%
\bibitem [{\citenamefont {Weickgenannt}\ \emph
  {et~al.}(2021{\natexlab{a}})\citenamefont {Weickgenannt}, \citenamefont
  {Speranza}, \citenamefont {Sheng}, \citenamefont {Wang},\ and\ \citenamefont
  {Rischke}}]{Weickgenannt:2021cuo}%
  \BibitemOpen
  \bibfield  {author} {\bibinfo {author} {\bibfnamefont {N.}~\bibnamefont
  {Weickgenannt}}, \bibinfo {author} {\bibfnamefont {E.}~\bibnamefont
  {Speranza}}, \bibinfo {author} {\bibfnamefont {X.-l.}\ \bibnamefont {Sheng}},
  \bibinfo {author} {\bibfnamefont {Q.}~\bibnamefont {Wang}},\ and\ \bibinfo
  {author} {\bibfnamefont {D.~H.}\ \bibnamefont {Rischke}},\ }\href
  {https://doi.org/10.1103/PhysRevD.104.016022} {\bibfield  {journal} {\bibinfo
   {journal} {Phys. Rev. D}\ }\textbf {\bibinfo {volume} {104}},\ \bibinfo
  {pages} {016022} (\bibinfo {year} {2021}{\natexlab{a}})},\ \Eprint
  {https://arxiv.org/abs/2103.04896} {arXiv:2103.04896 [nucl-th]} \BibitemShut
  {NoStop}%
\bibitem [{\citenamefont {Weickgenannt}\ \emph
  {et~al.}(2021{\natexlab{b}})\citenamefont {Weickgenannt}, \citenamefont
  {Speranza}, \citenamefont {Sheng}, \citenamefont {Wang},\ and\ \citenamefont
  {Rischke}}]{Weickgenannt:2020aaf}%
  \BibitemOpen
  \bibfield  {author} {\bibinfo {author} {\bibfnamefont {N.}~\bibnamefont
  {Weickgenannt}}, \bibinfo {author} {\bibfnamefont {E.}~\bibnamefont
  {Speranza}}, \bibinfo {author} {\bibfnamefont {X.-l.}\ \bibnamefont {Sheng}},
  \bibinfo {author} {\bibfnamefont {Q.}~\bibnamefont {Wang}},\ and\ \bibinfo
  {author} {\bibfnamefont {D.~H.}\ \bibnamefont {Rischke}},\ }\href
  {https://doi.org/10.1103/PhysRevLett.127.052301} {\bibfield  {journal}
  {\bibinfo  {journal} {Phys. Rev. Lett.}\ }\textbf {\bibinfo {volume} {127}},\
  \bibinfo {pages} {052301} (\bibinfo {year} {2021}{\natexlab{b}})},\ \Eprint
  {https://arxiv.org/abs/2005.01506} {arXiv:2005.01506 [hep-ph]} \BibitemShut
  {NoStop}%
\bibitem [{\citenamefont {Wagner}\ \emph {et~al.}(2022)\citenamefont {Wagner},
  \citenamefont {Weickgenannt},\ and\ \citenamefont
  {Rischke}}]{Wagner:2022amr}%
  \BibitemOpen
  \bibfield  {author} {\bibinfo {author} {\bibfnamefont {D.}~\bibnamefont
  {Wagner}}, \bibinfo {author} {\bibfnamefont {N.}~\bibnamefont
  {Weickgenannt}},\ and\ \bibinfo {author} {\bibfnamefont {D.~H.}\ \bibnamefont
  {Rischke}},\ }\href {https://doi.org/10.1103/PhysRevD.106.116021} {\bibfield
  {journal} {\bibinfo  {journal} {Phys. Rev. D}\ }\textbf {\bibinfo {volume}
  {106}},\ \bibinfo {pages} {116021} (\bibinfo {year} {2022})},\ \Eprint
  {https://arxiv.org/abs/2210.06187} {arXiv:2210.06187 [nucl-th]} \BibitemShut
  {NoStop}%
\bibitem [{\citenamefont {Weickgenannt}\ \emph
  {et~al.}(2022{\natexlab{a}})\citenamefont {Weickgenannt}, \citenamefont
  {Wagner}, \citenamefont {Speranza},\ and\ \citenamefont
  {Rischke}}]{Weickgenannt:2022zxs}%
  \BibitemOpen
  \bibfield  {author} {\bibinfo {author} {\bibfnamefont {N.}~\bibnamefont
  {Weickgenannt}}, \bibinfo {author} {\bibfnamefont {D.}~\bibnamefont
  {Wagner}}, \bibinfo {author} {\bibfnamefont {E.}~\bibnamefont {Speranza}},\
  and\ \bibinfo {author} {\bibfnamefont {D.~H.}\ \bibnamefont {Rischke}},\
  }\href {https://doi.org/10.1103/PhysRevD.106.096014} {\bibfield  {journal}
  {\bibinfo  {journal} {Phys. Rev. D}\ }\textbf {\bibinfo {volume} {106}},\
  \bibinfo {pages} {096014} (\bibinfo {year} {2022}{\natexlab{a}})},\ \Eprint
  {https://arxiv.org/abs/2203.04766} {arXiv:2203.04766 [nucl-th]} \BibitemShut
  {NoStop}%
\bibitem [{\citenamefont {Weickgenannt}\ and\ \citenamefont
  {Blaizot}(2024{\natexlab{a}})}]{Weickgenannt:2023nge}%
  \BibitemOpen
  \bibfield  {author} {\bibinfo {author} {\bibfnamefont {N.}~\bibnamefont
  {Weickgenannt}}\ and\ \bibinfo {author} {\bibfnamefont {J.-P.}\ \bibnamefont
  {Blaizot}},\ }\href {https://doi.org/10.1103/PhysRevD.109.056012} {\bibfield
  {journal} {\bibinfo  {journal} {Phys. Rev. D}\ }\textbf {\bibinfo {volume}
  {109}},\ \bibinfo {pages} {056012} (\bibinfo {year} {2024}{\natexlab{a}})},\
  \Eprint {https://arxiv.org/abs/2311.15817} {arXiv:2311.15817 [hep-ph]}
  \BibitemShut {NoStop}%
\bibitem [{\citenamefont {Weickgenannt}\ and\ \citenamefont
  {Blaizot}(2024{\natexlab{b}})}]{Weickgenannt:2023bss}%
  \BibitemOpen
  \bibfield  {author} {\bibinfo {author} {\bibfnamefont {N.}~\bibnamefont
  {Weickgenannt}}\ and\ \bibinfo {author} {\bibfnamefont {J.-P.}\ \bibnamefont
  {Blaizot}},\ }\href {https://doi.org/10.1103/PhysRevD.109.056019} {\bibfield
  {journal} {\bibinfo  {journal} {Phys. Rev. D}\ }\textbf {\bibinfo {volume}
  {109}},\ \bibinfo {pages} {056019} (\bibinfo {year} {2024}{\natexlab{b}})},\
  \Eprint {https://arxiv.org/abs/2312.05917} {arXiv:2312.05917 [hep-ph]}
  \BibitemShut {NoStop}%
\bibitem [{\citenamefont {Weickgenannt}\ and\ \citenamefont
  {Blaizot}(2024{\natexlab{c}})}]{Weickgenannt:2024ibf}%
  \BibitemOpen
  \bibfield  {author} {\bibinfo {author} {\bibfnamefont {N.}~\bibnamefont
  {Weickgenannt}}\ and\ \bibinfo {author} {\bibfnamefont {J.-P.}\ \bibnamefont
  {Blaizot}},\ }\href@noop {} {\  (\bibinfo {year} {2024}{\natexlab{c}})},\
  \Eprint {https://arxiv.org/abs/2409.11045} {arXiv:2409.11045 [hep-ph]}
  \BibitemShut {NoStop}%
\bibitem [{\citenamefont {Wagner}\ \emph {et~al.}(2024)\citenamefont {Wagner},
  \citenamefont {Shokri},\ and\ \citenamefont {Rischke}}]{Wagner:2024fhf}%
  \BibitemOpen
  \bibfield  {author} {\bibinfo {author} {\bibfnamefont {D.}~\bibnamefont
  {Wagner}}, \bibinfo {author} {\bibfnamefont {M.}~\bibnamefont {Shokri}},\
  and\ \bibinfo {author} {\bibfnamefont {D.~H.}\ \bibnamefont {Rischke}},\
  }\href {https://doi.org/10.1103/PhysRevResearch.6.043103} {\bibfield
  {journal} {\bibinfo  {journal} {Phys. Rev. Res.}\ }\textbf {\bibinfo {volume}
  {6}},\ \bibinfo {pages} {043103} (\bibinfo {year} {2024})},\ \Eprint
  {https://arxiv.org/abs/2405.00533} {arXiv:2405.00533 [nucl-th]} \BibitemShut
  {NoStop}%
\bibitem [{\citenamefont {Peng}\ \emph {et~al.}(2021)\citenamefont {Peng},
  \citenamefont {Zhang}, \citenamefont {Sheng},\ and\ \citenamefont
  {Wang}}]{Peng:2021ago}%
  \BibitemOpen
  \bibfield  {author} {\bibinfo {author} {\bibfnamefont {H.-H.}\ \bibnamefont
  {Peng}}, \bibinfo {author} {\bibfnamefont {J.-J.}\ \bibnamefont {Zhang}},
  \bibinfo {author} {\bibfnamefont {X.-L.}\ \bibnamefont {Sheng}},\ and\
  \bibinfo {author} {\bibfnamefont {Q.}~\bibnamefont {Wang}},\ }\href
  {https://doi.org/10.1088/0256-307X/38/11/116701} {\bibfield  {journal}
  {\bibinfo  {journal} {Chin. Phys. Lett.}\ }\textbf {\bibinfo {volume} {38}},\
  \bibinfo {pages} {116701} (\bibinfo {year} {2021})},\ \Eprint
  {https://arxiv.org/abs/2107.00448} {arXiv:2107.00448 [hep-th]} \BibitemShut
  {NoStop}%
\bibitem [{\citenamefont {Cartwright}\ \emph {et~al.}(2023)\citenamefont
  {Cartwright}, \citenamefont {Amano}, \citenamefont {Kaminski}, \citenamefont
  {Noronha},\ and\ \citenamefont {Speranza}}]{Cartwright:2021qpp}%
  \BibitemOpen
  \bibfield  {author} {\bibinfo {author} {\bibfnamefont {C.}~\bibnamefont
  {Cartwright}}, \bibinfo {author} {\bibfnamefont {M.~G.}\ \bibnamefont
  {Amano}}, \bibinfo {author} {\bibfnamefont {M.}~\bibnamefont {Kaminski}},
  \bibinfo {author} {\bibfnamefont {J.}~\bibnamefont {Noronha}},\ and\ \bibinfo
  {author} {\bibfnamefont {E.}~\bibnamefont {Speranza}},\ }\href
  {https://doi.org/10.1103/PhysRevD.108.046014} {\bibfield  {journal} {\bibinfo
   {journal} {Phys. Rev. D}\ }\textbf {\bibinfo {volume} {108}},\ \bibinfo
  {pages} {046014} (\bibinfo {year} {2023})},\ \Eprint
  {https://arxiv.org/abs/2112.10781} {arXiv:2112.10781 [hep-th]} \BibitemShut
  {NoStop}%
\bibitem [{\citenamefont {Hu}(2022)}]{Hu:2021pwh}%
  \BibitemOpen
  \bibfield  {author} {\bibinfo {author} {\bibfnamefont {J.}~\bibnamefont
  {Hu}},\ }\href {https://doi.org/10.1103/PhysRevD.105.076009} {\bibfield
  {journal} {\bibinfo  {journal} {Phys. Rev. D}\ }\textbf {\bibinfo {volume}
  {105}},\ \bibinfo {pages} {076009} (\bibinfo {year} {2022})},\ \Eprint
  {https://arxiv.org/abs/2111.03571} {arXiv:2111.03571 [hep-ph]} \BibitemShut
  {NoStop}%
\bibitem [{\citenamefont {Li}\ \emph {et~al.}(2021)\citenamefont {Li},
  \citenamefont {Stephanov},\ and\ \citenamefont {Yee}}]{Li:2020eon}%
  \BibitemOpen
  \bibfield  {author} {\bibinfo {author} {\bibfnamefont {S.}~\bibnamefont
  {Li}}, \bibinfo {author} {\bibfnamefont {M.~A.}\ \bibnamefont {Stephanov}},\
  and\ \bibinfo {author} {\bibfnamefont {H.-U.}\ \bibnamefont {Yee}},\ }\href
  {https://doi.org/10.1103/PhysRevLett.127.082302} {\bibfield  {journal}
  {\bibinfo  {journal} {Phys. Rev. Lett.}\ }\textbf {\bibinfo {volume} {127}},\
  \bibinfo {pages} {082302} (\bibinfo {year} {2021})},\ \Eprint
  {https://arxiv.org/abs/2011.12318} {arXiv:2011.12318 [hep-th]} \BibitemShut
  {NoStop}%
\bibitem [{\citenamefont {Shi}\ \emph {et~al.}(2021)\citenamefont {Shi},
  \citenamefont {Gale},\ and\ \citenamefont {Jeon}}]{Shi:2020htn}%
  \BibitemOpen
  \bibfield  {author} {\bibinfo {author} {\bibfnamefont {S.}~\bibnamefont
  {Shi}}, \bibinfo {author} {\bibfnamefont {C.}~\bibnamefont {Gale}},\ and\
  \bibinfo {author} {\bibfnamefont {S.}~\bibnamefont {Jeon}},\ }\href
  {https://doi.org/10.1103/PhysRevC.103.044906} {\bibfield  {journal} {\bibinfo
   {journal} {Phys. Rev. C}\ }\textbf {\bibinfo {volume} {103}},\ \bibinfo
  {pages} {044906} (\bibinfo {year} {2021})},\ \Eprint
  {https://arxiv.org/abs/2008.08618} {arXiv:2008.08618 [nucl-th]} \BibitemShut
  {NoStop}%
\bibitem [{\citenamefont {Hattori}\ \emph {et~al.}(2019)\citenamefont
  {Hattori}, \citenamefont {Hongo}, \citenamefont {Huang}, \citenamefont
  {Matsuo},\ and\ \citenamefont {Taya}}]{Hattori:2019lfp}%
  \BibitemOpen
  \bibfield  {author} {\bibinfo {author} {\bibfnamefont {K.}~\bibnamefont
  {Hattori}}, \bibinfo {author} {\bibfnamefont {M.}~\bibnamefont {Hongo}},
  \bibinfo {author} {\bibfnamefont {X.-G.}\ \bibnamefont {Huang}}, \bibinfo
  {author} {\bibfnamefont {M.}~\bibnamefont {Matsuo}},\ and\ \bibinfo {author}
  {\bibfnamefont {H.}~\bibnamefont {Taya}},\ }\href
  {https://doi.org/10.1016/j.physletb.2019.05.040} {\bibfield  {journal}
  {\bibinfo  {journal} {Phys. Lett. B}\ }\textbf {\bibinfo {volume} {795}},\
  \bibinfo {pages} {100} (\bibinfo {year} {2019})},\ \Eprint
  {https://arxiv.org/abs/1901.06615} {arXiv:1901.06615 [hep-th]} \BibitemShut
  {NoStop}%
\bibitem [{\citenamefont {Fukushima}\ and\ \citenamefont
  {Pu}(2021)}]{Fukushima:2020ucl}%
  \BibitemOpen
  \bibfield  {author} {\bibinfo {author} {\bibfnamefont {K.}~\bibnamefont
  {Fukushima}}\ and\ \bibinfo {author} {\bibfnamefont {S.}~\bibnamefont {Pu}},\
  }\href {https://doi.org/10.1016/j.physletb.2021.136346} {\bibfield  {journal}
  {\bibinfo  {journal} {Phys. Lett. B}\ }\textbf {\bibinfo {volume} {817}},\
  \bibinfo {pages} {136346} (\bibinfo {year} {2021})},\ \Eprint
  {https://arxiv.org/abs/2010.01608} {arXiv:2010.01608 [hep-th]} \BibitemShut
  {NoStop}%
\bibitem [{\citenamefont {Daher}\ \emph
  {et~al.}(2023{\natexlab{a}})\citenamefont {Daher}, \citenamefont {Das},
  \citenamefont {Florkowski},\ and\ \citenamefont {Ryblewski}}]{Daher:2022xon}%
  \BibitemOpen
  \bibfield  {author} {\bibinfo {author} {\bibfnamefont {A.}~\bibnamefont
  {Daher}}, \bibinfo {author} {\bibfnamefont {A.}~\bibnamefont {Das}}, \bibinfo
  {author} {\bibfnamefont {W.}~\bibnamefont {Florkowski}},\ and\ \bibinfo
  {author} {\bibfnamefont {R.}~\bibnamefont {Ryblewski}},\ }\href
  {https://doi.org/10.1103/PhysRevC.108.024902} {\bibfield  {journal} {\bibinfo
   {journal} {Phys. Rev. C}\ }\textbf {\bibinfo {volume} {108}},\ \bibinfo
  {pages} {024902} (\bibinfo {year} {2023}{\natexlab{a}})},\ \Eprint
  {https://arxiv.org/abs/2202.12609} {arXiv:2202.12609 [nucl-th]} \BibitemShut
  {NoStop}%
\bibitem [{\citenamefont {Daher}\ \emph
  {et~al.}(2023{\natexlab{b}})\citenamefont {Daher}, \citenamefont {Das},\ and\
  \citenamefont {Ryblewski}}]{Daher:2022wzf}%
  \BibitemOpen
  \bibfield  {author} {\bibinfo {author} {\bibfnamefont {A.}~\bibnamefont
  {Daher}}, \bibinfo {author} {\bibfnamefont {A.}~\bibnamefont {Das}},\ and\
  \bibinfo {author} {\bibfnamefont {R.}~\bibnamefont {Ryblewski}},\ }\href
  {https://doi.org/10.1103/PhysRevD.107.054043} {\bibfield  {journal} {\bibinfo
   {journal} {Phys. Rev. D}\ }\textbf {\bibinfo {volume} {107}},\ \bibinfo
  {pages} {054043} (\bibinfo {year} {2023}{\natexlab{b}})},\ \Eprint
  {https://arxiv.org/abs/2209.10460} {arXiv:2209.10460 [nucl-th]} \BibitemShut
  {NoStop}%
\bibitem [{\citenamefont {Sarwar}\ \emph {et~al.}(2023)\citenamefont {Sarwar},
  \citenamefont {Hasanujjaman}, \citenamefont {Bhatt}, \citenamefont {Mishra},\
  and\ \citenamefont {Alam}}]{Sarwar:2022yzs}%
  \BibitemOpen
  \bibfield  {author} {\bibinfo {author} {\bibfnamefont {G.}~\bibnamefont
  {Sarwar}}, \bibinfo {author} {\bibfnamefont {M.}~\bibnamefont
  {Hasanujjaman}}, \bibinfo {author} {\bibfnamefont {J.~R.}\ \bibnamefont
  {Bhatt}}, \bibinfo {author} {\bibfnamefont {H.}~\bibnamefont {Mishra}},\ and\
  \bibinfo {author} {\bibfnamefont {J.-e.}\ \bibnamefont {Alam}},\ }\href
  {https://doi.org/10.1103/PhysRevD.107.054031} {\bibfield  {journal} {\bibinfo
   {journal} {Phys. Rev. D}\ }\textbf {\bibinfo {volume} {107}},\ \bibinfo
  {pages} {054031} (\bibinfo {year} {2023})},\ \Eprint
  {https://arxiv.org/abs/2209.08652} {arXiv:2209.08652 [nucl-th]} \BibitemShut
  {NoStop}%
\bibitem [{\citenamefont {Yi}\ \emph {et~al.}(2022)\citenamefont {Yi},
  \citenamefont {Pu}, \citenamefont {Gao},\ and\ \citenamefont
  {Yang}}]{Yi:2021unq}%
  \BibitemOpen
  \bibfield  {author} {\bibinfo {author} {\bibfnamefont {C.}~\bibnamefont
  {Yi}}, \bibinfo {author} {\bibfnamefont {S.}~\bibnamefont {Pu}}, \bibinfo
  {author} {\bibfnamefont {J.-H.}\ \bibnamefont {Gao}},\ and\ \bibinfo {author}
  {\bibfnamefont {D.-L.}\ \bibnamefont {Yang}},\ }\href
  {https://doi.org/10.1103/PhysRevC.105.044911} {\bibfield  {journal} {\bibinfo
   {journal} {Phys. Rev. C}\ }\textbf {\bibinfo {volume} {105}},\ \bibinfo
  {pages} {044911} (\bibinfo {year} {2022})},\ \Eprint
  {https://arxiv.org/abs/2112.15531} {arXiv:2112.15531 [hep-ph]} \BibitemShut
  {NoStop}%
\bibitem [{\citenamefont {Wang}\ \emph {et~al.}(2021)\citenamefont {Wang},
  \citenamefont {Fang},\ and\ \citenamefont {Pu}}]{Wang:2021ngp}%
  \BibitemOpen
  \bibfield  {author} {\bibinfo {author} {\bibfnamefont {D.-L.}\ \bibnamefont
  {Wang}}, \bibinfo {author} {\bibfnamefont {S.}~\bibnamefont {Fang}},\ and\
  \bibinfo {author} {\bibfnamefont {S.}~\bibnamefont {Pu}},\ }\href
  {https://doi.org/10.1103/PhysRevD.104.114043} {\bibfield  {journal} {\bibinfo
   {journal} {Phys. Rev. D}\ }\textbf {\bibinfo {volume} {104}},\ \bibinfo
  {pages} {114043} (\bibinfo {year} {2021})},\ \Eprint
  {https://arxiv.org/abs/2107.11726} {arXiv:2107.11726 [nucl-th]} \BibitemShut
  {NoStop}%
\bibitem [{\citenamefont {Wang}\ \emph {et~al.}(2022)\citenamefont {Wang},
  \citenamefont {Xie}, \citenamefont {Fang},\ and\ \citenamefont
  {Pu}}]{Wang:2021wqq}%
  \BibitemOpen
  \bibfield  {author} {\bibinfo {author} {\bibfnamefont {D.-L.}\ \bibnamefont
  {Wang}}, \bibinfo {author} {\bibfnamefont {X.-Q.}\ \bibnamefont {Xie}},
  \bibinfo {author} {\bibfnamefont {S.}~\bibnamefont {Fang}},\ and\ \bibinfo
  {author} {\bibfnamefont {S.}~\bibnamefont {Pu}},\ }\href
  {https://doi.org/10.1103/PhysRevD.105.114050} {\bibfield  {journal} {\bibinfo
   {journal} {Phys. Rev. D}\ }\textbf {\bibinfo {volume} {105}},\ \bibinfo
  {pages} {114050} (\bibinfo {year} {2022})},\ \Eprint
  {https://arxiv.org/abs/2112.15535} {arXiv:2112.15535 [hep-ph]} \BibitemShut
  {NoStop}%
\bibitem [{\citenamefont {Biswas}\ \emph
  {et~al.}(2023{\natexlab{a}})\citenamefont {Biswas}, \citenamefont {Daher},
  \citenamefont {Das}, \citenamefont {Florkowski},\ and\ \citenamefont
  {Ryblewski}}]{Biswas:2022bht}%
  \BibitemOpen
  \bibfield  {author} {\bibinfo {author} {\bibfnamefont {R.}~\bibnamefont
  {Biswas}}, \bibinfo {author} {\bibfnamefont {A.}~\bibnamefont {Daher}},
  \bibinfo {author} {\bibfnamefont {A.}~\bibnamefont {Das}}, \bibinfo {author}
  {\bibfnamefont {W.}~\bibnamefont {Florkowski}},\ and\ \bibinfo {author}
  {\bibfnamefont {R.}~\bibnamefont {Ryblewski}},\ }\href
  {https://doi.org/10.1103/PhysRevD.107.094022} {\bibfield  {journal} {\bibinfo
   {journal} {Phys. Rev. D}\ }\textbf {\bibinfo {volume} {107}},\ \bibinfo
  {pages} {094022} (\bibinfo {year} {2023}{\natexlab{a}})},\ \Eprint
  {https://arxiv.org/abs/2211.02934} {arXiv:2211.02934 [nucl-th]} \BibitemShut
  {NoStop}%
\bibitem [{\citenamefont {Cao}\ \emph {et~al.}(2022)\citenamefont {Cao},
  \citenamefont {Hattori}, \citenamefont {Hongo}, \citenamefont {Huang},\ and\
  \citenamefont {Taya}}]{Cao:2022aku}%
  \BibitemOpen
  \bibfield  {author} {\bibinfo {author} {\bibfnamefont {Z.}~\bibnamefont
  {Cao}}, \bibinfo {author} {\bibfnamefont {K.}~\bibnamefont {Hattori}},
  \bibinfo {author} {\bibfnamefont {M.}~\bibnamefont {Hongo}}, \bibinfo
  {author} {\bibfnamefont {X.-G.}\ \bibnamefont {Huang}},\ and\ \bibinfo
  {author} {\bibfnamefont {H.}~\bibnamefont {Taya}},\ }\href
  {https://doi.org/10.1093/ptep/ptac091} {\bibfield  {journal} {\bibinfo
  {journal} {PTEP}\ }\textbf {\bibinfo {volume} {2022}},\ \bibinfo {pages}
  {071D01} (\bibinfo {year} {2022})},\ \Eprint
  {https://arxiv.org/abs/2205.08051} {arXiv:2205.08051 [hep-th]} \BibitemShut
  {NoStop}%
\bibitem [{\citenamefont {Weickgenannt}\ \emph
  {et~al.}(2022{\natexlab{b}})\citenamefont {Weickgenannt}, \citenamefont
  {Wagner},\ and\ \citenamefont {Speranza}}]{Weickgenannt:2022jes}%
  \BibitemOpen
  \bibfield  {author} {\bibinfo {author} {\bibfnamefont {N.}~\bibnamefont
  {Weickgenannt}}, \bibinfo {author} {\bibfnamefont {D.}~\bibnamefont
  {Wagner}},\ and\ \bibinfo {author} {\bibfnamefont {E.}~\bibnamefont
  {Speranza}},\ }\href {https://doi.org/10.1103/PhysRevD.105.116026} {\bibfield
   {journal} {\bibinfo  {journal} {Phys. Rev. D}\ }\textbf {\bibinfo {volume}
  {105}},\ \bibinfo {pages} {116026} (\bibinfo {year} {2022}{\natexlab{b}})},\
  \Eprint {https://arxiv.org/abs/2204.01797} {arXiv:2204.01797 [nucl-th]}
  \BibitemShut {NoStop}%
\bibitem [{\citenamefont {Weickgenannt}\ \emph
  {et~al.}(2022{\natexlab{c}})\citenamefont {Weickgenannt}, \citenamefont
  {Wagner}, \citenamefont {Speranza},\ and\ \citenamefont
  {Rischke}}]{Weickgenannt:2022qvh}%
  \BibitemOpen
  \bibfield  {author} {\bibinfo {author} {\bibfnamefont {N.}~\bibnamefont
  {Weickgenannt}}, \bibinfo {author} {\bibfnamefont {D.}~\bibnamefont
  {Wagner}}, \bibinfo {author} {\bibfnamefont {E.}~\bibnamefont {Speranza}},\
  and\ \bibinfo {author} {\bibfnamefont {D.~H.}\ \bibnamefont {Rischke}},\
  }\href {https://doi.org/10.1103/PhysRevD.106.L091901} {\bibfield  {journal}
  {\bibinfo  {journal} {Phys. Rev. D}\ }\textbf {\bibinfo {volume} {106}},\
  \bibinfo {pages} {L091901} (\bibinfo {year} {2022}{\natexlab{c}})},\ \Eprint
  {https://arxiv.org/abs/2208.01955} {arXiv:2208.01955 [nucl-th]} \BibitemShut
  {NoStop}%
\bibitem [{\citenamefont {Weickgenannt}(2023)}]{Weickgenannt:2023btk}%
  \BibitemOpen
  \bibfield  {author} {\bibinfo {author} {\bibfnamefont {N.}~\bibnamefont
  {Weickgenannt}},\ }\href {https://doi.org/10.1103/PhysRevD.108.076011}
  {\bibfield  {journal} {\bibinfo  {journal} {Phys. Rev. D}\ }\textbf {\bibinfo
  {volume} {108}},\ \bibinfo {pages} {076011} (\bibinfo {year} {2023})},\
  \Eprint {https://arxiv.org/abs/2307.13561} {arXiv:2307.13561 [nucl-th]}
  \BibitemShut {NoStop}%
\bibitem [{\citenamefont {Biswas}\ \emph
  {et~al.}(2023{\natexlab{b}})\citenamefont {Biswas}, \citenamefont {Daher},
  \citenamefont {Das}, \citenamefont {Florkowski},\ and\ \citenamefont
  {Ryblewski}}]{Biswas:2023qsw}%
  \BibitemOpen
  \bibfield  {author} {\bibinfo {author} {\bibfnamefont {R.}~\bibnamefont
  {Biswas}}, \bibinfo {author} {\bibfnamefont {A.}~\bibnamefont {Daher}},
  \bibinfo {author} {\bibfnamefont {A.}~\bibnamefont {Das}}, \bibinfo {author}
  {\bibfnamefont {W.}~\bibnamefont {Florkowski}},\ and\ \bibinfo {author}
  {\bibfnamefont {R.}~\bibnamefont {Ryblewski}},\ }\href
  {https://doi.org/10.1103/PhysRevD.108.014024} {\bibfield  {journal} {\bibinfo
   {journal} {Phys. Rev. D}\ }\textbf {\bibinfo {volume} {108}},\ \bibinfo
  {pages} {014024} (\bibinfo {year} {2023}{\natexlab{b}})},\ \Eprint
  {https://arxiv.org/abs/2304.01009} {arXiv:2304.01009 [nucl-th]} \BibitemShut
  {NoStop}%
\bibitem [{\citenamefont {Xie}\ \emph {et~al.}(2023)\citenamefont {Xie},
  \citenamefont {Wang}, \citenamefont {Yang},\ and\ \citenamefont
  {Pu}}]{Xie:2023gbo}%
  \BibitemOpen
  \bibfield  {author} {\bibinfo {author} {\bibfnamefont {X.-Q.}\ \bibnamefont
  {Xie}}, \bibinfo {author} {\bibfnamefont {D.-L.}\ \bibnamefont {Wang}},
  \bibinfo {author} {\bibfnamefont {C.}~\bibnamefont {Yang}},\ and\ \bibinfo
  {author} {\bibfnamefont {S.}~\bibnamefont {Pu}},\ }\href
  {https://doi.org/10.1103/PhysRevD.108.094031} {\bibfield  {journal} {\bibinfo
   {journal} {Phys. Rev. D}\ }\textbf {\bibinfo {volume} {108}},\ \bibinfo
  {pages} {094031} (\bibinfo {year} {2023})},\ \Eprint
  {https://arxiv.org/abs/2306.13880} {arXiv:2306.13880 [hep-ph]} \BibitemShut
  {NoStop}%
\bibitem [{\citenamefont {Daher}\ \emph
  {et~al.}(2024{\natexlab{a}})\citenamefont {Daher}, \citenamefont
  {Florkowski},\ and\ \citenamefont {Ryblewski}}]{Daher:2024ixz}%
  \BibitemOpen
  \bibfield  {author} {\bibinfo {author} {\bibfnamefont {A.}~\bibnamefont
  {Daher}}, \bibinfo {author} {\bibfnamefont {W.}~\bibnamefont {Florkowski}},\
  and\ \bibinfo {author} {\bibfnamefont {R.}~\bibnamefont {Ryblewski}},\ }\href
  {https://doi.org/10.1103/PhysRevD.110.034029} {\bibfield  {journal} {\bibinfo
   {journal} {Phys. Rev. D}\ }\textbf {\bibinfo {volume} {110}},\ \bibinfo
  {pages} {034029} (\bibinfo {year} {2024}{\natexlab{a}})},\ \Eprint
  {https://arxiv.org/abs/2401.07608} {arXiv:2401.07608 [hep-ph]} \BibitemShut
  {NoStop}%
\bibitem [{\citenamefont {Ren}\ \emph {et~al.}(2024)\citenamefont {Ren},
  \citenamefont {Yang}, \citenamefont {Wang},\ and\ \citenamefont
  {Pu}}]{Ren:2024pur}%
  \BibitemOpen
  \bibfield  {author} {\bibinfo {author} {\bibfnamefont {X.}~\bibnamefont
  {Ren}}, \bibinfo {author} {\bibfnamefont {C.}~\bibnamefont {Yang}}, \bibinfo
  {author} {\bibfnamefont {D.-L.}\ \bibnamefont {Wang}},\ and\ \bibinfo
  {author} {\bibfnamefont {S.}~\bibnamefont {Pu}},\ }\href
  {https://doi.org/10.1103/PhysRevD.110.034010} {\bibfield  {journal} {\bibinfo
   {journal} {Phys. Rev. D}\ }\textbf {\bibinfo {volume} {110}},\ \bibinfo
  {pages} {034010} (\bibinfo {year} {2024})},\ \Eprint
  {https://arxiv.org/abs/2405.03105} {arXiv:2405.03105 [nucl-th]} \BibitemShut
  {NoStop}%
\bibitem [{\citenamefont {Daher}\ \emph
  {et~al.}(2024{\natexlab{b}})\citenamefont {Daher}, \citenamefont
  {Florkowski}, \citenamefont {Ryblewski},\ and\ \citenamefont
  {Taghinavaz}}]{Daher:2024bah}%
  \BibitemOpen
  \bibfield  {author} {\bibinfo {author} {\bibfnamefont {A.}~\bibnamefont
  {Daher}}, \bibinfo {author} {\bibfnamefont {W.}~\bibnamefont {Florkowski}},
  \bibinfo {author} {\bibfnamefont {R.}~\bibnamefont {Ryblewski}},\ and\
  \bibinfo {author} {\bibfnamefont {F.}~\bibnamefont {Taghinavaz}},\ }\href
  {https://doi.org/10.1103/PhysRevD.109.114001} {\bibfield  {journal} {\bibinfo
   {journal} {Phys. Rev. D}\ }\textbf {\bibinfo {volume} {109}},\ \bibinfo
  {pages} {114001} (\bibinfo {year} {2024}{\natexlab{b}})},\ \Eprint
  {https://arxiv.org/abs/2403.04711} {arXiv:2403.04711 [hep-ph]} \BibitemShut
  {NoStop}%
\bibitem [{\citenamefont {Fang}\ \emph {et~al.}(2024)\citenamefont {Fang},
  \citenamefont {Hattori},\ and\ \citenamefont {Hu}}]{Fang:2024sym}%
  \BibitemOpen
  \bibfield  {author} {\bibinfo {author} {\bibfnamefont {Z.}~\bibnamefont
  {Fang}}, \bibinfo {author} {\bibfnamefont {K.}~\bibnamefont {Hattori}},\ and\
  \bibinfo {author} {\bibfnamefont {J.}~\bibnamefont {Hu}},\ }\href@noop {} {\
  (\bibinfo {year} {2024})},\ \Eprint {https://arxiv.org/abs/2409.07096}
  {arXiv:2409.07096 [hep-ph]} \BibitemShut {NoStop}%
\bibitem [{\citenamefont {Dey}\ and\ \citenamefont {Das}(2024)}]{Dey:2024cwo}%
  \BibitemOpen
  \bibfield  {author} {\bibinfo {author} {\bibfnamefont {S.}~\bibnamefont
  {Dey}}\ and\ \bibinfo {author} {\bibfnamefont {A.}~\bibnamefont {Das}},\
  }\href@noop {} {\  (\bibinfo {year} {2024})},\ \Eprint
  {https://arxiv.org/abs/2410.04141} {arXiv:2410.04141 [nucl-th]} \BibitemShut
  {NoStop}%
\bibitem [{\citenamefont {She}\ \emph {et~al.}(2024)\citenamefont {She},
  \citenamefont {Qiu},\ and\ \citenamefont {Hou}}]{She:2024rnx}%
  \BibitemOpen
  \bibfield  {author} {\bibinfo {author} {\bibfnamefont {D.}~\bibnamefont
  {She}}, \bibinfo {author} {\bibfnamefont {Y.-W.}\ \bibnamefont {Qiu}},\ and\
  \bibinfo {author} {\bibfnamefont {D.}~\bibnamefont {Hou}},\ }\href@noop {} {\
   (\bibinfo {year} {2024})},\ \Eprint {https://arxiv.org/abs/2410.15142}
  {arXiv:2410.15142 [nucl-th]} \BibitemShut {NoStop}%
\bibitem [{\citenamefont {Gallegos}\ \emph {et~al.}(2021)\citenamefont
  {Gallegos}, \citenamefont {G\"ursoy},\ and\ \citenamefont
  {Yarom}}]{Gallegos:2021bzp}%
  \BibitemOpen
  \bibfield  {author} {\bibinfo {author} {\bibfnamefont {A.~D.}\ \bibnamefont
  {Gallegos}}, \bibinfo {author} {\bibfnamefont {U.}~\bibnamefont {G\"ursoy}},\
  and\ \bibinfo {author} {\bibfnamefont {A.}~\bibnamefont {Yarom}},\ }\href
  {https://doi.org/10.21468/SciPostPhys.11.2.041} {\bibfield  {journal}
  {\bibinfo  {journal} {SciPost Phys.}\ }\textbf {\bibinfo {volume} {11}},\
  \bibinfo {pages} {041} (\bibinfo {year} {2021})},\ \Eprint
  {https://arxiv.org/abs/2101.04759} {arXiv:2101.04759 [hep-th]} \BibitemShut
  {NoStop}%
\bibitem [{\citenamefont {Gallegos}\ \emph {et~al.}(2023)\citenamefont
  {Gallegos}, \citenamefont {Gursoy},\ and\ \citenamefont
  {Yarom}}]{Gallegos:2022jow}%
  \BibitemOpen
  \bibfield  {author} {\bibinfo {author} {\bibfnamefont {A.~D.}\ \bibnamefont
  {Gallegos}}, \bibinfo {author} {\bibfnamefont {U.}~\bibnamefont {Gursoy}},\
  and\ \bibinfo {author} {\bibfnamefont {A.}~\bibnamefont {Yarom}},\ }\href
  {https://doi.org/10.1007/JHEP05(2023)139} {\bibfield  {journal} {\bibinfo
  {journal} {JHEP}\ }\textbf {\bibinfo {volume} {05}},\ \bibinfo {pages}
  {139}},\ \Eprint {https://arxiv.org/abs/2203.05044} {arXiv:2203.05044
  [hep-th]} \BibitemShut {NoStop}%
\bibitem [{\citenamefont {Hongo}\ \emph {et~al.}(2021)\citenamefont {Hongo},
  \citenamefont {Huang}, \citenamefont {Kaminski}, \citenamefont {Stephanov},\
  and\ \citenamefont {Yee}}]{Hongo:2021ona}%
  \BibitemOpen
  \bibfield  {author} {\bibinfo {author} {\bibfnamefont {M.}~\bibnamefont
  {Hongo}}, \bibinfo {author} {\bibfnamefont {X.-G.}\ \bibnamefont {Huang}},
  \bibinfo {author} {\bibfnamefont {M.}~\bibnamefont {Kaminski}}, \bibinfo
  {author} {\bibfnamefont {M.}~\bibnamefont {Stephanov}},\ and\ \bibinfo
  {author} {\bibfnamefont {H.-U.}\ \bibnamefont {Yee}},\ }\href
  {https://doi.org/10.1007/JHEP11(2021)150} {\bibfield  {journal} {\bibinfo
  {journal} {JHEP}\ }\textbf {\bibinfo {volume} {11}},\ \bibinfo {pages}
  {150}},\ \Eprint {https://arxiv.org/abs/2107.14231} {arXiv:2107.14231
  [hep-th]} \BibitemShut {NoStop}%
\bibitem [{\citenamefont {Kumar}\ \emph {et~al.}(2024)\citenamefont {Kumar},
  \citenamefont {Yang},\ and\ \citenamefont {Gubler}}]{Kumar:2023ojl}%
  \BibitemOpen
  \bibfield  {author} {\bibinfo {author} {\bibfnamefont {A.}~\bibnamefont
  {Kumar}}, \bibinfo {author} {\bibfnamefont {D.-L.}\ \bibnamefont {Yang}},\
  and\ \bibinfo {author} {\bibfnamefont {P.}~\bibnamefont {Gubler}},\ }\href
  {https://doi.org/10.1103/PhysRevD.109.054038} {\bibfield  {journal} {\bibinfo
   {journal} {Phys. Rev. D}\ }\textbf {\bibinfo {volume} {109}},\ \bibinfo
  {pages} {054038} (\bibinfo {year} {2024})},\ \Eprint
  {https://arxiv.org/abs/2312.16900} {arXiv:2312.16900 [nucl-th]} \BibitemShut
  {NoStop}%
\bibitem [{\citenamefont {She}\ \emph {et~al.}(2022)\citenamefont {She},
  \citenamefont {Huang}, \citenamefont {Hou},\ and\ \citenamefont
  {Liao}}]{She:2021lhe}%
  \BibitemOpen
  \bibfield  {author} {\bibinfo {author} {\bibfnamefont {D.}~\bibnamefont
  {She}}, \bibinfo {author} {\bibfnamefont {A.}~\bibnamefont {Huang}}, \bibinfo
  {author} {\bibfnamefont {D.}~\bibnamefont {Hou}},\ and\ \bibinfo {author}
  {\bibfnamefont {J.}~\bibnamefont {Liao}},\ }\href
  {https://doi.org/10.1016/j.scib.2022.10.020} {\bibfield  {journal} {\bibinfo
  {journal} {Sci. Bull.}\ }\textbf {\bibinfo {volume} {67}},\ \bibinfo {pages}
  {2265} (\bibinfo {year} {2022})},\ \Eprint {https://arxiv.org/abs/2105.04060}
  {arXiv:2105.04060 [nucl-th]} \BibitemShut {NoStop}%
\bibitem [{\citenamefont {Montenegro}\ \emph {et~al.}(2017)\citenamefont
  {Montenegro}, \citenamefont {Tinti},\ and\ \citenamefont
  {Torrieri}}]{Montenegro:2017rbu}%
  \BibitemOpen
  \bibfield  {author} {\bibinfo {author} {\bibfnamefont {D.}~\bibnamefont
  {Montenegro}}, \bibinfo {author} {\bibfnamefont {L.}~\bibnamefont {Tinti}},\
  and\ \bibinfo {author} {\bibfnamefont {G.}~\bibnamefont {Torrieri}},\ }\href
  {https://doi.org/10.1103/PhysRevD.96.056012} {\bibfield  {journal} {\bibinfo
  {journal} {Phys. Rev. D}\ }\textbf {\bibinfo {volume} {96}},\ \bibinfo
  {pages} {056012} (\bibinfo {year} {2017})},\ \bibinfo {note} {[Addendum:
  Phys.Rev.D 96, 079901 (2017)]},\ \Eprint {https://arxiv.org/abs/1701.08263}
  {arXiv:1701.08263 [hep-th]} \BibitemShut {NoStop}%
\bibitem [{\citenamefont {Torrieri}\ and\ \citenamefont
  {Montenegro}(2023)}]{Torrieri:2022ogj}%
  \BibitemOpen
  \bibfield  {author} {\bibinfo {author} {\bibfnamefont {G.}~\bibnamefont
  {Torrieri}}\ and\ \bibinfo {author} {\bibfnamefont {D.}~\bibnamefont
  {Montenegro}},\ }\href {https://doi.org/10.1103/PhysRevD.107.076010}
  {\bibfield  {journal} {\bibinfo  {journal} {Phys. Rev. D}\ }\textbf {\bibinfo
  {volume} {107}},\ \bibinfo {pages} {076010} (\bibinfo {year} {2023})},\
  \Eprint {https://arxiv.org/abs/2207.00537} {arXiv:2207.00537 [hep-th]}
  \BibitemShut {NoStop}%
\bibitem [{\citenamefont {Florkowski}\ and\ \citenamefont
  {Hontarenko}(2024)}]{Florkowski:2024bfw}%
  \BibitemOpen
  \bibfield  {author} {\bibinfo {author} {\bibfnamefont {W.}~\bibnamefont
  {Florkowski}}\ and\ \bibinfo {author} {\bibfnamefont {M.}~\bibnamefont
  {Hontarenko}},\ }\href@noop {} {\  (\bibinfo {year} {2024})},\ \Eprint
  {https://arxiv.org/abs/2405.03263} {arXiv:2405.03263 [hep-ph]} \BibitemShut
  {NoStop}%
\bibitem [{\citenamefont {Becattini}\ \emph
  {et~al.}(2024{\natexlab{b}})\citenamefont {Becattini}, \citenamefont
  {Daher},\ and\ \citenamefont {Sheng}}]{Becattini:2023ouz}%
  \BibitemOpen
  \bibfield  {author} {\bibinfo {author} {\bibfnamefont {F.}~\bibnamefont
  {Becattini}}, \bibinfo {author} {\bibfnamefont {A.}~\bibnamefont {Daher}},\
  and\ \bibinfo {author} {\bibfnamefont {X.-L.}\ \bibnamefont {Sheng}},\ }\href
  {https://doi.org/10.1016/j.physletb.2024.138533} {\bibfield  {journal}
  {\bibinfo  {journal} {Phys. Lett. B}\ }\textbf {\bibinfo {volume} {850}},\
  \bibinfo {pages} {138533} (\bibinfo {year} {2024}{\natexlab{b}})},\ \Eprint
  {https://arxiv.org/abs/2309.05789} {arXiv:2309.05789 [nucl-th]} \BibitemShut
  {NoStop}%
\bibitem [{\citenamefont {Kiamari}\ \emph {et~al.}(2024)\citenamefont
  {Kiamari}, \citenamefont {Sadooghi},\ and\ \citenamefont
  {Jafari}}]{Kiamari:2023fbe}%
  \BibitemOpen
  \bibfield  {author} {\bibinfo {author} {\bibfnamefont {M.}~\bibnamefont
  {Kiamari}}, \bibinfo {author} {\bibfnamefont {N.}~\bibnamefont {Sadooghi}},\
  and\ \bibinfo {author} {\bibfnamefont {M.~S.}\ \bibnamefont {Jafari}},\
  }\href {https://doi.org/10.1103/PhysRevD.109.036024} {\bibfield  {journal}
  {\bibinfo  {journal} {Phys. Rev. D}\ }\textbf {\bibinfo {volume} {109}},\
  \bibinfo {pages} {036024} (\bibinfo {year} {2024})},\ \Eprint
  {https://arxiv.org/abs/2310.01874} {arXiv:2310.01874 [nucl-th]} \BibitemShut
  {NoStop}%
\bibitem [{\citenamefont {Tiwari}\ and\ \citenamefont
  {Patra}(2024)}]{Tiwari:2024trl}%
  \BibitemOpen
  \bibfield  {author} {\bibinfo {author} {\bibfnamefont {A.}~\bibnamefont
  {Tiwari}}\ and\ \bibinfo {author} {\bibfnamefont {B.~K.}\ \bibnamefont
  {Patra}},\ }\href@noop {} {\  (\bibinfo {year} {2024})},\ \Eprint
  {https://arxiv.org/abs/2408.11514} {arXiv:2408.11514 [hep-th]} \BibitemShut
  {NoStop}%
\bibitem [{\citenamefont {Drogosz}\ \emph
  {et~al.}(2024{\natexlab{a}})\citenamefont {Drogosz}, \citenamefont
  {Florkowski},\ and\ \citenamefont {Hontarenko}}]{Drogosz:2024gzv}%
  \BibitemOpen
  \bibfield  {author} {\bibinfo {author} {\bibfnamefont {Z.}~\bibnamefont
  {Drogosz}}, \bibinfo {author} {\bibfnamefont {W.}~\bibnamefont
  {Florkowski}},\ and\ \bibinfo {author} {\bibfnamefont {M.}~\bibnamefont
  {Hontarenko}},\ }\href {https://doi.org/10.1103/PhysRevD.110.096018}
  {\bibfield  {journal} {\bibinfo  {journal} {Phys. Rev. D}\ }\textbf {\bibinfo
  {volume} {110}},\ \bibinfo {pages} {096018} (\bibinfo {year}
  {2024}{\natexlab{a}})},\ \Eprint {https://arxiv.org/abs/2408.03106}
  {arXiv:2408.03106 [hep-ph]} \BibitemShut {NoStop}%
\bibitem [{\citenamefont {Drogosz}\ \emph
  {et~al.}(2024{\natexlab{b}})\citenamefont {Drogosz}, \citenamefont
  {Florkowski}, \citenamefont {Hontarenko},\ and\ \citenamefont
  {Ryblewski}}]{Drogosz:2024rbd}%
  \BibitemOpen
  \bibfield  {author} {\bibinfo {author} {\bibfnamefont {Z.}~\bibnamefont
  {Drogosz}}, \bibinfo {author} {\bibfnamefont {W.}~\bibnamefont {Florkowski}},
  \bibinfo {author} {\bibfnamefont {M.}~\bibnamefont {Hontarenko}},\ and\
  \bibinfo {author} {\bibfnamefont {R.}~\bibnamefont {Ryblewski}},\ }\href@noop
  {} {\  (\bibinfo {year} {2024}{\natexlab{b}})},\ \Eprint
  {https://arxiv.org/abs/2411.06249} {arXiv:2411.06249 [hep-ph]} \BibitemShut
  {NoStop}%
\bibitem [{\citenamefont {Drogosz}\ \emph
  {et~al.}(2024{\natexlab{c}})\citenamefont {Drogosz}, \citenamefont
  {Florkowski}, \citenamefont {\L{}ygan},\ and\ \citenamefont
  {Ryblewski}}]{Drogosz:2024lkx}%
  \BibitemOpen
  \bibfield  {author} {\bibinfo {author} {\bibfnamefont {Z.}~\bibnamefont
  {Drogosz}}, \bibinfo {author} {\bibfnamefont {W.}~\bibnamefont {Florkowski}},
  \bibinfo {author} {\bibfnamefont {N.}~\bibnamefont {\L{}ygan}},\ and\
  \bibinfo {author} {\bibfnamefont {R.}~\bibnamefont {Ryblewski}},\ }\href@noop
  {} {\  (\bibinfo {year} {2024}{\natexlab{c}})},\ \Eprint
  {https://arxiv.org/abs/2411.06154} {arXiv:2411.06154 [hep-ph]} \BibitemShut
  {NoStop}%
\bibitem [{\citenamefont {Schenke}(2021)}]{Schenke:2021mxx}%
  \BibitemOpen
  \bibfield  {author} {\bibinfo {author} {\bibfnamefont {B.}~\bibnamefont
  {Schenke}},\ }\href {https://doi.org/10.1088/1361-6633/ac14c9} {\bibfield
  {journal} {\bibinfo  {journal} {Rept. Prog. Phys.}\ }\textbf {\bibinfo
  {volume} {84}},\ \bibinfo {pages} {082301} (\bibinfo {year} {2021})},\
  \Eprint {https://arxiv.org/abs/2102.11189} {arXiv:2102.11189 [nucl-th]}
  \BibitemShut {NoStop}%
\bibitem [{\citenamefont {Denicol}\ \emph {et~al.}(2012)\citenamefont
  {Denicol}, \citenamefont {Niemi}, \citenamefont {Molnar},\ and\ \citenamefont
  {Rischke}}]{Denicol:2012cn}%
  \BibitemOpen
  \bibfield  {author} {\bibinfo {author} {\bibfnamefont {G.~S.}\ \bibnamefont
  {Denicol}}, \bibinfo {author} {\bibfnamefont {H.}~\bibnamefont {Niemi}},
  \bibinfo {author} {\bibfnamefont {E.}~\bibnamefont {Molnar}},\ and\ \bibinfo
  {author} {\bibfnamefont {D.~H.}\ \bibnamefont {Rischke}},\ }\href
  {https://doi.org/10.1103/PhysRevD.85.114047} {\bibfield  {journal} {\bibinfo
  {journal} {Phys. Rev. D}\ }\textbf {\bibinfo {volume} {85}},\ \bibinfo
  {pages} {114047} (\bibinfo {year} {2012})},\ \bibinfo {note} {[Erratum:
  Phys.Rev.D 91, 039902 (2015)]},\ \Eprint {https://arxiv.org/abs/1202.4551}
  {arXiv:1202.4551 [nucl-th]} \BibitemShut {NoStop}%
\bibitem [{\citenamefont {Denicol}\ \emph {et~al.}(2014)\citenamefont
  {Denicol}, \citenamefont {Jeon},\ and\ \citenamefont
  {Gale}}]{Denicol:2015transcoeff}%
  \BibitemOpen
  \bibfield  {author} {\bibinfo {author} {\bibfnamefont {G.~S.}\ \bibnamefont
  {Denicol}}, \bibinfo {author} {\bibfnamefont {S.}~\bibnamefont {Jeon}},\ and\
  \bibinfo {author} {\bibfnamefont {C.}~\bibnamefont {Gale}},\ }\href
  {https://doi.org/10.1103/PhysRevC.90.024912} {\bibfield  {journal} {\bibinfo
  {journal} {Phys. Rev. C}\ }\textbf {\bibinfo {volume} {90}},\ \bibinfo
  {pages} {024912} (\bibinfo {year} {2014})}\BibitemShut {NoStop}%
\bibitem [{\citenamefont {Denicol}\ \emph {et~al.}(2018)\citenamefont
  {Denicol}, \citenamefont {Gale}, \citenamefont {Jeon}, \citenamefont
  {Monnai}, \citenamefont {Schenke},\ and\ \citenamefont
  {Shen}}]{Denicol:2018wdp}%
  \BibitemOpen
  \bibfield  {author} {\bibinfo {author} {\bibfnamefont {G.~S.}\ \bibnamefont
  {Denicol}}, \bibinfo {author} {\bibfnamefont {C.}~\bibnamefont {Gale}},
  \bibinfo {author} {\bibfnamefont {S.}~\bibnamefont {Jeon}}, \bibinfo {author}
  {\bibfnamefont {A.}~\bibnamefont {Monnai}}, \bibinfo {author} {\bibfnamefont
  {B.}~\bibnamefont {Schenke}},\ and\ \bibinfo {author} {\bibfnamefont
  {C.}~\bibnamefont {Shen}},\ }\href
  {https://doi.org/10.1103/PhysRevC.98.034916} {\bibfield  {journal} {\bibinfo
  {journal} {Phys. Rev. C}\ }\textbf {\bibinfo {volume} {98}},\ \bibinfo
  {pages} {034916} (\bibinfo {year} {2018})},\ \Eprint
  {https://arxiv.org/abs/1804.10557} {arXiv:1804.10557 [nucl-th]} \BibitemShut
  {NoStop}%
\bibitem [{\citenamefont {Moln\'ar}\ \emph {et~al.}(2014)\citenamefont
  {Moln\'ar}, \citenamefont {Niemi}, \citenamefont {Denicol},\ and\
  \citenamefont {Rischke}}]{Transcoeff2014_p7}%
  \BibitemOpen
  \bibfield  {author} {\bibinfo {author} {\bibfnamefont {E.}~\bibnamefont
  {Moln\'ar}}, \bibinfo {author} {\bibfnamefont {H.}~\bibnamefont {Niemi}},
  \bibinfo {author} {\bibfnamefont {G.~S.}\ \bibnamefont {Denicol}},\ and\
  \bibinfo {author} {\bibfnamefont {D.~H.}\ \bibnamefont {Rischke}},\ }\href
  {https://doi.org/10.1103/PhysRevD.89.074010} {\bibfield  {journal} {\bibinfo
  {journal} {Phys. Rev. D}\ }\textbf {\bibinfo {volume} {89}},\ \bibinfo
  {pages} {074010} (\bibinfo {year} {2014})}\BibitemShut {NoStop}%
\bibitem [{\citenamefont {Florkowski}\ \emph {et~al.}(2022)\citenamefont
  {Florkowski}, \citenamefont {Ryblewski}, \citenamefont {Singh},\ and\
  \citenamefont {Sophys}}]{Florkowski:2021wvk}%
  \BibitemOpen
  \bibfield  {author} {\bibinfo {author} {\bibfnamefont {W.}~\bibnamefont
  {Florkowski}}, \bibinfo {author} {\bibfnamefont {R.}~\bibnamefont
  {Ryblewski}}, \bibinfo {author} {\bibfnamefont {R.}~\bibnamefont {Singh}},\
  and\ \bibinfo {author} {\bibfnamefont {G.}~\bibnamefont {Sophys}},\ }\href
  {https://doi.org/10.1103/PhysRevD.105.054007} {\bibfield  {journal} {\bibinfo
   {journal} {Phys. Rev. D}\ }\textbf {\bibinfo {volume} {105}},\ \bibinfo
  {pages} {054007} (\bibinfo {year} {2022})},\ \Eprint
  {https://arxiv.org/abs/2112.01856} {arXiv:2112.01856 [hep-ph]} \BibitemShut
  {NoStop}%
\bibitem [{\citenamefont {Singh}\ and\ \citenamefont
  {Alam}(2023)}]{Singh2023hdo}%
  \BibitemOpen
  \bibfield  {author} {\bibinfo {author} {\bibfnamefont {S.~K.}\ \bibnamefont
  {Singh}}\ and\ \bibinfo {author} {\bibfnamefont {J.}~\bibnamefont {Alam}},\
  }\href {https://doi.org/10.1140/epjc/s10052-023-11776-5} {\bibfield
  {journal} {\bibinfo  {journal} {The European Physical Journal C}\ }\textbf
  {\bibinfo {volume} {83}},\ \bibinfo {pages} {585} (\bibinfo {year}
  {2023})}\BibitemShut {NoStop}%
\bibitem [{\citenamefont {Karpenko}\ \emph {et~al.}(2014)\citenamefont
  {Karpenko}, \citenamefont {Huovinen},\ and\ \citenamefont
  {Bleicher}}]{Karpenko:2013wva}%
  \BibitemOpen
  \bibfield  {author} {\bibinfo {author} {\bibfnamefont {I.}~\bibnamefont
  {Karpenko}}, \bibinfo {author} {\bibfnamefont {P.}~\bibnamefont {Huovinen}},\
  and\ \bibinfo {author} {\bibfnamefont {M.}~\bibnamefont {Bleicher}},\ }\href
  {https://doi.org/10.1016/j.cpc.2014.07.010} {\bibfield  {journal} {\bibinfo
  {journal} {Comput. Phys. Commun.}\ }\textbf {\bibinfo {volume} {185}},\
  \bibinfo {pages} {3016} (\bibinfo {year} {2014})},\ \Eprint
  {https://arxiv.org/abs/1312.4160} {arXiv:1312.4160 [nucl-th]} \BibitemShut
  {NoStop}%
\bibitem [{\citenamefont {Monnai}\ \emph {et~al.}(2019)\citenamefont {Monnai},
  \citenamefont {Schenke},\ and\ \citenamefont {Shen}}]{Monnai2019eos}%
  \BibitemOpen
  \bibfield  {author} {\bibinfo {author} {\bibfnamefont {A.}~\bibnamefont
  {Monnai}}, \bibinfo {author} {\bibfnamefont {B.}~\bibnamefont {Schenke}},\
  and\ \bibinfo {author} {\bibfnamefont {C.}~\bibnamefont {Shen}},\ }\href
  {https://doi.org/10.1103/PhysRevC.100.024907} {\bibfield  {journal} {\bibinfo
   {journal} {Phys. Rev. C}\ }\textbf {\bibinfo {volume} {100}},\ \bibinfo
  {pages} {024907} (\bibinfo {year} {2019})}\BibitemShut {NoStop}%
\bibitem [{eos()}]{eosweblink}%
  \BibitemOpen
  \href@noop {} {}\bibinfo {howpublished}
  {\url{https://sites.google.com/view/qcdneos/}}\BibitemShut {NoStop}%
\bibitem [{\citenamefont {Shen}\ and\ \citenamefont
  {Alzhrani}(2020)}]{Shen:2020jwv}%
  \BibitemOpen
  \bibfield  {author} {\bibinfo {author} {\bibfnamefont {C.}~\bibnamefont
  {Shen}}\ and\ \bibinfo {author} {\bibfnamefont {S.}~\bibnamefont
  {Alzhrani}},\ }\href {https://doi.org/10.1103/PhysRevC.102.014909} {\bibfield
   {journal} {\bibinfo  {journal} {Phys. Rev. C}\ }\textbf {\bibinfo {volume}
  {102}},\ \bibinfo {pages} {014909} (\bibinfo {year} {2020})},\ \Eprint
  {https://arxiv.org/abs/2003.05852} {arXiv:2003.05852 [nucl-th]} \BibitemShut
  {NoStop}%
\bibitem [{\citenamefont {Ryu}\ \emph {et~al.}(2021)\citenamefont {Ryu},
  \citenamefont {Jupic},\ and\ \citenamefont {Shen}}]{Ryu:2021lnx}%
  \BibitemOpen
  \bibfield  {author} {\bibinfo {author} {\bibfnamefont {S.}~\bibnamefont
  {Ryu}}, \bibinfo {author} {\bibfnamefont {V.}~\bibnamefont {Jupic}},\ and\
  \bibinfo {author} {\bibfnamefont {C.}~\bibnamefont {Shen}},\ }\href
  {https://doi.org/10.1103/PhysRevC.104.054908} {\bibfield  {journal} {\bibinfo
   {journal} {Phys. Rev. C}\ }\textbf {\bibinfo {volume} {104}},\ \bibinfo
  {pages} {054908} (\bibinfo {year} {2021})},\ \Eprint
  {https://arxiv.org/abs/2106.08125} {arXiv:2106.08125 [nucl-th]} \BibitemShut
  {NoStop}%
\bibitem [{\citenamefont {Huovinen}\ and\ \citenamefont
  {Petersen}(2012)}]{Huovinen2012}%
  \BibitemOpen
  \bibfield  {author} {\bibinfo {author} {\bibfnamefont {P.}~\bibnamefont
  {Huovinen}}\ and\ \bibinfo {author} {\bibfnamefont {H.}~\bibnamefont
  {Petersen}},\ }\href {https://doi.org/10.1140/epja/i2012-12171-9} {\bibfield
  {journal} {\bibinfo  {journal} {The European Physical Journal A}\ }\textbf
  {\bibinfo {volume} {48}},\ \bibinfo {pages} {171} (\bibinfo {year}
  {2012})}\BibitemShut {NoStop}%
\bibitem [{\citenamefont {Karpenko}\ \emph {et~al.}(2015)\citenamefont
  {Karpenko}, \citenamefont {Huovinen}, \citenamefont {Petersen},\ and\
  \citenamefont {Bleicher}}]{Karpenko:2015xea}%
  \BibitemOpen
  \bibfield  {author} {\bibinfo {author} {\bibfnamefont {I.~A.}\ \bibnamefont
  {Karpenko}}, \bibinfo {author} {\bibfnamefont {P.}~\bibnamefont {Huovinen}},
  \bibinfo {author} {\bibfnamefont {H.}~\bibnamefont {Petersen}},\ and\
  \bibinfo {author} {\bibfnamefont {M.}~\bibnamefont {Bleicher}},\ }\href
  {https://doi.org/10.1103/PhysRevC.91.064901} {\bibfield  {journal} {\bibinfo
  {journal} {Phys. Rev. C}\ }\textbf {\bibinfo {volume} {91}},\ \bibinfo
  {pages} {064901} (\bibinfo {year} {2015})},\ \Eprint
  {https://arxiv.org/abs/1502.01978} {arXiv:1502.01978 [nucl-th]} \BibitemShut
  {NoStop}%
\bibitem [{\citenamefont {Schäfer}\ \emph {et~al.}(2022)\citenamefont
  {Schäfer}, \citenamefont {Karpenko}, \citenamefont {Wu}, \citenamefont
  {Hammelmann},\ and\ \citenamefont {Elfner}}]{Schafer2022}%
  \BibitemOpen
  \bibfield  {author} {\bibinfo {author} {\bibfnamefont {A.}~\bibnamefont
  {Schäfer}}, \bibinfo {author} {\bibfnamefont {I.}~\bibnamefont {Karpenko}},
  \bibinfo {author} {\bibfnamefont {X.-Y.}\ \bibnamefont {Wu}}, \bibinfo
  {author} {\bibfnamefont {J.}~\bibnamefont {Hammelmann}},\ and\ \bibinfo
  {author} {\bibfnamefont {H.}~\bibnamefont {Elfner}},\ }\href
  {https://doi.org/10.1140/epja/s10050-022-00872-x} {\bibfield  {journal}
  {\bibinfo  {journal} {The European Physical Journal A}\ }\textbf {\bibinfo
  {volume} {58}},\ \bibinfo {pages} {230} (\bibinfo {year} {2022})}\BibitemShut
  {NoStop}%
\bibitem [{sam()}]{samplerweblink}%
  \BibitemOpen
  \href@noop {} {}\bibinfo {howpublished}
  {\url{https://github.com/smash-transport/smash-hadron-sampler}}\BibitemShut
  {NoStop}%
\bibitem [{\citenamefont {Weil}\ \emph {et~al.}(2016)\citenamefont {Weil} \emph
  {et~al.}}]{SMASH2016prc}%
  \BibitemOpen
  \bibfield  {author} {\bibinfo {author} {\bibfnamefont {J.}~\bibnamefont
  {Weil}} \emph {et~al.},\ }\href {https://doi.org/10.1103/PhysRevC.94.054905}
  {\bibfield  {journal} {\bibinfo  {journal} {Phys. Rev. C}\ }\textbf {\bibinfo
  {volume} {94}},\ \bibinfo {pages} {054905} (\bibinfo {year}
  {2016})}\BibitemShut {NoStop}%
\bibitem [{\citenamefont {Wergieluk}\ \emph {et~al.}(2024)\citenamefont
  {Wergieluk} \emph {et~al.}}]{wergieluk_2024_10707746}%
  \BibitemOpen
  \bibfield  {author} {\bibinfo {author} {\bibfnamefont {A.}~\bibnamefont
  {Wergieluk}} \emph {et~al.},\ }\href
  {https://doi.org/10.5281/zenodo.10707746} {\bibinfo {title}
  {smash-transport/smash: Smash-3.1}} (\bibinfo {year} {2024})\BibitemShut
  {NoStop}%
\bibitem [{\citenamefont {Adams}\ \emph {et~al.}(2003)\citenamefont {Adams}
  \emph {et~al.}}]{STAR2003_pT}%
  \BibitemOpen
  \bibfield  {author} {\bibinfo {author} {\bibfnamefont {J.}~\bibnamefont
  {Adams}} \emph {et~al.} (\bibinfo {collaboration} {STAR Collaboration}),\
  }\href {https://doi.org/10.1103/PhysRevLett.91.172302} {\bibfield  {journal}
  {\bibinfo  {journal} {Phys. Rev. Lett.}\ }\textbf {\bibinfo {volume} {91}},\
  \bibinfo {pages} {172302} (\bibinfo {year} {2003})}\BibitemShut {NoStop}%
\bibitem [{\citenamefont {Bearden}\ \emph {et~al.}(2002)\citenamefont {Bearden}
  \emph {et~al.}}]{BRAHMS2002_dNdeta}%
  \BibitemOpen
  \bibfield  {author} {\bibinfo {author} {\bibfnamefont {I.~G.}\ \bibnamefont
  {Bearden}} \emph {et~al.} (\bibinfo {collaboration} {BRAHMS Collaboration}),\
  }\href {https://doi.org/10.1103/PhysRevLett.88.202301} {\bibfield  {journal}
  {\bibinfo  {journal} {Phys. Rev. Lett.}\ }\textbf {\bibinfo {volume} {88}},\
  \bibinfo {pages} {202301} (\bibinfo {year} {2002})}\BibitemShut {NoStop}%
\bibitem [{\citenamefont {Adams}\ \emph {et~al.}(2005)\citenamefont {Adams}
  \emph {et~al.}}]{STAR200_v2}%
  \BibitemOpen
  \bibfield  {author} {\bibinfo {author} {\bibfnamefont {J.}~\bibnamefont
  {Adams}} \emph {et~al.} (\bibinfo {collaboration} {STAR Collaboration and
  STAR-RICH Collaboration}),\ }\href
  {https://doi.org/10.1103/PhysRevC.72.014904} {\bibfield  {journal} {\bibinfo
  {journal} {Phys. Rev. C}\ }\textbf {\bibinfo {volume} {72}},\ \bibinfo
  {pages} {014904} (\bibinfo {year} {2005})}\BibitemShut {NoStop}%
\bibitem [{\citenamefont {Buzzegoli}(2022)}]{Matteo2022glw}%
  \BibitemOpen
  \bibfield  {author} {\bibinfo {author} {\bibfnamefont {M.}~\bibnamefont
  {Buzzegoli}},\ }\href {https://doi.org/10.1103/PhysRevC.105.044907}
  {\bibfield  {journal} {\bibinfo  {journal} {Phys. Rev. C}\ }\textbf {\bibinfo
  {volume} {105}},\ \bibinfo {pages} {044907} (\bibinfo {year}
  {2022})}\BibitemShut {NoStop}%
\bibitem [{\citenamefont {Wagner}(2024)}]{Wagner:2024fry}%
  \BibitemOpen
  \bibfield  {author} {\bibinfo {author} {\bibfnamefont {D.}~\bibnamefont
  {Wagner}},\ }\href@noop {} {\  (\bibinfo {year} {2024})},\ \Eprint
  {https://arxiv.org/abs/2409.07143} {arXiv:2409.07143 [nucl-th]} \BibitemShut
  {NoStop}%
\bibitem [{\citenamefont {Wu}\ \emph {et~al.}(2019)\citenamefont {Wu},
  \citenamefont {Pang}, \citenamefont {Huang},\ and\ \citenamefont
  {Wang}}]{Wu:2019eyi}%
  \BibitemOpen
  \bibfield  {author} {\bibinfo {author} {\bibfnamefont {H.-Z.}\ \bibnamefont
  {Wu}}, \bibinfo {author} {\bibfnamefont {L.-G.}\ \bibnamefont {Pang}},
  \bibinfo {author} {\bibfnamefont {X.-G.}\ \bibnamefont {Huang}},\ and\
  \bibinfo {author} {\bibfnamefont {Q.}~\bibnamefont {Wang}},\ }\href
  {https://doi.org/10.1103/PhysRevResearch.1.033058} {\bibfield  {journal}
  {\bibinfo  {journal} {Phys. Rev. Research.}\ }\textbf {\bibinfo {volume}
  {1}},\ \bibinfo {pages} {033058} (\bibinfo {year} {2019})},\ \Eprint
  {https://arxiv.org/abs/1906.09385} {arXiv:1906.09385 [nucl-th]} \BibitemShut
  {NoStop}%
\bibitem [{\citenamefont {Becattini}\ \emph {et~al.}(2015)\citenamefont
  {Becattini}, \citenamefont {Inghirami}, \citenamefont {Rolando},
  \citenamefont {Beraudo}, \citenamefont {Del~Zanna}, \citenamefont {De~Pace},
  \citenamefont {Nardi}, \citenamefont {Pagliara},\ and\ \citenamefont
  {Chandra}}]{Becattini:2015ska}%
  \BibitemOpen
  \bibfield  {author} {\bibinfo {author} {\bibfnamefont {F.}~\bibnamefont
  {Becattini}}, \bibinfo {author} {\bibfnamefont {G.}~\bibnamefont
  {Inghirami}}, \bibinfo {author} {\bibfnamefont {V.}~\bibnamefont {Rolando}},
  \bibinfo {author} {\bibfnamefont {A.}~\bibnamefont {Beraudo}}, \bibinfo
  {author} {\bibfnamefont {L.}~\bibnamefont {Del~Zanna}}, \bibinfo {author}
  {\bibfnamefont {A.}~\bibnamefont {De~Pace}}, \bibinfo {author} {\bibfnamefont
  {M.}~\bibnamefont {Nardi}}, \bibinfo {author} {\bibfnamefont
  {G.}~\bibnamefont {Pagliara}},\ and\ \bibinfo {author} {\bibfnamefont
  {V.}~\bibnamefont {Chandra}},\ }\href
  {https://doi.org/10.1140/epjc/s10052-015-3624-1} {\bibfield  {journal}
  {\bibinfo  {journal} {Eur. Phys. J. C}\ }\textbf {\bibinfo {volume} {75}},\
  \bibinfo {pages} {406} (\bibinfo {year} {2015})},\ \bibinfo {note} {[Erratum:
  Eur.Phys.J.C 78, 354 (2018)]},\ \Eprint {https://arxiv.org/abs/1501.04468}
  {arXiv:1501.04468 [nucl-th]} \BibitemShut {NoStop}%
\bibitem [{\citenamefont {Chatterjee}\ \emph {et~al.}(2016)\citenamefont
  {Chatterjee}, \citenamefont {Singh}, \citenamefont {Ghosh}, \citenamefont
  {Hasanujjaman}, \citenamefont {Alam},\ and\ \citenamefont
  {Sarkar}}]{CHATTERJEE2016269}%
  \BibitemOpen
  \bibfield  {author} {\bibinfo {author} {\bibfnamefont {S.}~\bibnamefont
  {Chatterjee}}, \bibinfo {author} {\bibfnamefont {S.~K.}\ \bibnamefont
  {Singh}}, \bibinfo {author} {\bibfnamefont {S.}~\bibnamefont {Ghosh}},
  \bibinfo {author} {\bibfnamefont {M.}~\bibnamefont {Hasanujjaman}}, \bibinfo
  {author} {\bibfnamefont {J.}~\bibnamefont {Alam}},\ and\ \bibinfo {author}
  {\bibfnamefont {S.}~\bibnamefont {Sarkar}},\ }\href
  {https://doi.org/https://doi.org/10.1016/j.physletb.2016.05.022} {\bibfield
  {journal} {\bibinfo  {journal} {Physics Letters B}\ }\textbf {\bibinfo
  {volume} {758}},\ \bibinfo {pages} {269} (\bibinfo {year}
  {2016})}\BibitemShut {NoStop}%
\bibitem [{\citenamefont {Abelev}\ \emph {et~al.}(2009)\citenamefont {Abelev}
  \emph {et~al.}}]{STAR2009mult}%
  \BibitemOpen
  \bibfield  {author} {\bibinfo {author} {\bibfnamefont {B.~I.}\ \bibnamefont
  {Abelev}} \emph {et~al.} (\bibinfo {collaboration} {STAR Collaboration}),\
  }\href {https://doi.org/10.1103/PhysRevC.79.034909} {\bibfield  {journal}
  {\bibinfo  {journal} {Phys. Rev. C}\ }\textbf {\bibinfo {volume} {79}},\
  \bibinfo {pages} {034909} (\bibinfo {year} {2009})}\BibitemShut {NoStop}%
\bibitem [{\citenamefont {Bo\ifmmode~\dot{z}\else \.{z}\fi{}ek}\ and\
  \citenamefont {Broniowski}(2013)}]{broniowski2013nbd}%
  \BibitemOpen
  \bibfield  {author} {\bibinfo {author} {\bibfnamefont {P.}~\bibnamefont
  {Bo\ifmmode~\dot{z}\else \.{z}\fi{}ek}}\ and\ \bibinfo {author}
  {\bibfnamefont {W.}~\bibnamefont {Broniowski}},\ }\href
  {https://doi.org/10.1103/PhysRevC.88.014903} {\bibfield  {journal} {\bibinfo
  {journal} {Phys. Rev. C}\ }\textbf {\bibinfo {volume} {88}},\ \bibinfo
  {pages} {014903} (\bibinfo {year} {2013})}\BibitemShut {NoStop}%
\bibitem [{\citenamefont {Harten}\ \emph {et~al.}(1983)\citenamefont {Harten},
  \citenamefont {Lax},\ and\ \citenamefont {Leer}}]{harten1983}%
  \BibitemOpen
  \bibfield  {author} {\bibinfo {author} {\bibfnamefont {A.}~\bibnamefont
  {Harten}}, \bibinfo {author} {\bibfnamefont {P.~D.}\ \bibnamefont {Lax}},\
  and\ \bibinfo {author} {\bibfnamefont {B.~v.}\ \bibnamefont {Leer}},\ }\href
  {https://doi.org/10.1137/1025002} {\bibfield  {journal} {\bibinfo  {journal}
  {SIAM Review}\ }\textbf {\bibinfo {volume} {25}},\ \bibinfo {pages} {35}
  (\bibinfo {year} {1983})},\ \Eprint
  {https://arxiv.org/abs/https://doi.org/10.1137/1025002}
  {https://doi.org/10.1137/1025002} \BibitemShut {NoStop}%
\bibitem [{\citenamefont {Schneider}\ \emph {et~al.}(1993)\citenamefont
  {Schneider}, \citenamefont {Katscher}, \citenamefont {Rischke}, \citenamefont
  {Waldhauser}, \citenamefont {Maruhn},\ and\ \citenamefont
  {Munz}}]{SCHNEIDER199392}%
  \BibitemOpen
  \bibfield  {author} {\bibinfo {author} {\bibfnamefont {V.}~\bibnamefont
  {Schneider}}, \bibinfo {author} {\bibfnamefont {U.}~\bibnamefont {Katscher}},
  \bibinfo {author} {\bibfnamefont {D.}~\bibnamefont {Rischke}}, \bibinfo
  {author} {\bibfnamefont {B.}~\bibnamefont {Waldhauser}}, \bibinfo {author}
  {\bibfnamefont {J.}~\bibnamefont {Maruhn}},\ and\ \bibinfo {author}
  {\bibfnamefont {C.-D.}\ \bibnamefont {Munz}},\ }\href
  {https://doi.org/https://doi.org/10.1006/jcph.1993.1056} {\bibfield
  {journal} {\bibinfo  {journal} {Journal of Computational Physics}\ }\textbf
  {\bibinfo {volume} {105}},\ \bibinfo {pages} {92} (\bibinfo {year}
  {1993})}\BibitemShut {NoStop}%
\bibitem [{\citenamefont {Font}\ \emph {et~al.}(1994)\citenamefont {Font},
  \citenamefont {Ibáñez}, \citenamefont {Marquina},\ and\ \citenamefont
  {Martí}}]{FONT1994_sigvel}%
  \BibitemOpen
  \bibfield  {author} {\bibinfo {author} {\bibfnamefont {J.~A.}\ \bibnamefont
  {Font}}, \bibinfo {author} {\bibfnamefont {J.~M.}\ \bibnamefont {Ibáñez}},
  \bibinfo {author} {\bibfnamefont {A.}~\bibnamefont {Marquina}},\ and\
  \bibinfo {author} {\bibfnamefont {J.~M.}\ \bibnamefont {Martí}},\
  }\href@noop {} {\bibfield  {journal} {\bibinfo  {journal} {Astronomy and
  Astrophysics}\ }\textbf {\bibinfo {volume} {282}},\ \bibinfo {pages} {304}
  (\bibinfo {year} {1994})}\BibitemShut {NoStop}%
\bibitem [{\citenamefont {Rischke}\ \emph {et~al.}(1995)\citenamefont
  {Rischke}, \citenamefont {Bernard},\ and\ \citenamefont
  {Maruhn}}]{RISCHKE1995346}%
  \BibitemOpen
  \bibfield  {author} {\bibinfo {author} {\bibfnamefont {D.~H.}\ \bibnamefont
  {Rischke}}, \bibinfo {author} {\bibfnamefont {S.}~\bibnamefont {Bernard}},\
  and\ \bibinfo {author} {\bibfnamefont {J.~A.}\ \bibnamefont {Maruhn}},\
  }\href {https://doi.org/https://doi.org/10.1016/0375-9474(95)00355-1}
  {\bibfield  {journal} {\bibinfo  {journal} {Nuclear Physics A}\ }\textbf
  {\bibinfo {volume} {595}},\ \bibinfo {pages} {346} (\bibinfo {year}
  {1995})}\BibitemShut {NoStop}%
\end{thebibliography}%

\begin{widetext}

\appendix

\section{Initial condition}
\label{sec:ic}

We use initial conditions defined in Refs.~\cite{Shen:2020jwv,Ryu:2021lnx}, which are based on the Glauber collision geometry with local energy-momentum conservation. However, in contrast to these references, we apply the optical Glauber model to compute thickness functions and wounded nucleon densities. We denote the wounded nucleon densities per unit area in the transverse plane for nuclei $A$ and $B$ as $n_A$ and  $n_B$, respectively. If the impact parameter is $b$, then $n_A$ and $n_B$ are given by:
\begin{align}
n_{A}(x,y)&=T_{A}\left(x-\frac{b}{2},y\right)\left[ 1-\left( 1-\frac{\sigma^{\textrm{}}_{\textrm{NN}}T_{B}(x+b/2,y)}{N_{B}} \right)^{N_{B}}\right],\\
 n_{B}(x,y)&=T_{B}\left(x+\frac{b}{2},y\right)\left[ 1-\left( 1-\frac{\sigma^{\textrm{}}_{\textrm{NN}}T_{A}(x-b/2,y)}{N_{A}} \right)^{N_{A}}\right],
\end{align}
 where $\sigma^{\textrm{}}_{\textrm{NN}}$ denotes the nucleon-nucleon inelastic cross-section, and $N_I$ denotes the total number of nucleons in nucleus $I\in\{A,B\}$. The nuclear thickness functions $T_I(x,y)$ are calculated by integrating the nuclear density $\varrho_I(x,y,z)$ along the longitudinal $z$-axis (the beam direction in the collision) 
\begin{equation}
T_I(x,y) = \int_{-\infty}^{\infty}\varrho_I(x,y,z')\, dz'.
\end{equation}
The nuclear density is assumed to have the Woods-Saxon form 
\begin{equation}
\varrho_{I}(x,y,z) = \frac{\varrho_0}{1+e^{\frac{\sqrt{x^2+y^2+z^2}-R}{\delta}}},
\end{equation}
where $R$ is the nuclear radius and $\delta$ is the surface thickness. For Au+Au collisions considered in this work, the values used are $R=6.37 \text{\,fm}$ and $\delta = 0.535 \text{\,fm}$. The normalization constant $\varrho_0$ is set such that
\begin{equation}
\int \varrho_i(\vec{r})\, d^3\vec{r} = N_I.
\end{equation}
If $y_{\text{beam}} = \cosh^{-1}\left(\frac{\sqrt{s_{\rm NN}}}{2m_N}\right)$ denotes the beam rapidity, then the energy and longitudinal momentum associated with a nucleon are $m_N\cosh (y_{\text{beam}})$ and $m_N\sinh(y_{\text{beam}})$, respectively. Assuming that nuclei $A$ and $B$ move along $z$-axis in positive and negative directions, respectively, the net energy and net longitudinal momentum deposited per unit area in the transverse plane are:
\begin{align*}
\frac{d^2{\cal E}}{dxdy} &= [n_A(x,y) + n_B(x,y)] \ m_N\cosh (y_{\text{beam}}) \equiv M(x,y)  \cosh(Y(x,y)),\\
\frac{d^2{\cal P}_z}{dxdy} &= [n_A(x,y) - n_B(x,y)] \ m_N\cosh (y_{\text{beam}}) \equiv M(x,y) \sinh(Y(x,y)).
\end{align*}
The local invariant mass, $M$, and the local center-of-mass rapidity, $Y$, are defined in terms of $n_A$ and $n_B$ as:
\begin{equation}
M(x,y) = m_N\sqrt{n_A^2+n_B^2+2n_An_B\cosh (2y_{\text{beam}})}\quad , \quad Y(x,y) = \tanh^{-1}\left[ \frac{n_A-n_B}{n_A+n_B}\tanh (y_{\text{beam}})\right].
\end{equation}
To match the deposited energy and momentum with those obtained from the energy-momentum tensor, we follow the ansatz in Ref.~\cite{Ryu:2021lnx} for the energy-momentum tensor at the initial proper time $\tau=\tau_0$ hypersurface:
\begin{align}
T^{\tau\tau}(\tau_0;x,y,\eta_s) &= \varepsilon(\tau_0;x,y,\eta_s)\cosh (f Y), \\
T^{\tau x}(\tau_0;x,y,\eta_s) &=T^{\tau y}(\tau_0;x,y,\eta_s)=0, \\
T^{\tau\eta_s}(\tau_0;x,y,\eta_s) &= \frac{1}{\tau_0}\varepsilon(\tau_0;x,y,\eta_s)\sinh (f Y),
\end{align}
where $f$ is a parameter controlling the fraction of longitudinal momentum converted into fluid velocity. The parameter $f$ is set equal to 0.15. For symmetric collisions, the initial energy density is given by
\begin{equation}
\varepsilon (\tau_0;x,y,\eta_s) = \varepsilon_\perp (x,y) \, F(\eta_s),
\end{equation}
where $F(\eta_s)$ is defined as~\cite{Shen:2020jwv}
\begin{align}
F(\eta_s) &= \exp\Bigg[ -\frac{(|\eta_s -(1-f)Y|-\eta_0)^2}{2\sigma_\eta^2} \Theta(|\eta_s -(1-f)Y|-\eta_0)\Bigg]
\end{align}
with $\Theta(x)$ denoting the Heaviside function.
 
By substituting the above expressions into the matching condition at the initial time, we obtain $\varepsilon_\perp(x,y) =\mathcal{N}_\varepsilon M(x,y)$, where $\mathcal{N}_\varepsilon$ is a normalization constant expressed in terms of $\eta_0$ and $\sigma_\eta$. The choice of initial energy-momentum tensor results in zero transverse velocity, and the $\eta_s$-component of the velocity is determined numerically by solving the implicit equation
$$v^{\eta_s}(\tau_0;x,y,\eta_s) = \frac{T^{\tau\eta_s}(\tau_0;x,y,\eta_s)}{T^{\tau\tau}(\tau_0;x,y,\eta_s)+P_{\rm eq}(\varepsilon,n)}.$$
For the initial baryon density, we consider the following profile 
\begin{equation}
 n (\tau_0;x,y,\eta_s) = \mathcal{N}_n\left[ g_A(\eta_s)n_A(x,y) + g_B(\eta_s)n_B(x,y)\right],
 \end{equation}
 where $g_A(\eta_s)$ and $g_B(\eta_s)$ are given by~\cite{Denicol:2018wdp}:
 \begin{align}
 g_A(\eta_s) &=\Theta (\eta_s-\eta_{B,0})
 \exp\left[-\frac{(\eta_s-\eta_{B,0})^2}{2\sigma _{B,{\texttt{out}}}^2}\right]+\Theta (\eta_{B,0}-\eta_s)\exp \left[-\frac{(\eta_s-\eta_{B,0})^2}{2\sigma_{B,{\texttt{in}}}^2}\right],
 \end{align}
 \begin{align}
 g_B(\eta_s) &= \Theta (\eta_s +\eta_{B,0})\exp \left[  -\frac{(\eta_s+\eta_{B,0})^2}{2\sigma _{B,{\texttt{in}}}^2}\right] + \Theta (-\eta_{B,0}-\eta_s)\exp \left[  -\frac{(\eta_s+\eta_{B,0})^2}{2\sigma _{B,{\texttt{out}}}^2}\right].
 \end{align}
 The normalization constant $\mathcal{N}_n$ is chosen by demanding that
\begin{equation}
\int \, \tau_0 \, dx\, dy\, d\eta_s \ n (\tau_0;x,y,\eta_s) = N_{\text{part}} \quad \Rightarrow \quad  \int \,  d\eta_s \ n (\tau_0;x,y,\eta_s) = \frac{1}{\tau_0}\left[ n_A(x,y)+n_B(x,y)\right],\\
\end{equation}
    where $N_{\text{part}}$ denotes the number of participants. The above condition gives
    \begin{equation}
    \mathcal{N}_n = \frac{1}{\tau_0}\sqrt{\frac{2}{\pi}}\frac{1}{\sigma _{B,{\texttt{in}}}+\sigma _{B,{\texttt{out}}}}.
    \end{equation}
    The parameter values used in the initial condition model are provided in Table~\ref{tab:ic}.
		\begin{table}[t]
		\centering
		\caption{Parameters for the initial condition used in this work. The values are taken from Ref.~\cite{Shen:2020jwv}.}
		\begin{tabular}{c c c c c c c}
			\hline
			$\sqrt{s_{\rm NN}}$ & $\tau_0$ & $\sigma_\eta$ & $\sigma_{B,\text{out}}$ & $\sigma_{B,\text{in}}$ & $\eta_0$ & $\eta_{B,0}$ \\
			(GeV) & (fm) &  & & & & \\
			\hline
			200 & 1.0 & 0.6 & 0.1 & 2.0 & 2.5 & 3.5\\
			\hline
		\end{tabular}
		\label{tab:ic}
	\end{table}
 
\section{Mean impact parameter determination}
\label{sec:impactpar}

To estimate the mean impact parameter across various centralities, we simulate $7\times 10^4$ nuclear collisions within an impact parameter range of 0--15 fm, following the procedure in Ref.~\cite{CHATTERJEE2016269}. First, we sample the positions $(x,y,z)$ of $N_A$ nucleons for nucleus $A$ from a Woods-Saxon distribution using acceptance-rejection sampling. We compute the center of mass coordinates as follows:
$$\bar{x} = \frac{1}{N_A}\sum_{i=1}^{N_A}x_i \qquad , \qquad \bar{y} = \frac{1}{N_A}\sum_{i=1}^{N_A}y_i.$$
Next, we adjust the coordinates of the nucleons such that the center of mass coincides with the coordinate system's origin. We repeat this procedure for nucleus $B$. Afterward, we sample the impact parameter from a linear distribution, $P(b)\propto b$. We then shift the $x$-coordinates of the nucleons in nucleus $A$ by $-b/2$ and those in nucleus $B$ by $b/2$ to the right, so that the distance between the centers of the nuclei equals $b$. 

\begin{figure}[t]
  \includegraphics[width=0.45 \textwidth]{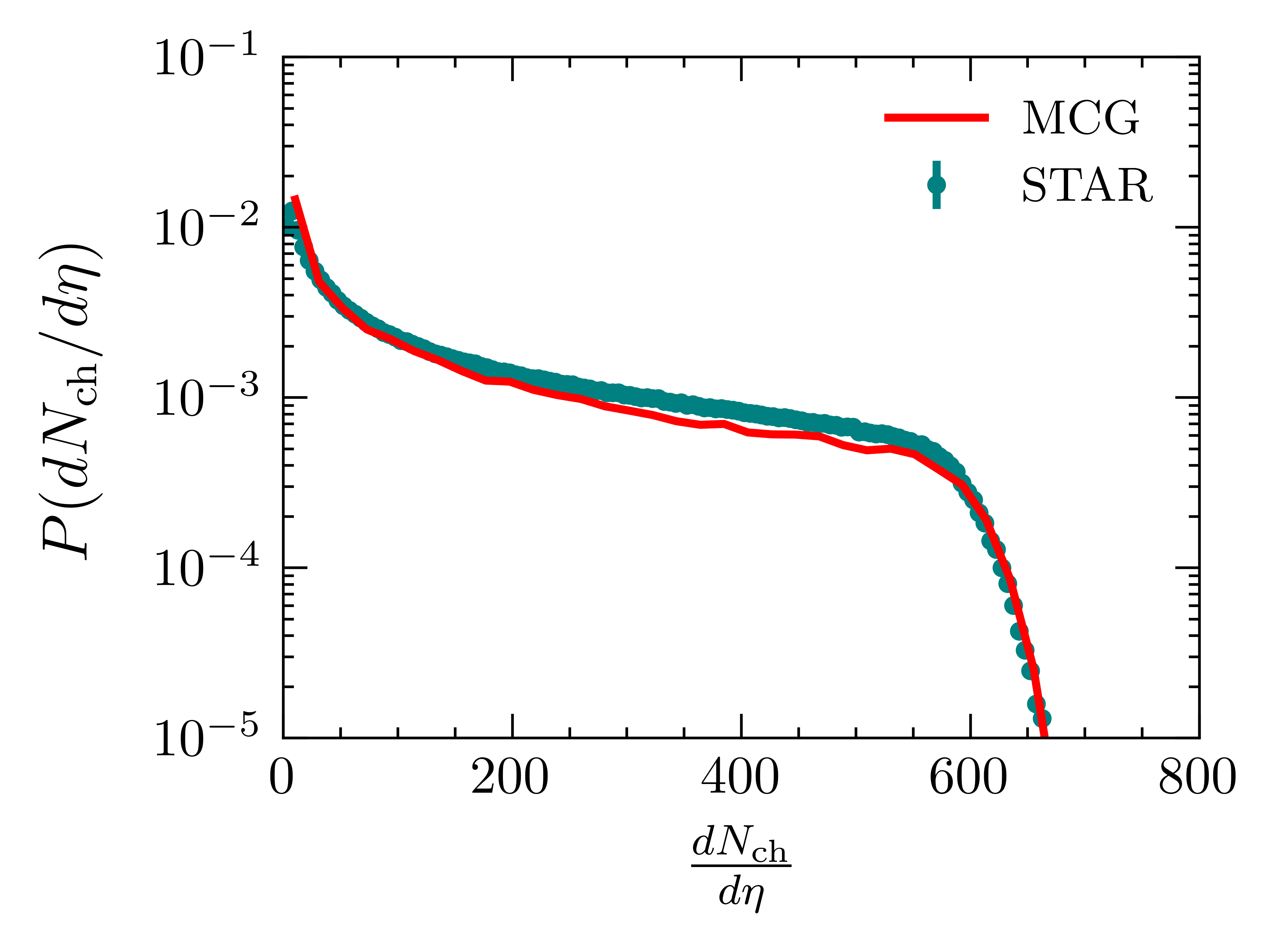}
  \caption{Comparison of the charged particle multiplicity distribution from the two-component Monte-Carlo Glauber model (solid red line) with STAR data~\cite{STAR2009mult} (green symbols).}
  \label{fig:cent}
\end{figure}

To determine binary collisions, we loop over the nucleons in nucleus $A$ and, for each, check whether there is a nucleon in nucleus $B$ within a transverse separation of $\sqrt{\sigma^{\textrm{}}_{\textrm{NN}}/\pi}$. We count all such nucleons from $B$ for each nucleon in $A$, which yields the number of binary collisions associated with each nucleon in $A$. If there is at least one binary collision, the nucleon from $A$ is classified as ``wounded''. We repeat this process for nucleus $B$, then compute the total number of binary collisions ($N_{\text{coll}}$) and participants ($N_{\text{part}}$) by summing the contributions from both $A$ and $B$.

In the two component Monte-Carlo Glauber model, the number of particles produced at mid-rapidity is given by
$$\left.\frac{dN_{ch}}{d\eta}\right|_{\eta=0} = n_{pp}\left[\alpha_h\ N_{\text{coll}}+(1-\alpha_h) \frac{N_{\text{part}}}{2}\right],$$
where $n_{pp}$ is the mean multiplicity per participant pair, and $\alpha_h$ represents the fraction of particle production due to hard scattering processes. To reproduce the tail of the multiplicity distribution, we sample $M= \alpha_h\ N_{\text{coll}}+(1-\alpha_h) \frac{N_{\text{part}}}{2}$ random variables from a negative binomial distribution (NBD) with mean $n_{pp}$ and variance $\sim\frac{1}{k}$~\cite{broniowski2013nbd}. The sampled values from the NBD are summed to obtain the model prediction for charged particle multiplicity. We tune the parameters to fit the distribution from the experiment, as shown in Fig.~\ref{fig:cent}, with the parameter values provided in Table~\ref{tab:mcg}. 

We denote the charged particle multiplicity distribution (normalized to 1) as $P(dN_{\rm ch}/d\eta)$. If  $(dN_{\text{ch}}/d\eta)_{\text{max}}$ denotes the maximum multiplicity, then the centrality class $0-C_1\%$ corresponds to the multiplicity range $(dN_{\text{ch}}/d\eta)_{1}$-$(dN_{\text{ch}}/d\eta)_{\text{max}}$ such that
\[
  \int_{(dN_{\text{ch}}/d\eta)_{1}}^{(dN_{\text{ch}}/d\eta)_{\text{max}}} P(N') \, dN' = \frac{C_1}{100}.
\]
Similarly, multiplicity range $(dN_{\text{ch}}/d\eta)_{2}$-$(dN_{\text{ch}}/d\eta)_{\text{max}}$ for the centrality class $0-C_2\%$, for $C_2>C_1$, is 
\[
  \int_{(dN_{\text{ch}}/d\eta)_{2}}^{(dN_{\text{ch}}/d\eta)_{\text{max}}} P(N') \, dN' = \frac{C_2}{100},
\]
so that the multiplicity range corresponding to $C_1-C_2\%$ class is $(dN_{\text{ch}}/d\eta)_{2}$-$(dN_{\text{ch}}/d\eta)_{1}$ (note that $(dN_{\text{ch}}/d\eta)_{1} > (dN_{\text{ch}}/d\eta)_{2}$). The mean impact parameter in $C_1-C_2\%$ centrality class is computed by averaging the $b$ values of events with multiplicities between $(dN_{\text{ch}}/d\eta)_{2}$-$(dN_{\text{ch}}/d\eta)_{1}$. This exercise is repeated across all centrality classes, ultimately yielding the mean impact parameter as a function of centrality. The values of mean impact parameter for the different centrality classes considered in this work are listed in Table~\ref{tab:cent}.

    \begin{table}[t]
      \centering
      \caption{Parameters for impact parameter determination using the two-component Monte-Carlo Glauber model}
      \begin{tabular}{c c c c c}
    	\hline
    	$\sqrt{s_\textrm{NN}}$ & $\sigma^{\textrm{}}_{\textrm{NN}}$ (mb) & $n_{pp}$ & $\alpha_h$ & $k$\\
    			\hline 
    			200 & 42.0  & 1.80 & 0.11 & 5.0\\
    	\hline
      \end{tabular}
      \label{tab:mcg}
    \end{table}

	\begin{table}[t]
		\centering
		\caption{The values of mean impact parameter for the different centrality classes considered.}
		\begin{tabular}{c c}
			\hline
			 Centrality & $b$ \\
			 (\%) & (fm) \\ 
			  5-10 & 3.8  \\
			  10-20 & 5.6 \\
			  20-30 & 7.2 \\
			 30-40 & 8.6 \\
			  40-50 & 9.7 \\
              20-50 & 8.4 \\
			\hline
		\end{tabular}
		\label{tab:cent}
	\end{table}

\section{Numerical implementation of spin equation of motion }
\label{sec:numspin}
	
The spin equations of motion in arbitrary coordinates can be expressed as:
\beq
D_\alpha S^{\alpha, \beta\gamma}=0,
\eeq
where $D_\alpha$ represents the covariant derivative. 

Expanding the covariant derivative using Christoffel symbols, the above equation becomes
\beq
\label{eq:SconC}
\partial_\alpha S^{\alpha, \beta\gamma} + \Gamma^\alpha_{\rho \alpha}S^{\rho, \beta\gamma}+ \Gamma^\beta_{\delta \alpha}S^{\alpha, \delta\gamma}+ \Gamma^\gamma_{\omega \alpha}S^{\alpha, \beta\omega}= 0.
\eeq
Since the hydrodynamic code is written in the Milne coordinates $(\tau, x,y,\eta_s)$, the only non-vanishing Christoffel symbols are $\Gamma^{\eta_s}_{\mbox{ }\tau \eta_s}=\Gamma^{\eta_s}_{\mbox{ }\eta_s\tau }=\frac{1}{\tau}$ and $\Gamma^{\tau}_{\mbox{ }\eta_s\eta_s }=\tau$. Hence, expressing the Eqs.~(\ref{eq:SconC}) component-wise, we obtain:
%
		\begin{align}
			\partial_\tau (\tau S^{\tau, \tau x}) +\partial_x (\tau S^{x, \tau x}) +\partial_y(\tau S^{y, \tau x})+\partial_{\eta_s}(\tau S^{\eta_s, \tau x})+\tau^2 S^{\eta_s, \eta_s x}=0\label{eqn:spin_appdx_first},\\
			\partial_\tau(\tau S^{\tau, \tau y}) +\partial_x (\tau S^{x, \tau y}) +\partial_y (\tau S^{y, \tau y})+\partial_{\eta_s} (\tau S^{\eta_s, \tau y})+ \tau^2 S^{\eta_s, \eta_s y}=0,\\
			\partial_\tau (\tau S^{\tau, \tau \eta_s}) +\partial_x (\tau S^{x, \tau \eta_s}) +\partial_y (\tau S^{y, \tau \eta_s})+\partial_{\eta_s} ( \tau S^{\eta_s, \tau \eta_s})+ S^{\tau, \tau \eta_s}=0,\\
			\partial_\tau (\tau S^{\tau, xy}) +\partial_x (\tau S^{x, xy} )+\partial_y (\tau S^{y, xy})+\partial_{\eta_s} (\tau S^{\eta_s, xy})=0,\\
			\partial_\tau (\tau S^{\tau, x\eta_s}) +\partial_x (\tau S^{x, x\eta_s}) +\partial_y (\tau S^{y, x\eta_s})+\partial_{\eta_s} (\tau S^{\eta_s, x\eta_s})+ S^{\tau, x\eta_s}+S^{\eta_s, x\tau}=0, \\
			\partial_\tau (\tau S^{\tau, y\eta_s}) +\partial_x (\tau S^{x, y\eta_s}) +\partial_y (\tau S^{y, y\eta_s})+\partial_{\eta_s} (\tau S^{\eta_s, y\eta_s})+ S^{\tau, y\eta_s}+S^{\eta_s, y\tau}=0. \label{eqn:spin_appdx_last}
		\end{align}
%
Following methodology of Ref.~\cite{Karpenko:2013wva}, we define scaled spin current as
	\begin{align*}
		\widetilde{S}^{\alpha,\beta \gamma} &= S^{\alpha,\beta \gamma} \qquad \text{\,\,\,\,\,for } \qquad \alpha,\beta,\gamma \ne \eta_s, \\
		\widetilde{S}^{\alpha,\beta \gamma} &= \tau S^{\alpha,\beta \gamma} \qquad \text{\,\,if any index out of} \ \alpha,\beta, \gamma \text{ is } \eta_s, \\
		\widetilde{S}^{\alpha,\beta \gamma} &= \tau^2 S^{\alpha,\beta \gamma} \qquad \text{if any two indices out of} \ \alpha,\beta, \gamma \text{ is } \eta_s.
	\end{align*}
Defining  $\tilde{\partial}_{\nu}=\left\{ \frac{\partial}{\partial \tau},\frac{\partial}{\partial x},\frac{\partial}{\partial y},\frac{1}{\tau}\frac{\partial}{\partial \eta_s}\right\}$, we may write Eqs.~\eqref{eqn:spin_appdx_first}--\eqref{eqn:spin_appdx_last} as
%
\begin{align}
&\widetilde{\partial}_\tau (\tau \widetilde{S}^{\tau, \tau x}) +\widetilde{\partial}_x (\tau \widetilde{S}^{x, \tau x}) +\widetilde{\partial}_y(\tau \widetilde{S}^{y, \tau x})+\widetilde{\partial}_{\eta_s}(\tau \widetilde{S}^{\eta_s, \tau x})
=\widetilde{S}^{\eta_s, x\eta_s },  \label{eqn:spin_appdx_scaled_first}\\
&\widetilde{\partial}_\tau(\tau \widetilde{S}^{\tau, \tau y}) +\widetilde{\partial}_x (\tau \widetilde{S}^{x, \tau y}) +\widetilde{\partial}_y (\tau \widetilde{S}^{y, \tau y})+\widetilde{\partial}_{\eta_s} (\tau \widetilde{S}^{\eta_s, \tau y})  
=\widetilde{S}^{\eta_s, y\eta_s}, \\
&\widetilde{\partial}_\tau (\tau \widetilde{S}^{\tau, \tau \eta_s}) +\widetilde{\partial}_x (\tau \widetilde{S}^{x, \tau \eta_s}) +\widetilde{\partial}_y (\tau \widetilde{S}^{y, \tau \eta_s})+\widetilde{\partial}_{\eta_s} ( \tau \widetilde{S}^{\eta_s, \tau \eta_s})=0, \\
&\widetilde{\partial}_\tau (\tau \widetilde{S}^{\tau, xy}) +\widetilde{\partial}_x (\tau \widetilde{S}^{x, xy} )+\widetilde{\partial}_y (\tau \widetilde{S}^{y, xy})+\widetilde{\partial}_{\eta_s} (\tau \widetilde{S}^{\eta_s, xy})=0, \\
&\widetilde{\partial}_\tau (\tau \widetilde{S}^{\tau, x\eta_s}) +\widetilde{\partial}_x (\tau \widetilde{S}^{x, x\eta_s}) +\widetilde{\partial}_y (\tau \widetilde{S}^{y, x\eta_s})+\widetilde{\partial}_{\eta_s} (\tau \widetilde{S}^{\eta_s, x\eta_s})
=\widetilde{S}^{\eta_s, \tau x}, \\
&\widetilde{\partial}_\tau (\tau \widetilde{S}^{\tau, y\eta_s}) +\widetilde{\partial}_x (\tau \widetilde{S}^{x, y\eta_s}) +\widetilde{\partial}_y (\tau \widetilde{S}^{y, y\eta_s})+\widetilde{\partial}_{\eta_s} (\tau \widetilde{S}^{\eta_s, y\eta_s})
=\widetilde{S}^{\eta_s, \tau y}. \label{eqn:spin_appdx_scaled_last}
\end{align}
%
It should be noted that, apart from the source terms, the set of equations~\eqref{eqn:spin_appdx_scaled_first}--\eqref{eqn:spin_appdx_scaled_last} is in the conservative form
	\beq
	\label{spineqnnumerical}
	\tilde{\partial}_\tau W +\tilde{\partial_x} F+\tilde{\partial}_y G +\tilde{\partial}_{\eta_s} H = 0.
	\eeq
One can employ any algorithm from the class of Godunov-type algorithms which are designed to solve such equations. Following Ref.~\cite{Karpenko:2013wva}, we choose the relativistic Harten-Lax-van Leer-Einfeldt (HLLE) approximate Riemann solver~\cite{harten1983,SCHNEIDER199392}. The numerical solution of Eqs.~\eqref{eqn:spin_appdx_scaled_first}--\eqref{eqn:spin_appdx_scaled_last} proceeds with the following steps:
\begin{enumerate}
\item We initialize the background hydrodynamics using the initial condition model and the equation of state described in Sec.~\ref{sec:bkgnum} and Appendix~\ref{sec:ic}. If the initial time for spin evolution, $\tau_0^s$, is chosen to coincide with the initial time for background hydrodynamics, $\tau_0$, then the background hydrodynamics is evolved separately for three timesteps, and the data for $\beta^{\mu}=\frac{u^\mu}{T}$ is stored in a data file. These data are then used to compute the rank-2 tensor $D_\mu\beta_\nu$, from which the thermal vorticity $\varpi_{\mu\nu}$ and thermal shear $\xi_{\mu\nu}$ are calculated. We use a forward finite difference scheme for time derivatives and a central finite difference scheme for spatial derivatives.
\item If $\tau_0^s$ is different from the initial time for background hydrodynamics, we evolve the background and spin hydrodynamics simultaneosuly, with initial spin current as 0. Due to the ideal nature of spin equations, the spin current will be zero during this evolution. At $\tau=\tau_0^s$, we compute $\varpi_{\mu\nu}$ and $\xi_{\mu\nu}$ and put the spin polarization tensor, $\omega$, to either thermal vorticity or one of the initial condition models in Eqs.~\eqref{eqn:ic1} and~\eqref{eqn:ic2}.
                 \item We solve the equations on a three dimensional grid $N_x\times N_y\times N_{\eta_s}=201\times 201 \times 121$ in the following range:
        $$(x_{\text{min}},x_{\text{max}}) = (-20,20), \quad (y_{\text{min}},y_{\text{max}}) = (-20,20), \quad (\eta_{s,\text{min}},\eta_{s,\text{max}}) = (-12,12).$$
        Each grid point with coordinates $(x_i,y_j,z_k)$ is treated as a fluid cell identified as $(i,j,k)$. The interface between fluid cells at $(i,j,k)$ and $(i+1,j,k)$ is denoted by $\left(i+\frac{1}{2},j,k\right)$. We choose $\Delta \tau = 0.05$, which satisfies the Courant-Friedrichs-Lewy (CFL) criterion for the stability of the algorithm.
		\item The conserved quantity of a fluid cell is updated by computing the net flux entering the cell independently from each of the three directions. For example, this translates to the following prescription for Eq.~(\ref{spineqnnumerical}):
		\begin{equation}
            \label{eq:updateW}
			W^{n+1}_{ijk} = W^n_{ijk} + \frac{\Delta t}{\Delta x} \left( F^n_{i-\frac{1}{2},jk}-F^n_{i+\frac{1}{2},jk}\right) + \frac{\Delta t}{\Delta y} \left( G^n_{i,j-\frac{1}{2},k}-G^n_{i,j+\frac{1}{2},k}\right) + \frac{\Delta t}{\tau\Delta \eta_s} \left( H^n_{ij,k-\frac{1}{2}}-H^n_{ij,k+\frac{1}{2}}\right).
		\end{equation}
        In case of spin hydrodynamics, the conserved quantities are $\mathbb{S}=(S^{\tau,\tau x},S^{\tau,\tau y},S^{\tau,\tau \eta_s},S^{\tau, xy},S^{\tau,x\eta_s},S^{\tau,y\eta_s })$.
		\item The numerical fluxes in the previous step are computed from the approximate solution of the local Riemann problem, according to HLLE prescription, at the interface between two fluid cells in each direction. The Riemann problem arises due to the reconstruction of the conserved quantities at the left and right boundaries of a fluid cell using the \texttt{minmod} slope limiter. For example, along the $x$-direction after reconstruction, if $W_{i+\frac{1}{2},jk}$ denotes the value at the right boundary of the fluid cell at $(i,j,k)$, and $W_{(i+1)-\frac{1}{2},jk}$ denotes the value at the left boundary of the fluid cell at $(i+1,j,k)$, then the flux that leaves the fluid cell at $(i,j,k)$ and enters the fluid cell at $(i+1,j,k)$ is given by
        $$F_{i+\frac{1}{2},jk}=\frac{b_r F(W_{i+\frac{1}{2},jk})-b_l F(W_{(i+1)-\frac{1}{2},jk})+b_l b_r (W_{(i+1)-\frac{1}{2},jk}-W_{i+\frac{1}{2},jk})}{b_r-b_l},$$
        where $b_l$ and $b_r$ are signal velocities, to be described in Appendix~\ref{sec:signal}.
        \item The computation of fluxes and source terms at each timestep requires a transformation from conservative quantities ($S^{\tau,\mu\nu}$) to primitive quantities ($\omega^{\mu \nu}$). Due to the choice of the spin current, its components can be mapped linearly to the primitive quantities $\varOmega = (\omega^{\tau x},\omega^{\tau y},\omega^{\tau \eta_s},\omega^{ xy},\omega^{x\eta_s},\omega^{y\eta_s })$, for example, the transformation between $S^{\tau,\mu\nu}$ and $\omega^{\mu \nu}$ can be written as
        $$\mathbb{S} = \mathbb{A}\varOmega,$$
        where $\mathbb{A}$ is a $6\times 6$ matrix given by (after omitting explicit $\tau$)
        \begin{align*}
        &\mathbb{A}_{16} = \mathbb{A}_{25} = \mathbb{A}_{34} = \mathbb{A}_{43} = \mathbb{A}_{52} = \mathbb{A}_{61} =0, \quad \mathbb{A}_{11} = (\mathcal{A}_1-\mathcal{A}_3)u^\tau +\frac{\mathcal{A}_2}{2}u^\tau [(u^x)^2-(u^\tau)^2], \\
        &\mathbb{A}_{22} = (\mathcal{A}_1-\mathcal{A}_3)u^\tau +\frac{\mathcal{A}_2}{2}u^\tau [(u^y)^2-(u^\tau)^2], \quad \mathbb{A}_{33} = (\mathcal{A}_1-\mathcal{A}_3)u^\tau +\frac{\mathcal{A}_2}{2}u^\tau [(u^{\eta_s})^2-(u^\tau)^2], \\
        &\mathbb{A}_{44} = \mathcal{A}_1 u^\tau +\frac{\mathcal{A}_2}{2}u^\tau [(u^x)^2+(u^y)^2], \quad \mathbb{A}_{55} = \mathcal{A}_1 u^\tau +\frac{\mathcal{A}_2}{2}u^\tau [(u^x)^2+(u^{\eta_s})^2], \\
        &\mathbb{A}_{66} = \mathcal{A}_1 u^\tau +\frac{\mathcal{A}_2}{2}u^\tau [(u^y)^2+(u^{\eta_s})^2], \\
        &\mathbb{A}_{12} = \mathbb{A}_{21} = \mathbb{A}_{56} = \mathbb{A}_{65} = \frac{\mathcal{A}_2}{2}u^\tau u^x u^y, \quad \mathbb{A}_{13} = \mathbb{A}_{31} = -\mathbb{A}_{46}=-\mathbb{A}_{64}=\frac{\mathcal{A}_2}{2}u^\tau u^x u^{\eta_s}, \\
        &\mathbb{A}_{14} = -\mathbb{A}_{41} = \mathbb{A}_{63}= - \mathbb{A}_{36} = -\frac{\mathcal{A}_3}{2}u^y -\frac{\mathcal{A}_2}{2}(u^\tau)^2 u^y, \quad \mathbb{A}_{15} = -\mathbb{A}_{51}=\mathbb{A}_{26}  =-\mathbb{A}_{62} =-\frac{\mathcal{A}_3}{2}u^{\eta_s} -\frac{\mathcal{A}_2}{2}(u^\tau)^2 u^{\eta_s}, \\
        &\mathbb{A}_{23} = -\mathbb{A}_{32}=\mathbb{A}_{45}  =-\mathbb{A}_{54} = \frac{\mathcal{A}_2}{2}u^\tau u^y u^{\eta_s}, \quad \mathbb{A}_{24} = -\mathbb{A}_{42}=\mathbb{A}_{35}  =-\mathbb{A}_{53} =\frac{\mathcal{A}_3}{2}u^x +\frac{\mathcal{A}_2}{2}(u^\tau)^2 u^x.
        \end{align*}
        \item The primitive quantities are then obtained by inverting the matrix equation in the above step i.e.
        $$\varOmega = \mathbb{A}^{-1}\mathbb{S}.$$
        The values of primitive quantities get very large for fluid cells towards the boundaries due to small temperatures. To keep this under check, we compute $\sqrt{S^{\tau,\mu\nu}\omega_{\mu\nu}}$ after transformation from conserved quantities and whenever its value increases a certain threshold, we re-scale the primitive quantities. We have checked that our results are insensitive towards such a re-scaling, as the fraction of cells whose values gets larger is very small.
        \item Consider the equation
        $$\tilde{\partial}_\tau W +\tilde{\partial_x} F+\tilde{\partial}_y G +\tilde{\partial}_{\eta_s} H = X.$$
        To handle the source terms, we first evolve $W^n$ according to Eq.~(\ref{eq:updateW}) to obtain $W^{*n+1}$. The final value of $W$ is obtained as 
        $$W_{ijk}^{n+1} = W_{ijk}^{*n+1}+\Delta \tau \ X^n_{ijk}. $$
	\end{enumerate}

\begin{figure}[t]
		\includegraphics[width=0.4 \textwidth]{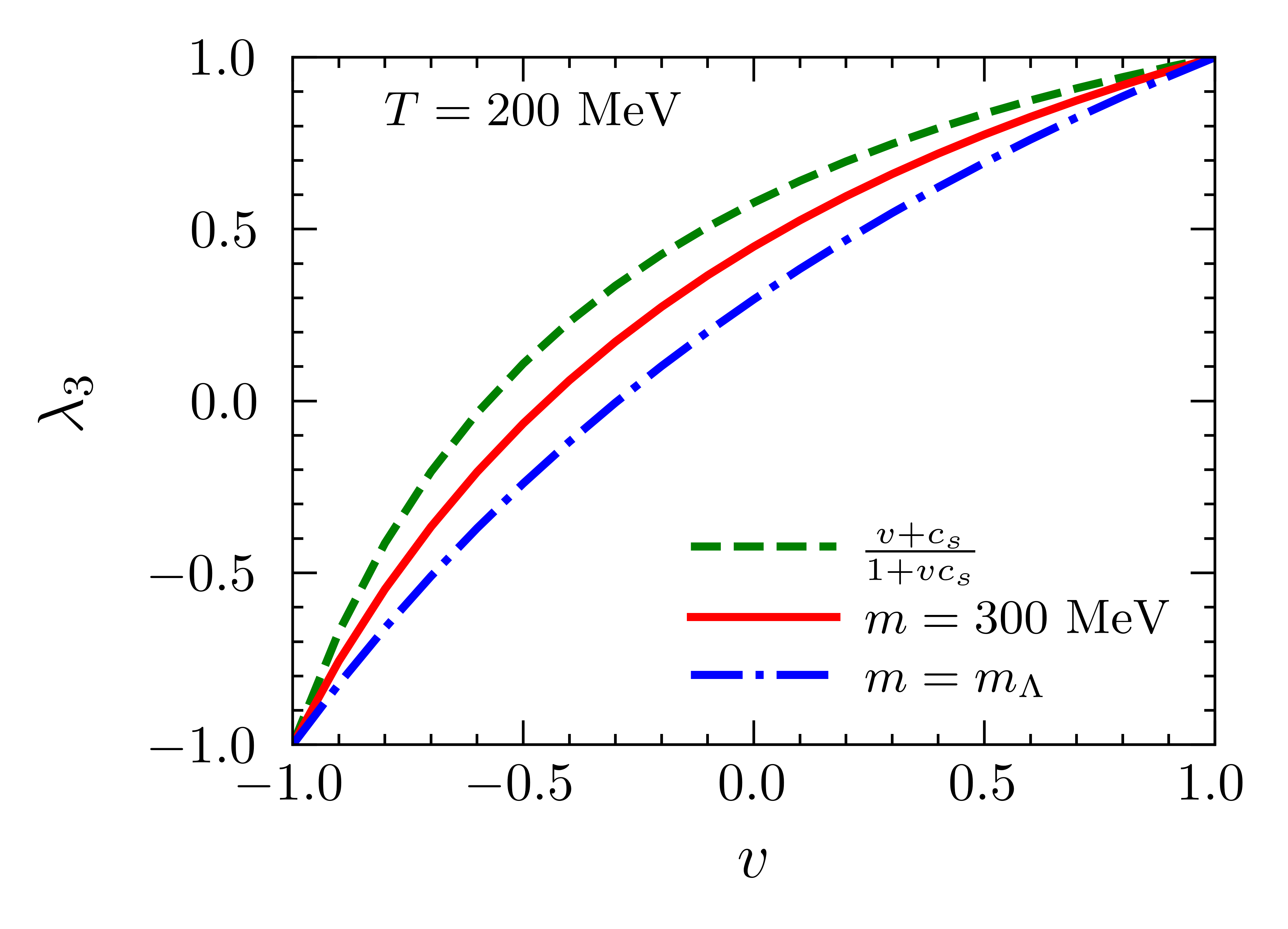}
		\caption{Third eigenvalue for spin hydrodynamics (see text for description) plotted as a function of fluid velocity for different masses (solid red and dash-dotted blue lines) when temperature $T$ is fixed at 200 MeV. The corresponding speed of background hydrodynamics (green dashed line) for constant $c_s=\frac{1}{\sqrt{3}}$ is also shown for comparison.}
\label{fig:signal}
	\end{figure}

    \section{\label{sec:signal} Signal velocity}
    One of the inputs needed to solve the equations of relativistic hydrodynamics using the relativistic HLLE algorithm is the signal velocity, which gives the speed at which information propagates from a point. Signal velocities are often estimated from the eigenvalues of the Jacobian matrix of the flux function~\cite{FONT1994_sigvel} at the interface between two fluid cells. If $u^\mu=(\gamma,\gamma v_x,\gamma v_y,\gamma v_z)$ denotes the four-velocity of a relativistic uncharged fluid with $\gamma = \frac{1}{\sqrt{1-v_x^2-v_y^2-v_z^2}}$ being the Lorentz factor, then the state vector is
    $$\vec{w} = (\varepsilon, v_x, v_y, v_z),$$
    and the fluxes are:
    \begin{align*}
    F^0 &= \left[(\varepsilon+P_{\rm eq})\gamma^2 -P_{\rm eq},(\varepsilon+P_{\rm eq} )\gamma ^2 v_x,(\varepsilon+P_{\rm eq} )\gamma ^2 v_y,(\varepsilon+P_{\rm eq} )\gamma ^2 v_z\right],\\
    F^1 &= \left[(\varepsilon+P_{\rm eq} )\gamma ^2 v_x,(\varepsilon+P_{\rm eq})\gamma^2v_x^2 +P_{\rm eq},(\varepsilon+P_{\rm eq} )\gamma ^2 v_xv_y,(\varepsilon+P_{\rm eq} )\gamma ^2 v_xv_z\right],\\
    F^2 &= \left[(\varepsilon+P_{\rm eq} )\gamma ^2 v_y,(\varepsilon+P_{\rm eq})\gamma^2v_xv_y,(\varepsilon+P_{\rm eq} )\gamma ^2 v_y^2+P_{\rm eq},(\varepsilon+P_{\rm eq} )\gamma ^2 v_yv_z\right],\\
    F^3 &= \left[(\varepsilon+P_{\rm eq} )\gamma ^2 v_z,(\varepsilon+P_{\rm eq})\gamma^2v_xv_z,(\varepsilon+P_{\rm eq} )\gamma ^2 v_yv_z,(\varepsilon+P_{\rm eq} )\gamma ^2 v_z^2+P_{\rm eq}\right].
    \end{align*}
    The Jacobian matrix, $J^\alpha$, is defined as
    $$J^\alpha = \frac{\partial F^\alpha}{\partial\vec{w}}.$$
    The hyperbolic condition for conservation equations is given by the following expression
    $$\text{det}(J^1-\lambda J^0)=0.$$
    Here $\lambda$ denotes the characteristic speeds of different waves. For a fluid flowing along $x$-direction, i.e.,  $v_y=v_z=0$ and $v_x=v \ne 0$, the above equation gives
    \begin{equation}
    \lambda = v,v , \frac{v+c_s}{1+vc_s}, \frac{v-c_s}{1-vc_s}.
    \end{equation}
    In the numerical solution of background hydrodynamics, a convenient choice for the right ($b_r$) and left ($b_l$) signal velocities at an interface is
    \begin{equation}
    \label{eq:bkgsignal}
    b_r = \max \left( 0,\frac{\bar{v}+\bar{c}_s}{1+\bar{v}\bar{c}_s},\frac{v+c_s}{1+vc_s}\right), \quad b_l = \min \left( 0,\frac{\bar{v}-\bar{c}_s}{1-\bar{v}\bar{c}_s},\frac{v-c_s}{1-vc_s}\right),
    \end{equation}
    where $\bar{v}$ is energy-weighted average of the left and right velocities at the interface, and similarly for $\bar{c}_s$~\cite{RISCHKE1995346}. As signal velocities are estimates for speeds of wave propagation, one can also consider constant $c_s=\frac{1}{\sqrt{3}}$ for simplicity~\cite{Karpenko:2013wva}.
    
    A similar analysis is possible for spin hydrodynamics. The state vector is
    $$\vec{w} = \left(\omega^{tx},\omega^{ty},\omega^{tz},\omega^{xy},\omega^{xz},\omega^{yz}\right)$$
    and the fluxes are
    $$F^\alpha = \left(S^{\alpha,tx},S^{\alpha,ty},S^{\alpha,tz},S^{\alpha,xy},S^{\alpha,xz},S^{\alpha,yz}\right).$$
    We compute the matrices $J^\alpha$ and solve the equation
    $$\text{det}(J^1-\lambda J^0)=0,$$
    to obtain characteristic speeds of waves for the case of a fluid flowing along the $x$-direction as
    \begin{equation}
    \lambda = v,v,\frac{v(1-\widetilde{\alpha}^2)+\widetilde{\alpha}(1-v^2)}{1-v^2\widetilde{\alpha}^2},\frac{v(1-\widetilde{\alpha}^2)-\widetilde{\alpha}(1-v^2)}{1-v^2\widetilde{\alpha}^2}\equiv \lambda_1,\lambda_2,\lambda_3,\lambda_4,
    \label{eq:spinsignal}
    \end{equation}
    where $\widetilde{\alpha}$ is given by
    $$\widetilde{\alpha} = \frac{|\mathcal{A}_3|}{2\sqrt{|S_1|S_2}}.$$
    In the above expression, $S_1=\mathcal{A}_1-\frac{\mathcal{A}_2}{2}-\mathcal{A}_3$ and $S_2=\mathcal{A}_1$. The characteristic speeds depend not only on the fluid velocity but also on the temperature and mass of the particles. We plot the third eigenvalue, $\lambda_3$, of Eq.~(\ref{eq:spinsignal}) in Fig.~\ref{fig:signal} for $T=200$ MeV and compare it with the corresponding quantity from background hydrodynamics for constant speed of sound, $c_s^2=\frac{1}{3}$. Since the characteristic speed for spin hydrodynamics is always smaller than that of background hydrodynamics in Eq.~(\ref{eq:bkgsignal}), irrespective of temperature, mass and fluid velocity, we use the same signal velocity as in background hydrodynamics. This choice is made not only for simplicity but also because we use the same grid for solving both spin and background hydrodynamics. The signal velocities in Eq.~(\ref{eq:bkgsignal}), along with the CFL criterion, already ensure that waves from two neighboring fluid cells do not overlap, thus maintaining algorithm stability.

   \begin{figure}[t]
		\includegraphics[width=0.4\textwidth]{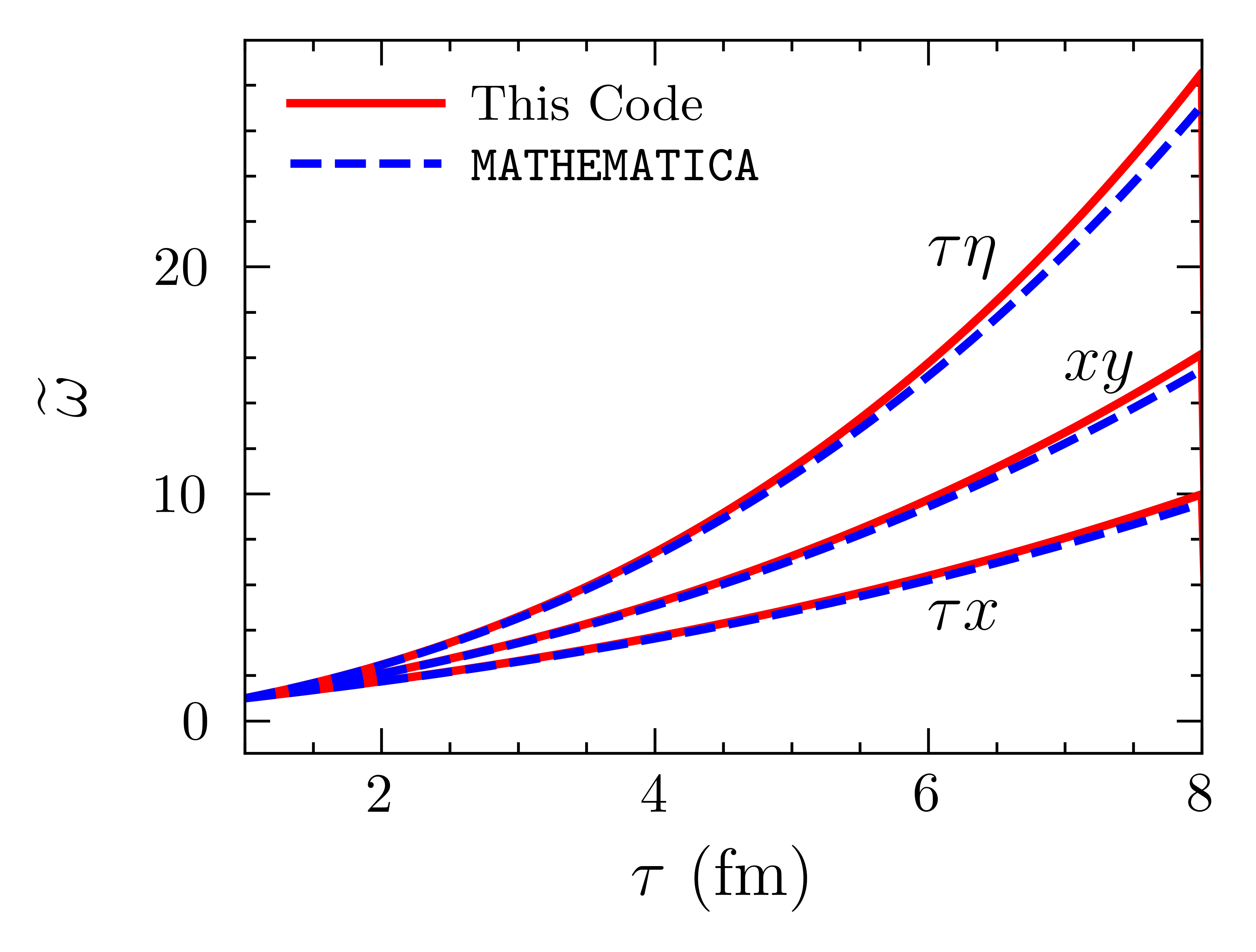}
		\caption{Comparison of code output with $\texttt{MATHEMATICA}$ solution for different components of the spin polarization tensor.}
        \label{fig:matcomp}
	\end{figure}
        
    \section{Test of the code}
    Since no analytical solution is available for relativistic spin hydrodynamics, we perform an indirect test of the code. Assuming boost invariance, we have $u^\tau = 1, \ u^x=u^y=u^{\eta_s}=0$. Under these conditions, Eqs.~(\ref{eqn:spin_appdx_scaled_first})--(\ref{eqn:spin_appdx_scaled_last}) reduce to a simplified form:
    \begin{align*}
    	\tau S_1\dot{\widetilde{\omega}}^{\tau x} + \left(S_1+\tau S'_1 \dot{T}+\frac{\mathcal{A}_3}{2}\right)\widetilde{\omega}^{\tau x} =0, \\
    	\tau S_1\dot{\widetilde{\omega}}^{\tau y} + \left(S_1+\tau S'_1 \dot{T}+\frac{\mathcal{A}_3}{2}\right)\widetilde{\omega}^{\tau y} =0,\\
    	\tau S_1\dot{\widetilde{\omega}}^{\tau \eta_s} + \left(S_1+\tau S'_1 \dot{T}\right)\widetilde{\omega}^{\tau \eta_s} =0, \\
    	\tau \mathcal{A}_1\dot{\widetilde{\omega}}^{x y} + \left(\mathcal{A}_1+\tau \mathcal{A}'_1 \dot{T}\right)\widetilde{\omega}^{x y} =0,\\
   		\tau \mathcal{A}_1\dot{\widetilde{\omega}}^{x\eta_s} + \left(\mathcal{A}_1+\tau \mathcal{A}'_1 \dot{T}-\frac{\mathcal{A}_3}{2}\right)\widetilde{\omega}^{x\eta_s} =0,\\
    	\tau \mathcal{A}_1\dot{\widetilde{\omega}}^{y\eta_s} + \left(\mathcal{A}_1+\tau \mathcal{A}'_1 \dot{T}-\frac{\mathcal{A}_3}{2}\right)\widetilde{\omega}^{y\eta_s} =0.
    \end{align*}
    The primes in above equations denote derivative with respect to $T$. If we further assume Bjorken expansion for the background, the temperature evolution is given by $\dot{T}=-\frac{T}{3\tau}$. These equations can be solved using $\texttt{MATHEMATICA}$'s $\texttt{NDSolve}$ function. We compare the numerical solution from our code with $\texttt{MATHEMATICA}$ solution in Fig.~\ref{fig:matcomp}, and the two solutions show good agreement.
    
\end{widetext}    
\end{document}